# Investigation on uncertain RRab variable stars from ASAS-SN catalogue

BARSANTI CHIARA[1], BONELLI AURORA[1], BRUNO MORGANA[1], IVANTOC VICTORIA[1],
TESTA MARTA[1] AND ZANONI MARTINA[1]
BENNA CARLO[2], GARDIOL DANIELE[2] AND PETTITI GIUSEPPE[2]

1) IIS Curie Vittorini, Corso Allamano 130, 10095, Grugliasco (TO), Italy, TOIS03400P@istruzione.it

2) INAF-Osservatorio Astrofisico di Torino, via Osservatorio 20, I-10025 Pino Torinese (TO), Italy, giuseppe.pettiti@inaf.it

**Abstract:** an analysis of a group of seven variables stars, classed by ASAS-SN as uncertain RRab, is performed comparing their position in a H-R diagram with respect to a sample of variables of the same type built from public astronomical databases.

## 1   Introduction

The All-Sky Automated Survey for Supernovae (ASAS-SN) is a survey managed by Ohio State University's Department of Astronomy; its main goal is the search for supernovae. ASAS-SN has monitored the whole sky since 2014, using telescopes deployed in Hawaii, Chile, Texas, South Africa and China. This program is able to achieve a limiting magnitude of $V \lesssim 17$ and to provide V filter data with a regular interval of 2-3 days. As outcome of this survey, a catalogue of 412,000 variable stars was published (Jayasinghe et al. 2018b). From the ASAS-SN Catalog of Variable Stars: II, we have extracted all the 57 stars classified as uncertain RR Lyrae type a-b (from here on referred to as *Unc. RRab*), with the purpose of verifying the classification proposed by ASAS-SN.

## 2    Data analysis

In order to confirm the nature of the *Unc. RRab* stars, we compared their position in the H-R diagram to the positions of a sample of known RRab type variables (from here on referred to as *Sample group*). We created the *Sample group* starting from the General Catalogue of Variable Stars (GCVS) 5.1, extracting from the catalogue all the stars classed as RRab (Samus et al. 2017). Then we added to the *Sample group* the estimation of the following parameters as available from the European Space Agency (ESA) mission Gaia: effective temperature ($T_{eff}$), interstellar extinction in the G filter ($A_G$), magnitude in the G filter ($G_{mag}$) and the star distance (d). These parameters were downloaded from the Gaia DR2 (Gaia Collaboration, 2018) and Gaia 1.3 billion stars (Bailer-Jones et al. 2018) databases. The initial *Sample group* derived from the GCVS consisted of 5889 stars, but for our analysys we selected only those stars for which $T_{eff}$ and $A_G$ are available. Moreover, we excluded from our sample the variables that are not classed as RRab in the AAVSO database.

Our final *Sample group* of RRab variables was therefore reduced to 1484 stars.

The data of *Unc. RRab* stars extracted from ASAS-SN database and of the *Sample group* are shown in Appendix 1 and 2 respectively.

Only for 7 of 57 *Unc. RRab* stars for which $T_{eff}$ and $A_G$ data are available from Gaia database, shown in Table 1, it was possible to perform a direct comparison with the *Sample group* derived from the GCVS/AAVSO database.

| ASAS-SN Name | R.A. (J2000.0) | Decl. (J2000.0) | $G_{mag}$ | $T_{eff}$ (K) | $A_G$ | d (pc) | $d_{min}$ (pc) | $d_{max}$ (pc) |
|---|---|---|---|---|---|---|---|---|
| ASASSN-V J164146.37+172103.9 | 250.44321 | 17.35108 | 14.877 | 6707 | 1.1673 | 7623.34 | 6403.16 | 9235.60 |
| ASASSN-V J091010.02-680737.9 | 137.54174 | -68.12720 | 13.635 | 6993 | 1.3260 | 3359.43 | 3197.32 | 3538.39 |
| ASASSN-V J111550.35-622141.7 | 168.95979 | -62.36158 | 14.799 | 5250 | 1.0108 | 1311.48 | 1271.79 | 1353.69 |
| ASASSN-V J211849.36+321343.2 | 319.70566 | 32.22866 | 14.309 | 6463 | 1.1390 | 1763.77 | 1696.80 | 1836.13 |
| ASASSN-V J084042.84-455307.7 | 130.17852 | -45.88546 | 12.609 | 5748 | 2.0180 | 1931.96 | 1822.63 | 2054.92 |
| ASASSN-V J075537.08-330027.2 | 118.90449 | -33.00756 | 15.287 | 5143 | 0.7813 | 1278.63 | 1125.34 | 1479.01 |
| ASASSN-V J053827.67-021055.8 | 84.61532 | -2.18220 | 14.901 | 5665 | 1.7165 | 3853.38 | 3362.59 | 4485.78 |

Table 1 - Selected *Unc. RRab* from ASAS-SN

Absolute magnitudes for both the selected *Unc. RRab* from ASAS-SN listed in Table 1 and the *Sample group* were calculated as shown in equation 1:

$$M = m - A + 5 - 5 \cdot Log_{10}(d) \qquad [1]$$

where:
- m is the apparent magnitude in G filter;
- A is the interstellar extinction, in mag, as available from Gaia DR2 catalogue;
- d is the distance, in parsec (pc), as available from Distances to 1.33 billion stars in Gaia DR2 catalogue.

The minimum ($d_{min}$) and the maximum ($d_{max}$) distance of the selected *Unc. RRab* were used to estimate the error on the absolute magnitude.

## 3    Results

The calculated absolute magnitudes in the Gaia G filter ($M_G$) of the selected *Unc. RRab*, together with a 3σ error, are reported in Table 2.

| ASAS-SN Name | $M_G$ |
|---|---|
| ASASSN-V J164146.37+172103.9 | -0.70 ± 0.42 |
| ASASSN-V J091010.02-680737.9 | -0.32 ± 0.11 |
| ASASSN-V J111550.35-622141.7 | 3.20 ± 0.07 |
| ASASSN-V J211849.36+321343.2 | 1.94 ± 0.09 |
| ASASSN-V J084042.84-455307.7 | -0.84 ± 0.13 |
| ASASSN-V J075537.08-330027.2 | 3.97 ± 0.32 |
| ASASSN-V J053827.67-021055.8 | 0.26 ± 0.33 |

Table 2 - Absolute magnitudes of selected *Unc. RRab* from ASAS-SN

The absolute magnitude in the Gaia G filter ($M_G$) vs. the effective temperature ($T_{eff}$) is shown in Figure 1. The blue dots represent the 1484 stars of the *Sample Group* and the red ones are the selected *Unc. RRab* of Table 1. The distances and the absolute magnitudes of the stars of the *Sample Group* are reported in Appendix 2.

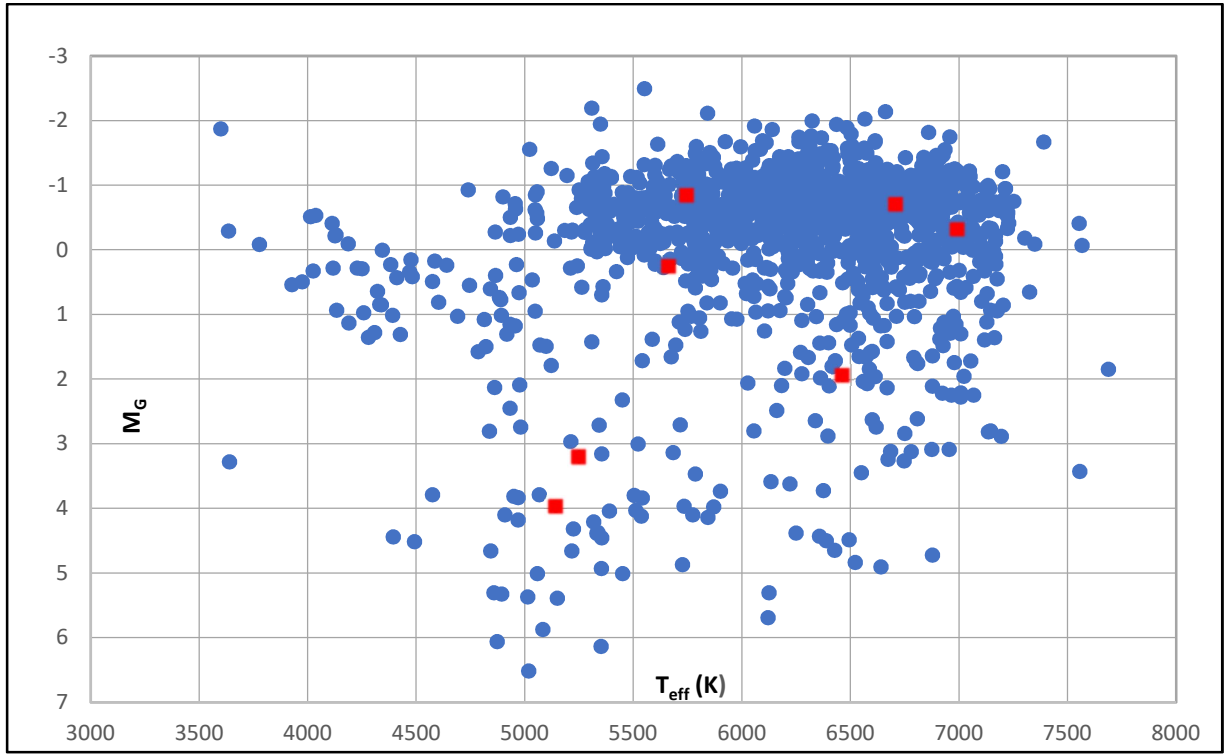

Figure 1 - *Unc. RRab* stars position (red squares) compared to the *Sample group* (blue dots).

The variables of the *Sample group* can be subdivided into three main subgroups, highlighted in Figure 2 and defined by the following rules:
1. $M_G < +1.2$ and $T_{eff} > 5{,}100$ K
2. $M_G < +1.2$ and $T_{eff} < 5{,}100$ K
3. $M_G \geq +1.2$

We assume that only the stars of subgroup 1, that are the 86% of the *Sample Group* and show compatible magnitudes and temperatures, represent the RR Lyrae type ab variables. The position in the H-R diagram of the stars which belong to subgroups 2 and 3 are not compatible with a typical RRab variable. This could be due to an uncorrect estimate of temperature, magnitude or extinction or could indicate that the variable belongs to a different type.

The mean absolute magnitude of the stars of subgroup 1 in the filter G is -0.57 ± 0.58 mag. and brighter than expected. In particular, this result is not consistent with the absolute magnitude in the G band, at metallicity of [Fe/H] = -1.5 dex and period of P = 0.5238 days, $M_G$ = 0.63 ± 0.08 mag. estimated on the basis of Gaia DR2 parallaxes (Muraveva et al. 2018).
The inconsistency we found could be due to an uncorrect estimate of the variable distance or interstellar extinction. At a first look, the values of extinction $A_G$ used in our job are significantly greater and are the most likely cause of the difference in the calculated mean absolute magnitude. A deeper analysys of the causes of this difference will be performed in a future job.

Despite of this offset in the absolute magnitude $M_G$, a reliable comparison of the color-luminosity characteristics of the selected *Unc. RRab* with the *Sample Group* is possible, because the values for extinction applied to the two groups of stars are similar and the offset acts on the absolute magnitude of all the stars with a comparable effect.

Based on their position in the H-R diagram with respect to the *Sample Group*, we therefore classed the seven selected *Unc. RRab* as 'Very Likely', 'Likely', Unlikely' and 'Very Unlikely' RRab variable stars. The outcome of our assessment is provided in Table 3.

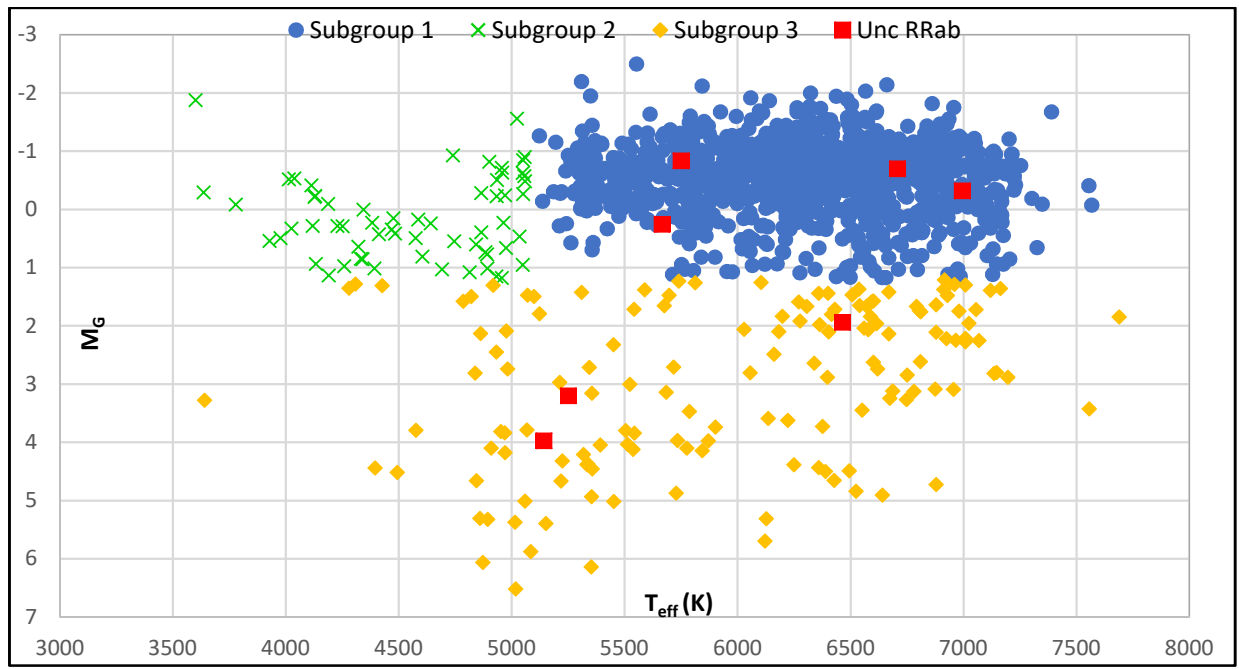

Figure 2 - *Unc. RRab* stars position compared to the ubggroups.

| ASAS-SN Name | $G_{mag}$ | $M_G$ | $T_{eff}$ (K) | RRab type |
|---|---|---|---|---|
| ASASSN-V J164146.37+172103.9 | 14.877 | -0.70 ± 0.42 | 6707 | Very likely |
| ASASSN-V J091010.02-680737.9 | 13.635 | -0.32 ± 0.11 | 6993 | Very likely |
| ASASSN-V J111550.35-622141.7 | 14.799 | 3.20 ± 0.07 | 5250 | Very unlikely |
| ASASSN-V J211849.36+321343.2 | 14.309 | 1.94 ± 0.09 | 6463 | Unlikely |
| ASASSN-V J084042.84-455307.7 | 12.609 | -0.84 ± 0.13 | 5748 | Very likely |
| ASASSN-V J075537.08-330027.2 | 15.287 | 3.97 ± 0.32 | 5143 | Very unlikely |
| ASASSN-V J053827.67-021055.8 | 14.901 | 0.26 ± 0.33 | 5665 | Likely |

Table 3 - Assessment of the selected *Unc. RRab*

The stars we classed as subgroup 2 and 3 are therefore very unlikely members of the RRab Group and further analyses and observations are necessary to confirm if they belong to the RRab or a different type.

## 4 Conclusions

Our analysis of seven RRab defined as uncertain by ASAS-SN catalogue highlighted that four of them show color-luminosity characteristics compatible with the proposed variable type.

The characteristics of the remaining three variables do not match with the typical magnitude-temperature characteristics of the RRab; these stars are unlikey members of this group unless further astrometric, photometric and spectroscopic observations refine their current available observables.

### Acknowledgements
- This activity has made use of the SIMBAD database, operated at CDS, Strasbourg, France.
- This work has made use of data from the European Space Agency (ESA) mission Gaia (https://www.cosmos.esa.int/gaia), processed by the Gaia Data Processing and Analysis Consortium (DPAC, https://www.cosmos.esa.int/web/gaia/dpac/consortium). Funding for the DPAC has been provided by national institutions, in particular the institutions participating in the Gaia Multilateral Agreement.
- We acknowledge with thanks the variable star observations from the *AAVSO International Database* contributed by observers worldwide and used in this research.
- This work was carried out in the context of educational and training activities provided by Italian law 'Alternanza Scuola Lavoro', July 13$^{th}$, 2015 n.107, Art.1, paragraphs 33-43.

# Appendix 1
## Uncertain RRab stars from ASAS-SN database

| # | ASAS-SN Name | R.A. (J2000) | Decl. (J2000) | $G_{mag}$ | $T_{eff}$ |
|---|---|---|---|---|---|
| 1. | ASASSN-V J164146.37+172103.9 | 250.44321 | 17.35108 | 14.877 | 6707 |
| 2. | ASASSN-V J091010.02-680737.9 | 137.54174 | -68.12720 | 13.635 | 6993 |
| 3. | ASASSN-V J135549.31-361025.0 | 208.95546 | -36.17361 | 16.384 | |
| 4. | ASASSN-V J090400.25-201223.3 | 136.00104 | -20.20647 | 17.032 | 6284 |
| 5. | ASASSN-V J170133.53-001945.6 | 255.38971 | -0.32933 | 17.082 | |
| 6. | ASASSN-V J181133.60-254506.5 | 272.89000 | -25.75181 | 17.356 | |
| 7. | ASASSN-V J175351.09-291113.6 | 268.46288 | -29.18711 | 17.726 | |
| 8. | ASASSN-V J175330.28-283525.4 | 268.37617 | -28.59038 | 16.667 | |
| 9. | ASASSN-V J163440.14-735901.7 | 248.66726 | -73.98381 | 12.858 | 5815 |
| 10. | ASASSN-V J175436.10-294849.7 | 268.65042 | -29.81381 | 17.105 | |
| 11. | ASASSN-V J161657.47-095044.1 | 244.23946 | -9.84558 | 17.249 | |
| 12. | ASASSN-V J111550.35-622141.7 | 168.95979 | -62.36158 | 14.799 | 5250 |
| 13. | ASASSN-V J211849.36+321343.2 | 319.70566 | 32.22866 | 14.309 | 6463 |
| 14. | ASASSN-V J174749.97-360441.9 | 266.95821 | -36.07831 | 16.834 | 4475 |
| 15. | ASASSN-V J180036.62-300001.5 | 270.15258 | -30.00042 | 17.131 | |
| 16. | ASASSN-V J105626.60-584800.1 | 164.11082 | -58.80004 | 15.786 | 4825 |
| 17. | ASASSN-V J181013.48-314345.1 | 272.55617 | -31.72919 | 16.588 | 4870 |
| 18. | ASASSN-V J180334.04-295647.4 | 270.89183 | -29.94650 | 16.550 | 5025 |
| 19. | ASASSN-V J181126.90-292319.0 | 272.86208 | -29.38861 | 15.987 | |
| 20. | ASASSN-V J175836.90-290053.2 | 269.65375 | -29.01478 | 16.301 | |
| 21. | ASASSN-V J175948.68-350132.0 | 269.95283 | -35.02556 | 15.966 | 4930 |
| 22. | ASASSN-V J180117.84-345023.6 | 270.32433 | -34.83989 | 16.388 | 5145 |
| 23. | ASASSN-V J161908.22-543725.2 | 244.78423 | -54.62367 | 15.586 | 4758 |
| 24. | ASASSN-V J122627.97-651225.6 | 186.61653 | -65.20712 | 15.351 | 5300 |
| 25. | ASASSN-V J125235.39-622835.0 | 193.14744 | -62.47640 | 13.031 | 5283 |
| 26. | ASASSN-V J175217.19-352820.2 | 268.07162 | -35.47228 | 16.280 | 5096 |
| 27. | ASASSN-V J084042.84-455307.7 | 130.17852 | -45.88546 | 12.609 | 5748 |
| 28. | ASASSN-V J184625.35-061620.1 | 281.60563 | -6.27225 | 14.631 | 4843 |
| 29. | ASASSN-V J075537.08-330027.2 | 118.90449 | -33.00756 | 15.287 | 5143 |
| 30. | ASASSN-V J181200.66-300727.6 | 273.00275 | -30.12433 | 16.248 | 5096 |
| 31. | ASASSN-V J143923.80-610231.2 | 219.84918 | -61.04201 | 15.678 | 4471 |
| 32. | ASASSN-V J182949.84-150254.8 | 277.45767 | -15.04856 | 17.230 | |
| 33. | ASASSN-V J071819.77-153430.3 | 109.58237 | -15.57509 | 13.726 | |
| 34. | ASASSN-V J181219.43-150304.3 | 273.08096 | -15.05119 | 20.1952 | |
| 35. | ASASSN-V J155606.06-173349.0 | 239.02525 | -17.56361 | 17.133 | |
| 36. | ASASSN-V J174823.30-231258.2 | 267.09708 | -23.21617 | | |
| 37. | ASASSN-V J151821.88-082742.2 | 229.59117 | -8.46173 | 12.107 | 6578 |
| 38. | ASASSN-V J064052.26-881519.8 | 100.21776 | -88.25550 | 11.809 | |
| 39. | ASASSN-V J053827.67-021055.8 | 84.61532 | -2.18220 | 14.901 | 5665 |
| 40. | ASASSN-V J155214.80-562325.7 | 238.06167 | -56.39047 | 17.370 | |
| 41. | ASASSN-V J181445.95-245438.0 | 273.69146 | -24.91056 | 16.834 | 4592 |
| 42. | ASASSN-V J174904.66-342536.9 | 267.26942 | -34.42692 | 16.464 | 4450 |
| 43. | ASASSN-V J023903.00+594355.0 | 39.76250 | 59.73194 | | |
| 44. | ASASSN-V J062240.54-523730.7 | 95.66890 | -52.62520 | | |
| 45. | ASASSN-V J194826.16-463921.6 | 297.10900 | -46.65600 | 20.253 | |
| 46. | ASASSN-V J154828.80+280941.8 | 237.12000 | 28.16160 | | |
| 47. | ASASSN-V J180454.99-272928.6 | 271.22913 | -27.49128 | 18.581 | |
| 48. | ASASSN-V J181647.90-240641.8 | 274.19958 | -24.11161 | 18.825 | |
| 49. | ASASSN-V J211011.28+682450.0 | 317.54700 | 68.41390 | 17.737 | |
| 50. | ASASSN-V J175750.16-280145.4 | 269.45900 | -28.02928 | 18.644 | |
| 51. | ASASSN-V J180453.60-293901.3 | 271.22333 | -29.65036 | 18.606 | |
| 52. | ASASSN-V J174112.07-242626.7 | 265.30029 | -24.44075 | 18.298 | |
| 53. | ASASSN-V J174431.49-243335.2 | 266.13121 | -24.55978 | 18.783 | |
| 54. | ASASSN-V J115509.29-615426.3 | 178.78871 | -61.90731 | 18.686 | |
| 55. | ASASSN-V J192606.24-071313.9 | 291.52600 | -7.22052 | 17.114 | |
| 56. | ASASSN-V J230937.09-163318.2 | 347.40455 | -16.55506 | 17.349 | |
| 57. | ASASSN-V J105819.62-604739.0 | 164.58175 | -60.79417 | 17.108 | |

# Appendix 2
## RRab Sample group from GCVS 5.1 and AAVSO database

| # | RRab | R.A. (J2000) | Decl. (J2000) | $G_{mag}$ | $T_{eff}$ | $A_G$ (mag) | d (pc) | $M_G$ |
|---|---|---|---|---|---|---|---|---|
| 1 | GM And | 0.0152606435 | 35.3628004916 | 12.8818 | 6309 | 1.7060 | 3565.24 | -1.58 |
| 2 | BP Tuc | 0.0682413439 | -72.7787507153 | 15.4792 | 6937 | 1.4920 | 6826.31 | -0.18 |
| 3 | V0420 Peg | 0.0758233085 | 19.5487358116 | 15.9644 | 5953 | 1.8265 | 4120.29 | 1.06 |
| 4 | RU Scl | 0.7004463925 | -24.9452991926 | 10.3921 | 6643 | 1.5145 | 905.08 | -0.91 |
| 5 | WW Scl | 1.5081592683 | -36.9041890229 | 13.4751 | 6798 | 1.5350 | 3579.76 | -0.83 |
| 6 | IQ Peg | 1.5237759565 | 29.3201838833 | 16.2099 | 6103 | 1.8357 | 4207.74 | 1.25 |
| 7 | CD Scl | 1.5867779158 | -35.2868604632 | 13.5082 | 7136 | 1.3582 | 3562.99 | -0.61 |
| 8 | RY Psc | 2.9212605278 | -1.7487030798 | 12.3676 | 6688 | 1.6620 | 2153.68 | -0.96 |
| 9 | UX Tuc | 3.5397159225 | -72.6097936996 | 13.9839 | 7092 | 1.4440 | 4397.75 | -0.68 |
| 10 | OV And | 5.1841745376 | 40.8280006504 | 11.3319 | 4745 | 0.6470 | 1066.20 | 0.55 |
| 11 | FI Psc | 5.8450249284 | 13.7613429149 | 13.4774 | 6499 | 1.5067 | 3000.99 | -0.42 |
| 12 | AV Tuc | 6.4064581247 | -73.3192117024 | 16.1553 | 6985 | 1.0045 | 6319.20 | 1.15 |
| 13 | VX Tuc | 6.5911595464 | -59.2284690360 | 15.1096 | 7223 | 1.1790 | 7117.69 | -0.33 |
| 14 | V0492 And | 8.2158488902 | 26.3040737582 | 14.6241 | 6712 | 0.7787 | 3663.45 | 1.03 |
| 15 | V1041 Cas | 8.2257093108 | 47.1470113177 | 12.7749 | 6237 | 1.5885 | 2517.65 | -0.82 |
| 16 | RX Cet | 8.4095012753 | -15.4874455767 | 11.3365 | 6833 | 1.5365 | 1447.55 | -1.00 |
| 17 | FV Cet | 8.4907185064 | -13.5219807897 | 13.5720 | 5927 | 1.8625 | 3041.83 | -0.71 |
| 18 | UY Oct | 8.5678011436 | -85.4776867085 | 15.4627 | 6600 | 1.6507 | 5909.79 | -0.05 |
| 19 | FW Cet | 10.1026499310 | -21.5761979536 | 13.3716 | 6816 | 1.6710 | 3543.76 | -1.05 |
| 20 | AB Hyi | 10.4680784734 | -79.6966683276 | 16.7805 | 6559 | 1.4710 | 4520.38 | 2.03 |
| 21 | IK And | 11.2509590785 | 43.2557864005 | 16.7975 | 6640 | 0.0195 | 2372.85 | 4.90 |
| 22 | FX Cet | 11.2754518065 | -18.9042629984 | 12.8869 | 6601 | 1.6493 | 2234.72 | -0.51 |
| 23 | AC Hyi | 11.9732067233 | -75.2695761404 | 15.3752 | 6124 | 1.8415 | 7004.03 | -0.69 |
| 24 | NR And | 12.0778149518 | 37.3710949523 | 13.9343 | 6346 | 1.6972 | 4795.60 | -1.17 |
| 25 | ZZ And | 12.3951962334 | 27.0221161626 | 13.1185 | 5992 | 1.7650 | 2819.78 | -0.90 |
| 26 | BK Cas | 13.2805952580 | 64.6009106955 | 13.2213 | 4411 | 2.2780 | 1268.81 | 0.43 |
| 27 | AG Tuc | 13.7222368676 | -66.7079865632 | 12.9410 | 6496 | 1.5155 | 2660.90 | -0.70 |
| 28 | IR And | 13.9696094958 | 40.6296393512 | 16.0102 | 6639 | 1.5647 | 4516.75 | 1.17 |
| 29 | UV Scl | 13.9941272836 | -26.3831850562 | 13.4743 | 6937 | 1.3422 | 4497.91 | -1.13 |
| 30 | IT And | 14.3356875898 | 41.6458755338 | 16.6283 | 6276 | 1.8030 | 3817.53 | 1.92 |
| 31 | GI Psc | 14.3924974782 | 7.7815173469 | 12.3321 | 6433 | 1.7073 | 2154.11 | -1.04 |
| 32 | AU Psc | 14.9763770934 | 32.7473722802 | 16.3421 | 6197 | 1.5513 | 3904.65 | 1.83 |
| 33 | GM Psc | 15.4772559996 | 15.9060660211 | 14.3441 | 6341 | 1.6453 | 4427.90 | -0.53 |
| 34 | TY Phe | 16.7546358326 | -41.9568545873 | 16.2338 | 6656 | 1.5615 | 5023.64 | 1.17 |
| 35 | AV Psc | 16.8418066159 | 32.2351469678 | 16.9860 | 6221 | 1.1400 | 2789.73 | 3.62 |
| 36 | AE Scl | 16.8576374062 | -32.3097685979 | 12.5323 | 6364 | 1.6897 | 2511.05 | -1.16 |
| 37 | CS Phe | 17.4560553556 | -44.3148537467 | 13.3102 | 6678 | 1.5688 | 3179.12 | -0.77 |
| 38 | TZ Phe | 17.4931517748 | -42.1288692644 | 12.7968 | 6860 | 1.6540 | 2936.62 | -1.20 |
| 39 | HU Cas | 17.7651755832 | 57.3460887325 | 12.4225 | 5431 | 1.9525 | 1445.42 | -0.33 |
| 40 | GU Cet | 18.4184967230 | 2.1610441196 | 13.6360 | 6871 | 1.5890 | 3848.98 | -0.88 |
| 41 | NX And | 18.9514830137 | 39.7515567075 | 14.3010 | 6892 | 1.6883 | 5077.36 | -0.92 |
| 42 | V0568 Cas | 19.3061509499 | 74.8555237309 | 12.5616 | 4899 | 2.4600 | 1530.58 | -0.82 |
| 43 | XX And | 19.3642262911 | 38.9505690223 | 10.6533 | 6360 | 1.4500 | 1376.22 | -1.49 |
| 44 | RW Scl | 20.7232638435 | -26.2926966550 | 13.7875 | 6815 | 1.4673 | 3539.44 | -0.42 |
| 45 | RX Hyi | 22.9873140424 | -79.3284438474 | 14.5947 | 6637 | 1.3653 | 5884.01 | -0.62 |
| 46 | RR Cet | 23.0340550198 | 1.3417307347 | 9.7496 | 6170 | 1.5310 | 646.16 | -0.83 |
| 47 | HT Psc | 24.4231467053 | 7.0554661996 | 13.4195 | 6359 | 1.5987 | 3082.31 | -0.62 |
| 48 | AT Tri | 24.5264076852 | 33.0100168368 | 14.5064 | 6600 | 1.7325 | 5365.12 | -0.87 |
| 49 | XX Phe | 24.8057801947 | -39.4768681484 | 14.7271 | 6843 | 1.6010 | 5665.09 | -0.64 |
| 50 | OX And | 28.1931134633 | 37.2114253823 | 15.0379 | 6977 | 1.2410 | 4377.96 | 0.59 |
| 51 | GG Cet | 28.2540669552 | -8.0728058537 | 14.5679 | 6979 | 1.1250 | 2187.17 | 1.74 |
| 52 | VV Tri | 30.2076086538 | 35.2328797393 | 15.9877 | 5696 | 1.7297 | 3610.23 | 1.47 |
| 53 | OZ And | 31.7008100903 | 36.4396614962 | 15.8169 | 6020 | 1.9183 | 4411.66 | 0.68 |
| 54 | SS For | 31.9665772254 | -26.8660330213 | 10.2786 | 6377 | 1.5527 | 834.04 | -0.88 |
| 55 | TU Ari | 32.2680347907 | 21.1911913243 | 14.2648 | 6310 | 1.7460 | 3154.97 | 0.02 |
| 56 | V0569 And | 33.6693808081 | 49.8886200597 | 12.4313 | 5769 | 1.6955 | 1758.55 | -0.49 |

| # | RRab | R.A. (J2000) | Decl. (J2000) | $G_{mag}$ | $T_{eff}$ | $A_G$ (mag) | d (pc) | $M_G$ |
|---|---|---|---|---|---|---|---|---|
| 57 | SY Ari | 34.3921123650 | 21.7164768074 | 12.7940 | 5833 | 1.8330 | 2442.31 | -0.98 |
| 58 | LV And | 34.8600001876 | 41.7657381048 | 15.4556 | 6120 | 1.5150 | 3983.18 | 0.94 |
| 59 | MQ And | 36.2779544594 | 43.3684448591 | 15.6658 | 4843 | 0.6840 | 1161.66 | 4.66 |
| 60 | FT And | 37.7132462204 | 38.2501542543 | 16.6582 | 6120 | 0.0250 | 1544.53 | 5.69 |
| 61 | MX And | 38.0129302237 | 42.1417989547 | 16.9655 | 6751 | 1.6587 | 3113.03 | 2.84 |
| 62 | MY And | 38.0395553545 | 43.0797946341 | 15.4680 | 6197 | 1.6320 | 4201.28 | 0.72 |
| 63 | HQ Cet | 38.9828784030 | 2.7747551895 | 13.8325 | 6756 | 1.5430 | 4188.81 | -0.82 |
| 64 | IZ Eri | 41.6210591175 | -13.9509301685 | 14.1867 | 6287 | 1.7060 | 4304.45 | -0.69 |
| 65 | PX Per | 41.8147660779 | 35.9271122122 | 15.8297 | 5899 | 1.8960 | 4197.38 | 0.82 |
| 66 | BK Eri | 42.4827186423 | -1.4200294660 | 12.5657 | 6559 | 1.2650 | 2090.32 | -0.30 |
| 67 | SZ Eri | 42.7874350737 | -20.0813359931 | 15.0875 | 6427 | 0.0743 | 1182.73 | 4.65 |
| 68 | V0454 Per | 46.6848924464 | 38.0377285243 | 16.9754 | 5353 | 0.4537 | 2082.39 | 4.93 |
| 69 | AI Hor | 46.9491503138 | -62.5686934741 | 14.4822 | 6339 | 1.6180 | 4403.81 | -0.35 |
| 70 | X Ari | 47.1286850808 | 10.4458949747 | 9.5826 | 5952 | 2.0043 | 536.25 | -1.07 |
| 71 | V0433 Per | 48.6466350964 | 43.2463635631 | 14.9857 | 5330 | 2.0600 | 4911.14 | -0.53 |
| 72 | UU Hor | 49.6057826387 | -49.6004725232 | 12.4610 | 6532 | 1.7950 | 2710.93 | -1.50 |
| 73 | UV Hor | 50.9483245362 | -48.0221492810 | 13.7554 | 6328 | 1.5380 | 3653.93 | -0.60 |
| 74 | UW Hor | 51.1221372031 | -48.9497470587 | 12.6515 | 6714 | 1.6740 | 2740.12 | -1.21 |
| 75 | W Ret | 51.2634793361 | -63.7793005032 | 15.7578 | 6814 | 1.4070 | 6182.26 | 0.40 |
| 76 | RU Ret | 55.6269624944 | -60.1300026182 | 15.0890 | 6897 | 1.3420 | 4672.51 | 0.40 |
| 77 | TW For | 55.6788009549 | -31.7307160704 | 14.8150 | 7168 | 1.3340 | 5527.34 | -0.23 |
| 78 | LM Eri | 55.7793229456 | -19.4400904182 | 13.5865 | 6714 | 1.6427 | 4199.37 | -1.17 |
| 79 | KL Eri | 56.3558284677 | -8.7926838083 | 13.9743 | 6450 | 1.5987 | 3567.28 | -0.39 |
| 80 | RZ Ret | 57.7564669219 | -61.5577381607 | 15.6298 | 6370 | 1.5117 | 5814.85 | 0.30 |
| 81 | XX Eri | 60.5824269889 | -12.1066451306 | 14.9625 | 6203 | 0.1657 | 6483.81 | 0.74 |
| 82 | XY Eri | 62.8200611948 | -13.8483580517 | 13.0144 | 6396 | 1.5243 | 3075.84 | -0.95 |
| 83 | NN Per | 63.6152367274 | 32.4693234077 | 15.1453 | 5332 | 2.3005 | 4472.44 | -0.41 |
| 84 | V0408 Tau | 63.8713460677 | 26.1259465970 | 14.6307 | 5323 | 1.8983 | 4542.38 | -0.55 |
| 85 | AC Eri | 64.0243459318 | -14.5666040672 | 14.2158 | 6295 | 1.6100 | 3859.05 | -0.33 |
| 86 | BO Tau | 64.1881108520 | 26.3687389858 | 12.9578 | 5582 | 1.9627 | 2074.09 | -0.59 |
| 87 | AG Eri | 65.5680424577 | -15.4274123246 | 14.8667 | 6395 | 1.5260 | 6218.28 | -0.63 |
| 88 | AL Eri | 67.0183551371 | -17.1758312708 | 13.5546 | 6860 | 1.4490 | 6092.12 | -1.82 |
| 89 | BE Eri | 69.5145087580 | -1.9958376100 | 13.0042 | 6367 | 1.6080 | 4233.80 | -1.74 |
| 90 | RX Eri | 72.4345481509 | -15.7411782740 | 9.6608 | 6365 | 1.5670 | 605.38 | -0.82 |
| 91 | U Lep | 74.0749390132 | -21.2172288442 | 10.7004 | 6541 | 1.5277 | 1042.55 | -0.92 |
| 92 | SW Dor | 75.5241579903 | -67.2824231458 | 13.8581 | 6488 | 1.5380 | 3788.22 | -0.57 |
| 93 | V1844 Ori | 75.9034995607 | -0.9991811763 | 15.0621 | 6240 | 1.7495 | 5961.20 | -0.56 |
| 94 | ZZ Dor | 76.3031392443 | -66.6814411903 | 15.5623 | 6746 | 1.4893 | 5380.76 | 0.42 |
| 95 | SZ Dor | 76.7098492093 | -68.8269424605 | 15.1763 | 6667 | 1.5278 | 5885.02 | -0.20 |
| 96 | V1846 Ori | 76.8598316832 | -0.2020116826 | 16.8737 | 7195 | 1.1775 | 3656.11 | 2.88 |
| 97 | SU Col | 76.9460132063 | -33.8651330445 | 12.5163 | 7225 | 1.3060 | 2280.47 | -0.58 |
| 98 | AC Men | 78.9584676428 | -70.6125967791 | 13.6820 | 6944 | 1.3963 | 4382.26 | -0.92 |
| 99 | RT Col | 79.2109192983 | -27.4734673164 | 12.7940 | 6263 | 1.5020 | 2824.60 | -0.96 |
| 100 | TY Dor | 81.0264090900 | -69.4197555681 | 13.8984 | 6437 | 1.6590 | 6870.43 | -1.95 |
| 101 | AO Lep | 81.0610002028 | -14.1008687739 | 12.3028 | 6229 | 1.6135 | 2057.59 | -0.88 |
| 102 | TY Cam | 83.3598624211 | 62.4874757570 | 12.9615 | 5612 | 2.0823 | 3188.42 | -1.64 |
| 103 | BL Col | 84.6267991870 | -35.9055384407 | 13.5729 | 6503 | 1.6105 | 5640.29 | -1.79 |
| 104 | BB Lep | 85.6257247442 | -16.3818260611 | 12.3416 | 6498 | 1.7552 | 1954.75 | -0.87 |
| 105 | BD Lep | 86.4863673531 | -14.6917462050 | 13.0744 | 6346 | 1.6660 | 2629.15 | -0.69 |
| 106 | BE Lep | 87.1584822097 | -19.4915782645 | 13.5488 | 6734 | 1.5160 | 3744.17 | -0.83 |
| 107 | BF Lep | 87.9039502041 | -14.5369334699 | 12.9615 | 5816 | 1.8618 | 2279.01 | -0.69 |
| 108 | AV Col | 89.2108284344 | -27.6670978690 | 12.5046 | 7186 | 1.3210 | 2286.20 | -0.61 |
| 109 | V0575 Aur | 91.8786114081 | 51.1146344392 | 13.9160 | 5475 | 2.1830 | 3339.60 | -0.89 |
| 110 | VW Dor | 91.9404862045 | -66.9774639760 | 11.8176 | 6823 | 1.5820 | 1641.23 | -0.84 |
| 111 | VX Dor | 92.4104535475 | -67.1115191856 | 13.7247 | 6656 | 1.4382 | 4757.58 | -1.10 |
| 112 | AV Men | 94.7827210338 | -78.5856957645 | 13.3701 | 6496 | 1.5240 | 3302.60 | -0.75 |
| 113 | RV Dor | 95.1908682329 | -66.7981509057 | 14.1548 | 6541 | 1.5380 | 4286.00 | -0.54 |
| 114 | RZ Cam | 98.4991036300 | 67.0252839428 | 13.0639 | 6073 | 1.5495 | 2229.67 | -0.23 |
| 115 | V0603 Car | 103.7920519319 | -62.1297680825 | 14.0671 | 6219 | 1.6695 | 4205.28 | -0.72 |
| 116 | V0714 Pup | 105.0026910805 | -37.5420691759 | 12.3424 | 5808 | 1.8133 | 1463.86 | -0.30 |

| # | RRab | R.A. (J2000) | Decl. (J2000) | $G_{mag}$ | $T_{eff}$ | $A_G$ (mag) | d (pc) | $M_G$ |
|---|---|---|---|---|---|---|---|---|
| 117 | MW Gem | 105.7166230708 | 17.5211517421 | 15.4964 | 6214 | 1.8275 | 5105.33 | 0.13 |
| 118 | UW Mon | 105.9155720341 | -0.1918929177 | 13.3331 | 5049 | 2.5620 | 2109.25 | -0.85 |
| 119 | GQ Gem | 107.5117771192 | 14.7847255064 | 14.4913 | 5779 | 1.8370 | 5081.49 | -0.88 |
| 120 | V0387 Gem | 108.0236307845 | 17.3800033946 | 13.6886 | 5924 | 1.8720 | 4991.40 | -1.67 |
| 121 | EP Lyn | 108.0784785522 | 47.3256724881 | 13.1975 | 6413 | 1.1727 | 2540.93 | 0.00 |
| 122 | V0426 Gem | 109.8561589358 | 22.9665365672 | 12.6590 | 6312 | 1.6673 | 2068.02 | -0.59 |
| 123 | DQ CMi | 110.1604109629 | 6.6820784706 | 12.0130 | 6500 | 1.6123 | 1967.05 | -1.07 |
| 124 | RR Gem | 110.3897121330 | 30.8831849599 | 11.4164 | 6020 | 1.6730 | 1392.44 | -0.98 |
| 125 | Y CMi | 111.7052122106 | 1.8963463355 | 13.3028 | 6376 | 0.1293 | 776.26 | 3.72 |
| 126 | EW Cam | 111.8666117221 | 72.7034740429 | 9.4090 | 6801 | 1.6482 | 604.68 | -1.15 |
| 127 | DT CMi | 112.0670459684 | 9.8240082506 | 14.0516 | 6447 | 1.7233 | 4288.89 | -0.83 |
| 128 | VX Lyn | 112.9659474076 | 39.1298272057 | 17.0365 | 6600 | 1.6525 | 3563.73 | 2.62 |
| 129 | HK Pup | 116.1950119695 | -13.0989751918 | 11.2815 | 6095 | 1.6250 | 1373.55 | -1.03 |
| 130 | AT CMi | 117.4041808728 | 1.9613572190 | 16.4445 | 6803 | 1.3360 | 4725.12 | 1.74 |
| 131 | ZZ Lyn | 117.5907495735 | 37.6999677997 | 16.3119 | 6791 | 1.5643 | 4144.52 | 1.66 |
| 132 | V0492 Cam | 120.0181398195 | 67.3792887317 | 14.3384 | 6618 | 1.6337 | 4936.87 | -0.76 |
| 133 | CC Pup | 120.3979202550 | -22.1693171417 | 13.3515 | 5634 | 1.9210 | 2890.90 | -0.87 |
| 134 | EY Lyn | 121.0897639376 | 48.3468161652 | 13.2924 | 6502 | 1.6335 | 3508.23 | -1.07 |
| 135 | Y Cnc | 121.1128332707 | 20.1306474922 | 15.1395 | 6489 | 1.6337 | 4589.94 | 0.20 |
| 136 | SS Cnc | 121.6066045193 | 23.2515637906 | 12.2440 | 6721 | 1.3943 | 1967.65 | -0.62 |
| 137 | DD Hya | 123.1325492365 | 2.8347249297 | 12.3827 | 6558 | 1.6845 | 1929.10 | -0.73 |
| 138 | FN Pup | 123.5801949718 | -20.6289489354 | 15.1929 | 5963 | 1.9700 | 6004.69 | -0.67 |
| 139 | KQ Cnc | 124.1498475567 | 18.3495322107 | 13.8220 | 6955 | 1.1430 | 4825.00 | -0.74 |
| 140 | NS UMa | 126.1029576217 | 65.7176048856 | 11.0796 | 5901 | 1.8013 | 1219.02 | -1.15 |
| 141 | AS Cnc | 126.4256613615 | 25.7190302952 | 12.5748 | 6495 | 1.4193 | 2632.46 | -0.95 |
| 142 | V0481 Hya | 126.4401546615 | -1.9094412534 | 14.4836 | 7150 | 1.5450 | 4513.97 | -0.33 |
| 143 | TT Cnc | 128.2299228136 | 13.1912347358 | 11.2028 | 6647 | 1.6300 | 1294.38 | -0.99 |
| 144 | V0487 Hya | 128.2376143720 | 2.9841369039 | 13.4080 | 6370 | 1.5490 | 3251.29 | -0.70 |
| 145 | ET Hya | 128.7675699439 | -8.8360990646 | 12.0175 | 6569 | 1.4787 | 1822.76 | -0.76 |
| 146 | KW Cnc | 130.1997998836 | 15.4145491730 | 14.3249 | 5522 | 1.5890 | 4333.38 | -0.45 |
| 147 | GL Hya | 130.2467646487 | 2.6228353137 | 13.3865 | 6937 | 1.6270 | 3304.89 | -0.84 |
| 148 | V0496 Hya | 131.1406821062 | -0.2174832088 | 13.4030 | 6501 | 1.5773 | 3210.08 | -0.71 |
| 149 | CQ Cnc | 131.3425659853 | 15.2747555760 | 13.1378 | 6866 | 1.6650 | 3667.55 | -1.35 |
| 150 | AK Lyn | 131.4796055351 | 39.2485911942 | 15.9657 | 6437 | 1.7197 | 4153.35 | 1.15 |
| 151 | EN Lyn | 131.5293290586 | 38.0479698321 | 13.3712 | 6712 | 1.7285 | 3579.57 | -1.13 |
| 152 | SV Vol | 132.1360276232 | -71.6540992299 | 12.1314 | 6017 | 1.8170 | 1773.19 | -0.93 |
| 153 | AL Lyn | 132.3044611979 | 38.8251809336 | 16.3900 | 7023 | 1.4480 | 3962.74 | 1.95 |
| 154 | SV Cnc | 132.5038845767 | 9.9964503473 | 14.6414 | 6827 | 1.4213 | 5208.96 | -0.36 |
| 155 | AF Cnc | 132.6153700852 | 10.7851293658 | 16.3835 | 6955 | 0.9230 | 2982.32 | 3.09 |
| 156 | EZ Cnc | 133.2402785132 | 23.7983898771 | 12.2989 | 6566 | 1.5860 | 1840.45 | -0.61 |
| 157 | DZ Oct | 133.7014380481 | -83.2825577670 | 12.9614 | 6146 | 1.5875 | 2602.96 | -0.70 |
| 158 | GO Hya | 133.7229084274 | 6.4368795899 | 12.2998 | 6774 | 1.7368 | 2160.80 | -1.11 |
| 159 | AI Cnc | 133.8320797984 | 12.3872581256 | 14.7926 | 6214 | 1.6932 | 4420.03 | -0.13 |
| 160 | PR UMa | 134.0065893623 | 62.2458477488 | 14.1184 | 5720 | 1.9040 | 3798.39 | -0.68 |
| 161 | AM Cnc | 134.0617957209 | 11.6223703850 | 14.4259 | 6931 | 1.4803 | 4724.51 | -0.43 |
| 162 | DG Hya | 134.5264978997 | -5.4403294029 | 12.1933 | 6779 | 1.3812 | 2480.19 | -1.16 |
| 163 | V0509 Hya | 134.6766749328 | -0.6163685849 | 17.0127 | 6874 | 1.4070 | 3194.15 | 3.08 |
| 164 | V0510 Hya | 134.9690080714 | -0.1002772325 | 15.5407 | 6343 | 0.7500 | 5655.00 | 1.03 |
| 165 | EY UMa | 135.5865010632 | 49.8192111640 | 15.0603 | 6772 | 1.3388 | 4735.01 | 0.34 |
| 166 | EW Oct | 137.2074155848 | -84.4868138071 | 13.3988 | 5741 | 1.7905 | 2876.31 | -0.69 |
| 167 | LW Car | 138.9089428549 | -69.3396708675 | 14.8049 | 5743 | 1.6390 | 7027.66 | -1.07 |
| 168 | RR Pyx | 139.7326782886 | -26.5605560655 | 13.9090 | 6372 | 1.6715 | 4623.25 | -1.09 |
| 169 | RW Cnc | 139.7751440720 | 29.0654782896 | 11.8787 | 6652 | 1.5520 | 1811.58 | -0.96 |
| 170 | FW Lyn | 139.9645429118 | 33.8733027178 | 13.6553 | 6588 | 1.4015 | 3817.62 | -0.66 |
| 171 | RT Pyx | 140.3400984091 | -26.5519761068 | 15.1940 | 6550 | 1.3800 | 6739.68 | -0.33 |
| 172 | IV Hya | 140.5858424523 | -13.6469345628 | 13.4363 | 6629 | 1.5725 | 3295.91 | -0.73 |
| 173 | AD UMa | 140.9110978872 | 55.7758590393 | 15.7365 | 6795 | 1.6258 | 4131.17 | 1.03 |
| 174 | V0527 Hya | 141.2328261598 | -13.1997860201 | 13.7322 | 7200 | 1.3080 | 4157.32 | -0.67 |
| 175 | V0590 Car | 141.3996783953 | -63.5977811281 | 13.1480 | 5954 | 1.6612 | 2700.52 | -0.67 |
| 176 | V0529 Hya | 141.7571619369 | -0.1595032396 | 15.0662 | 6055 | 1.5975 | 3549.19 | 0.72 |

| # | RRab | R.A. (J2000) | Decl. (J2000) | $G_{mag}$ | $T_{eff}$ | $A_G$ (mag) | d (pc) | $M_G$ |
|---|---|---|---|---|---|---|---|---|
| 177 | FX Lyn | 142.3204569762 | 39.6697893640 | 15.0688 | 6684 | 1.5010 | 4797.70 | 0.16 |
| 178 | WW Leo | 142.6089084808 | 7.2058724439 | 12.5247 | 6258 | 1.5055 | 2572.71 | -1.03 |
| 179 | RT Ant | 143.0635669457 | -25.1997285888 | 14.0519 | 6701 | 1.6230 | 3761.21 | -0.45 |
| 180 | V0531 Hya | 143.2933463400 | -0.1036755360 | 17.0497 | 6619 | 1.6165 | 3456.92 | 2.74 |
| 181 | V0537 Hya | 144.7571663214 | -1.0095741068 | 15.8425 | 7008 | 1.2640 | 2890.48 | 2.27 |
| 182 | HW Leo | 145.8700119860 | 29.4539563345 | 12.1109 | 6684 | 1.6003 | 1617.32 | -0.53 |
| 183 | BK Ant | 146.0617393588 | -39.6614139366 | 12.0369 | 5884 | 1.6670 | 1491.09 | -0.50 |
| 184 | V0689 Car | 146.2888401290 | -63.4572205274 | 12.3698 | 5859 | 1.8755 | 1968.57 | -0.98 |
| 185 | AD Sex | 147.2278065687 | -0.9632069398 | 15.2342 | 6869 | 1.2980 | 4563.05 | 0.64 |
| 186 | V0691 Car | 147.3541164296 | -72.7506205133 | 14.3038 | 5794 | 2.0780 | 4325.38 | -0.95 |
| 187 | QY UMa | 147.4788325101 | 51.7395889680 | 13.0432 | 7153 | 1.2075 | 3259.62 | -0.73 |
| 188 | ST Sex | 148.0844879270 | 2.9135361556 | 15.9699 | 6430 | 1.4350 | 3669.03 | 1.71 |
| 189 | V0694 Car | 148.6646787867 | -68.0656893637 | 13.7352 | 6210 | 1.6110 | 3801.08 | -0.78 |
| 190 | V0540 Hya | 149.0449013934 | -17.0021830650 | 13.7702 | 6845 | 1.5057 | 5037.20 | -1.25 |
| 191 | AB LMi | 151.1516525538 | 31.8804049141 | 13.3668 | 6124 | 1.6835 | 3382.20 | -0.96 |
| 192 | X LMi | 151.5280325616 | 39.3579086131 | 12.3897 | 6571 | 1.4760 | 2083.14 | -0.68 |
| 193 | DI Leo | 151.8577371345 | 13.9495371362 | 13.6363 | 7142 | 1.3637 | 4407.99 | -0.95 |
| 194 | RR Leo | 151.9310853209 | 23.9917561623 | 10.8942 | 6614 | 1.6055 | 964.47 | -0.63 |
| 195 | HO Leo | 152.0579855841 | 26.3757183263 | 13.4941 | 6681 | 1.3275 | 3987.78 | -0.84 |
| 196 | V0420 Vel | 153.0612232252 | -46.1697589086 | 10.9884 | 4469 | 0.5080 | 1066.57 | 0.34 |
| 197 | DM Leo | 153.3327208568 | 11.1026118243 | 13.8452 | 6060 | 1.4093 | 5149.30 | -1.12 |
| 198 | WZ Hya | 153.3505325280 | -13.1381655992 | 10.8597 | 6576 | 1.6890 | 942.21 | -0.70 |
| 199 | II Leo | 153.6423060453 | 6.5570874451 | 14.3357 | 7030 | 0.8210 | 5213.24 | -0.07 |
| 200 | AR Sex | 153.9454573958 | -0.2102739016 | 13.6637 | 6050 | 1.6850 | 4721.82 | -1.39 |
| 201 | Y LMi | 153.9645547635 | 32.8592259302 | 12.6857 | 6960 | 1.3333 | 2173.77 | -0.33 |
| 202 | WY Ant | 154.0206077868 | -29.7284487756 | 10.7732 | 6512 | 1.1417 | 1066.44 | -0.51 |
| 203 | V0513 Cam | 154.8881627303 | 83.6034784300 | 14.2909 | 5757 | 1.7645 | 4253.67 | -0.62 |
| 204 | AX Sex | 155.0807379100 | -1.0253034739 | 15.0933 | 6931 | 1.2387 | 5538.60 | 0.14 |
| 205 | RV Leo | 155.9675180980 | 9.7570735620 | 14.0114 | 7048 | 1.5675 | 4025.53 | -0.58 |
| 206 | V0543 Hya | 156.5349862289 | -23.2538790670 | 13.3583 | 6407 | 1.6037 | 3400.57 | -0.90 |
| 207 | RV Cha | 157.3383118094 | -81.1414928286 | 13.5038 | 6276 | 1.8880 | 3953.10 | -1.37 |
| 208 | HU Car | 157.7733727406 | -62.3925979618 | 14.0258 | 4956 | 2.3965 | 2833.80 | -0.63 |
| 209 | V0339 UMa | 157.8820191918 | 50.2499187697 | 13.7809 | 6845 | 1.2270 | 4102.52 | -0.51 |
| 210 | BM Sex | 159.3823966730 | -0.6644923416 | 17.0203 | 6126 | 0.1037 | 2099.82 | 5.31 |
| 211 | BO Sex | 160.6651001823 | -1.2744805379 | 16.6662 | 7556 | 0.6702 | 3267.44 | 3.42 |
| 212 | IO Leo | 160.7606285350 | 6.5794820898 | 14.3178 | 6732 | 1.3560 | 4870.98 | -0.48 |
| 213 | V0546 Hya | 161.0283923285 | -18.8774064488 | 13.3896 | 6714 | 1.6900 | 3289.42 | -0.89 |
| 214 | BQ Sex | 161.4816774908 | -7.3599546486 | 13.4745 | 6636 | 1.5015 | 3364.46 | -0.66 |
| 215 | EE Oct | 161.8621707125 | -84.1396446317 | 13.3264 | 5806 | 1.6812 | 2528.92 | -0.37 |
| 216 | BS Sex | 162.0020749562 | -5.5090547581 | 13.1308 | 6930 | 1.3010 | 3582.25 | -0.94 |
| 217 | SW Leo | 163.9812897296 | -2.9823146746 | 13.8891 | 6663 | 1.6275 | 3111.67 | -0.20 |
| 218 | TT Cha | 164.3991827054 | -80.4815201480 | 15.6328 | 5852 | 1.6730 | 8666.41 | -0.73 |
| 219 | KT UMa | 164.5308036778 | 56.1192048477 | 11.2883 | 6534 | 1.6760 | 1427.37 | -1.16 |
| 220 | TX Car | 164.7419835556 | -59.0825288883 | 12.5829 | 5948 | 1.8685 | 2198.07 | -1.00 |
| 221 | HQ Leo | 164.9184331267 | 6.6251793576 | 13.9869 | 6325 | 1.2392 | 5545.16 | -0.97 |
| 222 | AX Cen | 166.3898050949 | -54.9645940875 | 13.9748 | 5713 | 1.7190 | 3655.11 | -0.56 |
| 223 | IZ Leo | 166.3983624865 | 15.6363798288 | 13.2074 | 6248 | 1.7665 | 3334.49 | -1.17 |
| 224 | UW Cha | 166.4662667131 | -80.1442405185 | 14.9480 | 5804 | 1.8747 | 5413.37 | -0.59 |
| 225 | KK Leo | 167.1593878019 | -0.0872847644 | 14.9393 | 6719 | 1.0883 | 4510.49 | 0.58 |
| 226 | KL Leo | 167.3508607097 | -0.7534834608 | 16.3897 | 6402 | 1.2275 | 4094.11 | 2.10 |
| 227 | V0590 Cen | 168.1048642251 | -36.8795812405 | 12.2331 | 5753 | 1.9180 | 1834.34 | -1.00 |
| 228 | PV Car | 168.1319106763 | -75.1064768071 | 15.8215 | 5242 | 2.0560 | 5058.39 | 0.25 |
| 229 | V0345 UMa | 169.4559573394 | 33.6707929947 | 13.9139 | 6994 | 1.3390 | 3993.14 | -0.43 |
| 230 | NX UMa | 169.8659366784 | 58.3147636837 | 14.5749 | 7186 | 1.2845 | 6055.98 | -0.62 |
| 231 | KR Leo | 170.0812506438 | -2.3099347091 | 14.0082 | 7212 | 1.3203 | 5337.37 | -0.95 |
| 232 | AU UMa | 170.3085911897 | 44.2378504856 | 15.9120 | 6915 | 1.2673 | 4865.68 | 1.21 |
| 233 | BU Leo | 170.7922382532 | 15.6985613804 | 12.6115 | 6590 | 1.7708 | 2611.72 | -1.24 |
| 234 | RX Leo | 170.9931394061 | 26.6150213133 | 11.8104 | 6826 | 1.3400 | 1958.99 | -0.99 |
| 235 | AH Crt | 171.2556621566 | -10.2688603632 | 14.1962 | 6126 | 1.8268 | 5434.79 | -1.31 |
| 236 | KW Leo | 171.4874131644 | -0.1615900571 | 14.3083 | 6763 | 1.2520 | 6121.02 | -0.88 |

| # | RRab | R.A. (J2000) | Decl. (J2000) | $G_{mag}$ | $T_{eff}$ | $A_G$ (mag) | d (pc) | $M_G$ |
|---|---|---|---|---|---|---|---|---|
| 237 | AE Leo | 171.5507932771 | 17.6610260390 | 12.3295 | 7052 | 1.3185 | 2693.43 | -1.14 |
| 238 | AI Crt | 171.5576334136 | -14.0687547299 | 13.0805 | 7061 | 1.2470 | 2728.32 | -0.35 |
| 239 | LS Cen | 172.1298663212 | -62.4774102786 | 13.3144 | 5479 | 2.0487 | 2409.21 | -0.64 |
| 240 | AV UMa | 172.4188807532 | 42.7401739208 | 15.3036 | 7008 | 1.3720 | 4525.91 | 0.65 |
| 241 | TU UMa | 172.4520400506 | 30.0673278730 | 9.6308 | 6617 | 1.2460 | 628.42 | -0.61 |
| 242 | LL Leo | 172.7234168972 | 13.3245587872 | 13.3276 | 6242 | 1.5250 | 3518.43 | -0.93 |
| 243 | LN Leo | 173.0788021848 | 3.2870039096 | 14.0039 | 7223 | 1.2510 | 4532.20 | -0.53 |
| 244 | CZ UMa | 173.4174549920 | 50.3910477105 | 15.4440 | 7066 | 1.4100 | 5310.56 | 0.41 |
| 245 | GL Mus | 174.7453551133 | -74.4690076323 | 14.9887 | 4973 | 0.4663 | 5925.65 | 0.66 |
| 246 | AA Leo | 174.8093041673 | 10.3273429789 | 12.3200 | 6786 | 1.3735 | 1959.87 | -0.51 |
| 247 | BY Leo | 175.2080607643 | 19.9286480439 | 16.7969 | 6388 | 0.3170 | 2491.88 | 4.50 |
| 248 | AN Crt | 175.2782328718 | -10.6185396301 | 12.7205 | 6777 | 1.4913 | 2418.48 | -0.69 |
| 249 | V0354 Vir | 175.8842445346 | 2.6987760484 | 12.8386 | 6948 | 1.0083 | 2570.35 | -0.22 |
| 250 | GP Leo | 176.4397027849 | 11.8690235648 | 13.4857 | 6922 | 1.3757 | 3951.26 | -0.87 |
| 251 | BI Cen | 176.4775827176 | -59.3779010328 | 11.9751 | 5411 | 1.9580 | 1395.71 | -0.71 |
| 252 | CF Leo | 176.6357577650 | 16.2312455694 | 12.0501 | 6775 | 1.6217 | 1933.04 | -1.00 |
| 253 | DE Cha | 176.6484953702 | -76.5874369660 | 14.7704 | 5324 | 2.1910 | 4576.27 | -0.72 |
| 254 | X Crt | 177.2342562055 | -10.4412552157 | 11.3882 | 6670 | 1.5840 | 1622.70 | -1.25 |
| 255 | EF Leo | 177.2955912108 | 28.0071288297 | 15.2237 | 7000 | 1.2270 | 5445.85 | 0.32 |
| 256 | CH Leo | 177.3120493884 | 24.1667109947 | 14.4301 | 7092 | 1.5592 | 4808.05 | -0.54 |
| 257 | CD UMa | 177.8918240462 | 52.9725715328 | 14.6084 | 6498 | 1.6455 | 5032.32 | -0.55 |
| 258 | V0574 Cen | 178.2690767684 | -39.2317460169 | 14.3028 | 6117 | 1.7715 | 5025.33 | -0.97 |
| 259 | AE Cha | 178.3309185100 | -78.8546821461 | 14.8856 | 5787 | 1.7790 | 5456.63 | -0.58 |
| 260 | BD UMa | 179.2784680324 | 48.4070361040 | 13.3897 | 6839 | 1.4500 | 4620.44 | -1.38 |
| 261 | V0578 Cen | 179.2786795845 | -39.5253165498 | 14.4381 | 5829 | 1.8050 | 6207.57 | -1.33 |
| 262 | V0580 Cen | 179.3350647258 | -36.6727316368 | 14.4206 | 6643 | 1.7950 | 5342.11 | -1.01 |
| 263 | V0349 UMa | 179.3521126031 | 38.4261225347 | 14.2797 | 6577 | 1.6457 | 4975.51 | -0.85 |
| 264 | CP UMa | 179.6402511072 | 52.0297266921 | 16.5564 | 6588 | 1.6150 | 4174.91 | 1.84 |
| 265 | KX Cen | 179.8084928882 | -40.2392471995 | 13.7556 | 6310 | 1.8123 | 4242.05 | -1.19 |
| 266 | V0363 Vir | 180.0665012326 | 3.7189783452 | 14.1016 | 6669 | 0.5133 | 2721.15 | 1.41 |
| 267 | V0582 Cen | 180.2068927607 | -36.2001650678 | 14.9292 | 6392 | 1.6665 | 5080.54 | -0.27 |
| 268 | CK UMa | 180.4016700865 | 31.9033556677 | 14.0558 | 6574 | 1.5573 | 3329.15 | -0.11 |
| 269 | AH Cha | 180.4798395131 | -77.2096173404 | 15.2682 | 4917 | 0.1403 | 5823.74 | 1.30 |
| 270 | BW Com | 181.0695165649 | 18.8868200026 | 14.4717 | 6465 | 1.5060 | 3916.15 | 0.00 |
| 271 | V0364 Cen | 181.3438071917 | -47.7413794728 | 15.2310 | 6550 | 1.1390 | 7675.80 | -0.33 |
| 272 | WX Com | 181.4374495073 | 22.0226592218 | 16.1639 | 6810 | 1.5077 | 3800.18 | 1.76 |
| 273 | V0370 Vir | 181.5172217846 | -2.2159391765 | 15.1675 | 6513 | 1.5105 | 4567.61 | 0.36 |
| 274 | WZ Com | 181.8916750321 | 20.1026297811 | 14.9627 | 6600 | 1.5725 | 3874.24 | 0.45 |
| 275 | XY Com | 181.9444280133 | 21.6281828040 | 15.3338 | 6536 | 1.6133 | 4263.01 | 0.57 |
| 276 | V0374 Vir | 182.1307116753 | -0.7882557402 | 16.8435 | 6878 | 0.1635 | 2463.83 | 4.72 |
| 277 | AY CVn | 182.2324045707 | 40.8281723578 | 16.3325 | 6506 | 1.6857 | 4322.65 | 1.47 |
| 278 | XZ Com | 182.2740239573 | 23.7544817521 | 16.6422 | 7146 | 1.1327 | 3486.31 | 2.80 |
| 279 | V0376 Vir | 182.3154961180 | -1.9123072794 | 15.6367 | 6965 | 1.2048 | 2736.13 | 2.25 |
| 280 | AL Cha | 182.4061350627 | -78.6839202206 | 14.6703 | 5309 | 1.8557 | 10057.57 | -2.20 |
| 281 | V Com | 182.5662604302 | 27.4316853151 | 13.4702 | 6329 | 1.4520 | 3493.31 | -0.70 |
| 282 | AB UMa | 182.8107881819 | 47.8288317663 | 10.7829 | 6269 | 1.6460 | 988.80 | -0.84 |
| 283 | V0380 Vir | 183.0644332225 | -1.3536660195 | 15.3323 | 7204 | 1.2990 | 4329.92 | 0.85 |
| 284 | SY Crv | 183.0692608349 | -22.9092446286 | 13.9396 | 6423 | 1.6527 | 3904.96 | -0.67 |
| 285 | YZ Com | 183.2816222822 | 21.9472011368 | 16.0608 | 7118 | 1.3525 | 4606.96 | 1.39 |
| 286 | TU Com | 183.4456333204 | 30.9854415913 | 13.8108 | 6649 | 1.3980 | 4972.46 | -1.07 |
| 287 | CH Com | 183.4461741745 | 22.3448259867 | 15.9546 | 6592 | 1.5105 | 3732.76 | 1.58 |
| 288 | V0382 Vir | 183.9652189269 | -0.0985725038 | 16.6929 | 6494 | 0.0620 | 2688.35 | 4.48 |
| 289 | LM Cen | 183.9751650939 | -45.1284811978 | 14.1291 | 6119 | 1.6483 | 4498.45 | -0.78 |
| 290 | AA Com | 184.0420126465 | 25.0224868617 | 15.6755 | 6482 | 1.4490 | 4435.76 | 0.99 |
| 291 | SW Dra | 184.4442859003 | 69.5106206390 | 10.5160 | 6363 | 1.5590 | 958.33 | -0.95 |
| 292 | DQ Vir | 184.5967893653 | 5.5114729881 | 15.0833 | 6436 | 1.8635 | 5524.18 | -0.49 |
| 293 | GY Com | 184.8584977215 | 28.4576170256 | 15.9179 | 6363 | 1.1090 | 3682.97 | 1.98 |
| 294 | AC Com | 185.2221620798 | 22.1748214973 | 14.6240 | 7019 | 1.3580 | 4586.12 | -0.04 |
| 295 | LQ Cen | 185.5892286657 | -46.9601925307 | 14.6308 | 5947 | 1.8623 | 5515.67 | -0.94 |
| 296 | DF Vir | 185.6395655488 | 11.2944455376 | 15.9024 | 6605 | 1.4987 | 4678.80 | 1.05 |

| # | RRab | R.A. (J2000) | Decl. (J2000) | $G_{mag}$ | $T_{eff}$ | $A_G$ (mag) | d (pc) | $M_G$ |
|---|---|---|---|---|---|---|---|---|
| 297 | V0389 Vir | 185.7953215959 | -0.2827324454 | 16.6601 | 6923 | 1.4303 | 4007.36 | 2.22 |
| 298 | VX Vir | 185.8650836090 | -2.6679621397 | 14.4041 | 6270 | 1.8030 | 4689.50 | -0.75 |
| 299 | EG Vir | 185.9845205801 | 0.8586656690 | 15.0334 | 6794 | 1.5730 | 4815.22 | 0.05 |
| 300 | OT Dra | 186.2687112395 | 66.6443541231 | 13.8261 | 6817 | 1.6710 | 4321.52 | -1.02 |
| 301 | CL UMa | 186.6064834689 | 52.5571628133 | 15.8893 | 6751 | 1.6030 | 4965.04 | 0.81 |
| 302 | DS Vir | 186.8623820738 | 7.8139026697 | 15.2588 | 6990 | 1.0033 | 5476.01 | 0.56 |
| 303 | V0397 Vir | 186.9507235615 | -0.9795215968 | 16.6865 | 6878 | 1.6290 | 3891.32 | 2.11 |
| 304 | RR CVn | 187.2812367307 | 34.6472563118 | 12.5843 | 6635 | 1.2715 | 2134.26 | -0.33 |
| 305 | EM Vir | 187.3455808468 | 9.2659672673 | 16.1978 | 6270 | 1.2530 | 4699.58 | 1.58 |
| 306 | BN CVn | 187.4031372145 | 47.8214572409 | 12.2868 | 6738 | 1.5220 | 2381.47 | -1.12 |
| 307 | V0444 Cen | 187.4318910265 | -34.3671835203 | 12.4391 | 6597 | 1.3200 | 1820.95 | -0.18 |
| 308 | UZ CVn | 187.6154349253 | 40.5088965240 | 12.0798 | 6564 | 1.5240 | 1822.87 | -0.75 |
| 309 | BL Vir | 187.7635649325 | -4.3206159566 | 14.5068 | 6268 | 1.6035 | 4499.24 | -0.36 |
| 310 | BM Vir | 187.9818067148 | 0.9085749557 | 13.7882 | 6632 | 1.6658 | 4014.95 | -0.90 |
| 311 | RS Crv | 188.0005987048 | -23.9050223006 | 13.2793 | 6548 | 1.6760 | 3654.14 | -1.21 |
| 312 | S Com | 188.1901044546 | 27.0292818573 | 11.7242 | 6564 | 1.2873 | 2031.64 | -1.10 |
| 313 | AM Com | 188.3677757042 | 22.4777763181 | 15.5912 | 7104 | 1.4575 | 4654.85 | 0.79 |
| 314 | BO Vir | 188.5841144393 | 0.6569433243 | 14.6593 | 6833 | 1.4070 | 4383.57 | 0.04 |
| 315 | BV Hya | 188.6465773973 | -30.0961432602 | 14.6884 | 6291 | 1.6177 | 4288.34 | -0.09 |
| 316 | OO UMa | 188.7249151903 | 53.6331203715 | 14.3531 | 6619 | 1.5667 | 6299.16 | -1.21 |
| 317 | SV CVn | 188.9832850392 | 37.2068568736 | 12.6286 | 6910 | 1.5960 | 3084.61 | -1.41 |
| 318 | BQ Vir | 189.1138454568 | -2.4257280650 | 12.0496 | 6478 | 1.5700 | 1782.60 | -0.78 |
| 319 | BW Hya | 189.3699388433 | -29.9366140155 | 13.6388 | 6402 | 1.7313 | 4681.92 | -1.44 |
| 320 | FU Vir | 189.6095389739 | 13.0159970635 | 13.2101 | 6564 | 1.6900 | 3080.55 | -0.92 |
| 321 | BY Hya | 189.6780369678 | -29.2241519182 | 13.6765 | 6516 | 1.7602 | 3774.95 | -0.97 |
| 322 | DP Com | 189.7908317649 | 20.7548324671 | 15.6098 | 6669 | 1.5530 | 4419.18 | 0.83 |
| 323 | AP Com | 189.8280020941 | 22.0542229647 | 14.3985 | 6754 | 1.6168 | 5779.46 | -1.03 |
| 324 | DH Vir | 189.9855289757 | 6.0792271643 | 13.3998 | 6933 | 0.8885 | 3759.90 | -0.36 |
| 325 | DG Vir | 189.9981985226 | 11.5102305725 | 15.8551 | 6305 | 0.7993 | 4771.16 | 1.66 |
| 326 | V0450 Cen | 190.0192143595 | -34.2713896712 | 14.3966 | 6288 | 1.5397 | 4567.66 | -0.44 |
| 327 | SW CVn | 190.2292409478 | 37.0852350832 | 13.1268 | 6458 | 1.2113 | 2386.85 | 0.03 |
| 328 | DT Com | 190.6679602514 | 28.2234529919 | 16.3128 | 6927 | 1.5730 | 4490.21 | 1.48 |
| 329 | WY Crv | 190.9279729631 | -13.8537002920 | 13.2099 | 6333 | 1.4680 | 3005.27 | -0.65 |
| 330 | EU CVn | 190.9678726652 | 47.6418247672 | 13.0244 | 6260 | 1.6977 | 3364.87 | -1.31 |
| 331 | DV Com | 190.9766879281 | 28.0210518042 | 14.7961 | 6749 | 1.5327 | 4957.50 | -0.21 |
| 332 | YY Crv | 191.0679249370 | -12.0518492981 | 12.5000 | 6395 | 1.6650 | 2989.74 | -1.54 |
| 333 | BU Vir | 191.1237561646 | -2.0806029578 | 14.6031 | 6226 | 1.6460 | 4307.88 | -0.21 |
| 334 | V1134 Cen | 191.2253436376 | -40.2882753891 | 14.1889 | 6357 | 1.6207 | 4314.79 | -0.61 |
| 335 | V0457 Cen | 191.2629821951 | -30.6130789031 | 14.9484 | 6566 | 1.6270 | 5786.10 | -0.49 |
| 336 | AU Com | 191.4901042087 | 19.8378209749 | 15.6797 | 6182 | 0.1657 | 4824.65 | 2.10 |
| 337 | V0460 Cen | 191.8078333105 | -33.9918485886 | 15.5621 | 6464 | 1.5490 | 5028.55 | 0.51 |
| 338 | DS CVn | 191.8178351189 | 35.2017240581 | 14.8893 | 6428 | 1.5465 | 4762.71 | -0.05 |
| 339 | OU Dra | 191.8312741012 | 69.0983313395 | 13.7668 | 6402 | 1.6170 | 4330.84 | -1.03 |
| 340 | V1255 Cen | 191.9044070080 | -39.5449009354 | 16.4546 | 5739 | 1.6805 | 5114.06 | 1.23 |
| 341 | EF Com | 192.1175373594 | 25.3899857868 | 16.1865 | 7688 | 1.0510 | 4554.81 | 1.84 |
| 342 | V0468 Cen | 192.1367289851 | -30.4778125140 | 15.8804 | 6669 | 1.1065 | 3375.00 | 2.13 |
| 343 | EH Com | 192.1569915370 | 18.1736971422 | 15.2468 | 6211 | 1.5417 | 4340.19 | 0.52 |
| 344 | V0470 Cen | 192.2262529170 | -31.5589138209 | 14.0368 | 6263 | 1.8805 | 6033.24 | -1.75 |
| 345 | Z CVn | 192.4390918076 | 43.7736658598 | 11.8804 | 6838 | 1.4942 | 2311.17 | -1.43 |
| 346 | BX Vir | 192.5481721837 | 1.9548871433 | 15.5180 | 6603 | 1.6503 | 5382.75 | 0.21 |
| 347 | AX Com | 192.6552528658 | 18.3098901549 | 15.7265 | 6815 | 1.4887 | 4888.60 | 0.79 |
| 348 | EI Com | 192.6783961293 | 30.1984788932 | 16.4572 | 6878 | 1.6727 | 4262.12 | 1.64 |
| 349 | V0474 Cen | 192.8019578795 | -35.8252608259 | 15.4494 | 6957 | 1.7130 | 4772.84 | 0.34 |
| 350 | GI Vir | 192.8428091139 | 2.9139764408 | 13.8188 | 6819 | 1.3010 | 4199.07 | -0.60 |
| 351 | EM Com | 192.9095276389 | 30.5176121910 | 15.2756 | 7095 | 1.3005 | 5740.73 | 0.18 |
| 352 | V0423 Vir | 193.0223748341 | -2.0019907622 | 16.6571 | 6686 | 1.0100 | 3207.00 | 3.12 |
| 353 | AS Vir | 193.1911196679 | -10.2601051444 | 11.9894 | 6640 | 1.7663 | 1669.22 | -0.89 |
| 354 | V0480 Cen | 193.5922528138 | -37.1713554209 | 12.7337 | 6910 | 1.6513 | 2618.60 | -1.01 |
| 355 | V0481 Cen | 193.6379026071 | -30.7496922921 | 14.3413 | 6923 | 1.5520 | 4625.03 | -0.54 |
| 356 | V0428 Vir | 193.7584617916 | -2.0485067356 | 16.7499 | 6396 | 1.3897 | 3137.33 | 2.88 |

| # | RRab | R.A. (J2000) | Decl. (J2000) | $G_{mag}$ | $T_{eff}$ | $A_G$ (mag) | d (pc) | $M_G$ |
|---|---|---|---|---|---|---|---|---|
| 357 | V0429 Vir | 193.7751745323 | -0.0951272457 | 16.0287 | 6959 | 1.4360 | 4584.11 | 1.29 |
| 358 | V0431 Vir | 194.0038541745 | -0.2471896876 | 16.6963 | 6523 | 0.0200 | 2333.55 | 4.84 |
| 359 | V0432 Vir | 194.0441730458 | -11.2950237986 | 12.5619 | 6419 | 1.5530 | 2258.76 | -0.76 |
| 360 | IP Com | 194.1284268014 | 29.8931194661 | 14.7620 | 6736 | 0.9293 | 6829.35 | -0.34 |
| 361 | V0483 Cen | 194.1790384907 | -35.1169927488 | 14.2911 | 6860 | 1.5493 | 5818.84 | -1.08 |
| 362 | V0437 Vir | 194.5182924256 | 9.8454181978 | 13.9548 | 6150 | 1.3477 | 5148.50 | -0.95 |
| 363 | V0438 Vir | 194.6132305310 | -3.6017930914 | 14.2952 | 6860 | 1.0760 | 4817.28 | -0.19 |
| 364 | V0488 Cen | 194.7866039943 | -35.5392251176 | 15.4825 | 6944 | 0.8995 | 4895.95 | 1.13 |
| 365 | V1145 Cen | 194.9491239107 | -50.2629373618 | 13.3989 | 5317 | 2.0600 | 2920.49 | -0.99 |
| 366 | V0489 Cen | 195.4127408690 | -30.2866731728 | 14.3354 | 6135 | 1.4270 | 6572.98 | -1.18 |
| 367 | BF Com | 195.5566254694 | 24.2390277992 | 13.4050 | 7040 | 1.4290 | 4151.46 | -1.12 |
| 368 | RY Com | 196.2832628790 | 23.2784393143 | 12.5526 | 6597 | 1.6540 | 1992.14 | -0.60 |
| 369 | MO Com | 196.3101574567 | 28.6202031669 | 14.3154 | 6410 | 1.6533 | 4271.10 | -0.49 |
| 370 | V0449 Vir | 197.3404534430 | -0.7188976335 | 16.3105 | 6577 | 0.9665 | 4526.04 | 2.07 |
| 371 | V0452 Vir | 197.8105106643 | 11.1579323361 | 13.7983 | 6748 | 1.3920 | 4366.26 | -0.79 |
| 372 | UZ Com | 198.1114533394 | 30.3544406276 | 13.1997 | 6613 | 1.5867 | 3840.54 | -1.31 |
| 373 | NY Com | 198.6435843956 | 20.5069363396 | 16.2789 | 7054 | 0.9523 | 5269.46 | 1.72 |
| 374 | RS Com | 198.6649264403 | 17.1969067500 | 15.6213 | 6890 | 1.4680 | 5522.62 | 0.44 |
| 375 | HI Vir | 198.8944900358 | -7.6526252600 | 15.1852 | 6209 | 1.6200 | 4376.07 | 0.36 |
| 376 | V1167 Cen | 198.9705103479 | -36.0633215110 | 13.5744 | 7215 | 1.2667 | 4484.24 | -0.95 |
| 377 | HK Vir | 199.2687874427 | -8.9687625778 | 15.0235 | 7168 | 1.3115 | 5288.29 | 0.10 |
| 378 | ST Com | 199.4639363917 | 20.7807665556 | 11.4178 | 6597 | 1.6020 | 1424.16 | -0.95 |
| 379 | OP Com | 199.5520456041 | 17.3673140342 | 16.7831 | 7067 | 1.3050 | 4425.91 | 2.25 |
| 380 | OQ Com | 199.6746281083 | 17.7566943773 | 15.0469 | 6287 | 1.6090 | 6163.09 | -0.51 |
| 381 | FF Com | 199.6956385071 | 22.5193955747 | 16.0162 | 6912 | 1.3307 | 4592.90 | 1.38 |
| 382 | BG Mus | 199.7278126873 | -65.6193221451 | 14.1836 | 4956 | 2.5200 | 2993.98 | -0.72 |
| 383 | OR Com | 199.9771075401 | 19.8991135278 | 13.3023 | 6792 | 1.5300 | 3314.09 | -0.83 |
| 384 | AV Vir | 200.0481827281 | 9.1878812639 | 11.7956 | 6614 | 1.6290 | 1772.71 | -1.08 |
| 385 | BH Com | 200.4947178594 | 16.7049908759 | 12.9399 | 6985 | 1.3893 | 3275.93 | -1.03 |
| 386 | HN Vir | 200.7914747554 | -2.0820463643 | 15.3362 | 6600 | 1.1705 | 3305.45 | 1.57 |
| 387 | AM Vir | 200.8888547011 | -16.6660795158 | 11.4478 | 5914 | 1.8700 | 1335.96 | -1.05 |
| 388 | FH Com | 201.1071486149 | 16.0016607103 | 14.6004 | 6594 | 1.3080 | 5011.57 | -0.21 |
| 389 | BS Cha | 201.5609099403 | -78.4633593452 | 14.9969 | 5876 | 1.6300 | 4976.56 | -0.12 |
| 390 | V0473 Vir | 201.8023841701 | -1.2749643924 | 14.8940 | 6688 | 1.6337 | 3973.77 | 0.26 |
| 391 | V1181 Cen | 201.9391707944 | -37.7829346461 | 12.9954 | 6486 | 1.6995 | 3760.86 | -1.58 |
| 392 | BK Com | 202.0411884980 | 20.2261604878 | 14.2863 | 6658 | 1.6580 | 4814.43 | -0.78 |
| 393 | WW Vir | 202.0995165581 | -5.2857925001 | 13.4021 | 6411 | 1.8192 | 4210.23 | -1.54 |
| 394 | BN Com | 202.2847483530 | 17.3170112217 | 15.7673 | 6725 | 1.6573 | 4996.65 | 0.62 |
| 395 | V0476 Vir | 202.3436198866 | -5.8831270282 | 11.6469 | 7134 | 1.5577 | 1650.43 | -1.00 |
| 396 | FL CVn | 202.8159798134 | 40.9491064531 | 14.6975 | 6859 | 1.3300 | 4846.33 | -0.06 |
| 397 | BQ Com | 203.0559054874 | 26.4253355620 | 14.7633 | 6559 | 1.7280 | 6265.25 | -0.95 |
| 398 | RV UMa | 203.3253497180 | 53.9873889548 | 10.8599 | 6765 | 1.5780 | 1050.23 | -0.82 |
| 399 | WW CVn | 203.6288512735 | 29.3042393074 | 14.8609 | 7348 | 1.3525 | 5244.73 | -0.09 |
| 400 | V0482 Vir | 203.9676828604 | -0.6185278195 | 15.3227 | 6383 | 1.8917 | 4210.42 | 0.31 |
| 401 | AM Boo | 204.7270406495 | 23.7636103971 | 14.6550 | 6665 | 1.6720 | 5283.58 | -0.63 |
| 402 | BQ Boo | 204.8346906913 | 18.2014172064 | 15.0662 | 6976 | 1.6248 | 5557.91 | -0.28 |
| 403 | AN Boo | 204.9489167598 | 15.5926537054 | 14.9544 | 7168 | 1.4193 | 5417.08 | -0.13 |
| 404 | V0671 Cen | 205.0404287876 | -37.4411117619 | 12.3549 | 6422 | 1.8510 | 2439.07 | -1.43 |
| 405 | IS Dra | 205.1117288674 | 67.9343003471 | 13.1942 | 6432 | 1.3700 | 3148.63 | -0.67 |
| 406 | WZ CVn | 205.5850132213 | 27.8976725585 | 15.6411 | 6608 | 1.2760 | 6232.73 | 0.39 |
| 407 | FQ Cen | 205.8916041097 | -50.4123039004 | 15.6561 | 5587 | 1.8867 | 6024.89 | -0.13 |
| 408 | V0494 Vir | 206.3388814718 | -0.0298305968 | 14.2885 | 7229 | 1.3695 | 4649.78 | -0.42 |
| 409 | RV Oct | 206.6322602995 | -84.4017735000 | 10.9540 | 6299 | 1.8155 | 961.74 | -0.78 |
| 410 | V0498 Vir | 206.7267026136 | -3.1263714651 | 14.5372 | 6984 | 1.1238 | 6786.78 | -0.74 |
| 411 | AV Mus | 206.8026185879 | -70.6366208231 | 15.0237 | 5239 | 2.1377 | 5120.20 | -0.66 |
| 412 | V1288 Cen | 206.9690051725 | -31.0670056648 | 14.6015 | 6182 | 1.8290 | 4736.56 | -0.60 |
| 413 | SS CVn | 207.0664261725 | 39.9008369620 | 11.9030 | 7052 | 1.3240 | 1666.26 | -0.53 |
| 414 | AS Boo | 207.0762633519 | 24.7846718111 | 15.2803 | 7176 | 1.1070 | 5566.86 | 0.45 |
| 415 | FT CVn | 207.1771998631 | 41.9187843585 | 15.2724 | 6925 | 1.4555 | 6781.38 | -0.34 |
| 416 | RX CVn | 207.1776008783 | 41.3852060404 | 12.5840 | 6581 | 1.4635 | 2352.74 | -0.74 |

| # | RRab | R.A. (J2000) | Decl. (J2000) | $G_{mag}$ | $T_{eff}$ | $A_G$ (mag) | d (pc) | $M_G$ |
|---|---|---|---|---|---|---|---|---|
| 417 | AU Boo | 207.4577388247 | 16.1297156105 | 15.3429 | 6570 | 1.4958 | 5639.19 | 0.09 |
| 418 | HQ Boo | 207.4675332223 | 12.3747567459 | 15.0800 | 7170 | 1.1050 | 5637.51 | 0.22 |
| 419 | RT Cir | 208.0105064739 | -70.1710934295 | 14.9130 | 5338 | 2.0643 | 5876.45 | -1.00 |
| 420 | AX Boo | 208.0144387305 | 18.5610890416 | 15.4462 | 6259 | 1.6920 | 5069.72 | 0.23 |
| 421 | V0507 Vir | 208.1854853326 | -2.3080569537 | 14.0962 | 6536 | 1.6307 | 3474.75 | -0.24 |
| 422 | FY Hya | 208.3977056092 | -29.5802395603 | 12.7577 | 6714 | 1.5290 | 2529.47 | -0.79 |
| 423 | BB Boo | 208.4066917238 | 21.8463747999 | 15.6225 | 7176 | 1.0570 | 5307.75 | 0.94 |
| 424 | BC Boo | 208.4096993751 | 26.3503149950 | 16.2985 | 7008 | 1.3290 | 3565.84 | 2.21 |
| 425 | TX Vir | 208.7496588863 | -6.0177705504 | 13.4096 | 6926 | 1.1383 | 5594.75 | -1.47 |
| 426 | V0499 Cen | 208.8115332762 | -43.2402226332 | 11.2819 | 5926 | 1.8090 | 1076.66 | -0.69 |
| 427 | YY CVn | 208.8215666670 | 29.6923266845 | 15.5214 | 7107 | 1.2880 | 6511.32 | 0.17 |
| 428 | BF Boo | 209.5878124117 | 25.3067966574 | 14.9319 | 6459 | 1.5003 | 6408.86 | -0.60 |
| 429 | V0517 Vir | 209.7094729541 | -2.4062512498 | 14.2480 | 6345 | 1.8510 | 4715.63 | -0.97 |
| 430 | UU CVn | 210.0884013548 | 29.4182752999 | 15.8020 | 7146 | 1.4678 | 4789.62 | 0.93 |
| 431 | UV CVn | 210.2051557806 | 28.3307973947 | 14.1191 | 7122 | 1.0647 | 6353.26 | -0.96 |
| 432 | V0674 Cen | 210.8503109481 | -36.4056057467 | 11.3471 | 6498 | 1.6790 | 1196.78 | -0.72 |
| 433 | BI Boo | 211.1678730595 | 28.1950456844 | 15.8293 | 6605 | 1.5070 | 5666.11 | 0.56 |
| 434 | II Boo | 211.2928783273 | 14.0807936116 | 13.4359 | 6843 | 1.6650 | 4253.00 | -1.37 |
| 435 | AD Vir | 211.4046463284 | -7.2474644459 | 13.2262 | 5831 | 1.8140 | 2863.20 | -0.87 |
| 436 | QV Cen | 211.4840523845 | -57.7539447856 | 12.6562 | 5220 | 2.0295 | 1537.18 | -0.31 |
| 437 | CS Boo | 211.5074088790 | 24.5708182978 | 12.9563 | 6402 | 1.3630 | 2714.27 | -0.57 |
| 438 | BN Boo | 211.8377617417 | 29.0477315555 | 16.2571 | 7163 | 1.3620 | 5096.85 | 1.36 |
| 439 | IM Boo | 212.2733755201 | 47.8244570300 | 13.3307 | 6971 | 1.3115 | 3999.24 | -0.99 |
| 440 | QZ Cen | 212.3170891945 | -59.7900557969 | 12.9879 | 5048 | 2.3905 | 1757.26 | -0.63 |
| 441 | XX UMi | 213.3352024292 | 73.8961807398 | 14.6298 | 6743 | 1.4970 | 6566.06 | -0.95 |
| 442 | V0559 Hya | 213.4395883401 | -22.9116321667 | 12.5871 | 6479 | 1.6270 | 2459.32 | -0.99 |
| 443 | FT Boo | 213.4981346616 | 47.4450032895 | 13.8688 | 6387 | 1.6197 | 4300.56 | -0.92 |
| 444 | V0535 Vir | 213.5909100773 | -1.2572299083 | 14.9655 | 6594 | 1.5052 | 6051.32 | -0.45 |
| 445 | IQ Boo | 213.8562802761 | 28.2821836189 | 13.8317 | 6873 | 1.5200 | 3584.77 | -0.46 |
| 446 | IR Boo | 213.9440138254 | 16.9348360174 | 14.6228 | 6528 | 1.6367 | 5626.13 | -0.76 |
| 447 | CM Boo | 214.0130313693 | 20.0604640061 | 12.7413 | 6594 | 1.3873 | 2459.71 | -0.60 |
| 448 | V0348 Vir | 214.1069251608 | -17.0912461304 | 12.5127 | 6393 | 1.6435 | 2388.70 | -1.02 |
| 449 | V0540 Vir | 214.3776830239 | -10.3016053223 | 14.1923 | 6095 | 1.5748 | 4775.03 | -0.78 |
| 450 | XY UMi | 214.4820437898 | 71.6854115550 | 13.1843 | 6290 | 1.5510 | 3538.79 | -1.11 |
| 451 | V0513 Cen | 214.8798166009 | -38.8580712477 | 14.2878 | 6488 | 1.6370 | 6722.67 | -1.49 |
| 452 | IY Boo | 214.9134164061 | 25.7900331659 | 14.7515 | 6142 | 1.6450 | 4280.64 | -0.05 |
| 453 | V0547 Vir | 214.9680088572 | -6.3842965063 | 14.2196 | 5998 | 1.8510 | 4886.31 | -1.08 |
| 454 | V0514 Cen | 215.0084039762 | -40.4683340072 | 15.4355 | 6015 | 1.8640 | 4085.25 | 0.52 |
| 455 | V0549 Vir | 215.4999008970 | -1.0187193567 | 15.6226 | 6571 | 1.3830 | 3286.26 | 1.66 |
| 456 | V0367 UMa | 215.5639514121 | 60.4842885891 | 13.1249 | 6927 | 1.6870 | 3281.53 | -1.14 |
| 457 | FU Boo | 215.7241673997 | 19.5387945469 | 14.0253 | 6293 | 1.5130 | 5446.27 | -1.17 |
| 458 | V0551 Vir | 215.7732710456 | 1.9002390879 | 13.4363 | 7095 | 1.4040 | 3628.43 | -0.77 |
| 459 | V1344 Cen | 215.8766728547 | -57.8844811882 | 11.7541 | 5651 | 2.0303 | 1141.34 | -0.56 |
| 460 | LV Lib | 216.0196399434 | -11.4149119642 | 14.2846 | 6594 | 1.6152 | 3978.74 | -0.33 |
| 461 | KR Boo | 216.3724431791 | 20.9629668588 | 13.2526 | 6438 | 1.5870 | 2997.35 | -0.72 |
| 462 | V0726 Cen | 216.5446167702 | -30.4180600979 | 14.4133 | 6260 | 1.7253 | 3693.86 | -0.15 |
| 463 | ST Vir | 216.9128070102 | -0.9016114273 | 11.7732 | 6343 | 1.7287 | 1428.14 | -0.73 |
| 464 | V0561 Vir | 217.1693756375 | -1.4665095336 | 15.6886 | 7031 | 1.6180 | 4992.38 | 0.58 |
| 465 | Y Oct | 217.2685805117 | -88.6454677883 | 12.2347 | 6571 | 1.6195 | 2250.38 | -1.15 |
| 466 | V0522 Cen | 217.3651268043 | -37.2287971589 | 13.1133 | 6516 | 1.6195 | 3296.01 | -1.10 |
| 467 | V1353 Cen | 217.6055982485 | -36.0764719491 | 13.5158 | 6696 | 1.5245 | 3514.28 | -0.74 |
| 468 | KX Boo | 217.7537305062 | 20.4255196178 | 14.0804 | 6505 | 1.6765 | 4137.74 | -0.68 |
| 469 | V0529 Cen | 218.0791996140 | -38.1430736979 | 15.2661 | 6062 | 1.6160 | 5024.99 | 0.14 |
| 470 | DK Lup | 218.1033390507 | -44.7748105079 | 15.3396 | 5804 | 0.1657 | 6681.60 | 1.05 |
| 471 | SV Boo | 218.5264710610 | 39.1091857339 | 13.0509 | 6906 | 1.5818 | 3464.89 | -1.23 |
| 472 | V0734 Cen | 218.5452600537 | -30.1261500936 | 15.2970 | 5885 | 1.9220 | 5394.72 | -0.28 |
| 473 | V0574 Vir | 218.6270015896 | -8.3090696448 | 13.9374 | 6537 | 1.7513 | 4892.07 | -1.26 |
| 474 | DV Lup | 219.2000409953 | -42.6454135520 | 14.6648 | 6080 | 1.5890 | 5344.40 | -0.56 |
| 475 | V0582 Vir | 219.3079213917 | 2.7723045136 | 14.6767 | 6590 | 0.1435 | 7234.17 | 0.24 |
| 476 | LQ Boo | 219.4738348919 | 34.9900157903 | 13.3593 | 6329 | 1.5690 | 3772.75 | -1.09 |

| # | RRab | R.A. (J2000) | Decl. (J2000) | $G_{mag}$ | $T_{eff}$ | $A_G$ (mag) | d (pc) | $M_G$ |
|---|---|---|---|---|---|---|---|---|
| 477 | LT Boo | 219.6880541806 | 53.7828901230 | 14.4233 | 6206 | 1.6370 | 6228.03 | -1.19 |
| 478 | DO Vir | 219.6916304479 | -5.3253878746 | 14.0254 | 6348 | 1.5993 | 3596.51 | -0.35 |
| 479 | VX Boo | 219.7488959796 | 25.1450992404 | 13.5897 | 6060 | 1.7550 | 3842.90 | -1.09 |
| 480 | V0585 Vir | 219.8640280576 | -3.4601629171 | 12.4673 | 6103 | 1.5417 | 2256.91 | -0.84 |
| 481 | LL Lib | 220.4299827696 | -20.8405167170 | 14.1673 | 6125 | 1.8983 | 3612.40 | -0.52 |
| 482 | SZ Boo | 220.5556982506 | 28.2065556438 | 12.5850 | 6962 | 1.6420 | 2459.82 | -1.01 |
| 483 | LZ Boo | 220.6739983876 | 50.1065827424 | 12.8245 | 6320 | 1.4540 | 2388.44 | -0.52 |
| 484 | V0741 Cen | 220.8244630674 | -33.6743619973 | 16.0639 | 5224 | 0.5065 | 1769.90 | 4.32 |
| 485 | CY Vir | 221.2137213325 | -4.6783483566 | 13.7483 | 5848 | 1.8433 | 3252.27 | -0.66 |
| 486 | CZ Vir | 221.3708701789 | -4.0210670705 | 13.7193 | 6016 | 1.6955 | 3446.48 | -0.66 |
| 487 | TY Aps | 222.2083701791 | -71.3283038982 | 11.7925 | 6059 | 1.8485 | 1538.98 | -0.99 |
| 488 | VY Boo | 222.4370947725 | 24.4862742412 | 13.1664 | 6566 | 1.8175 | 4738.32 | -2.03 |
| 489 | MO Lib | 222.5887782675 | -9.0973107089 | 13.0282 | 6419 | 1.6375 | 2833.71 | -0.87 |
| 490 | MR Lib | 223.3143167934 | -14.5991022387 | 12.9300 | 6524 | 1.8380 | 2436.74 | -0.84 |
| 491 | VZ Boo | 223.3162461135 | 25.9554150665 | 15.2830 | 6878 | 1.2222 | 5940.96 | 0.19 |
| 492 | AR Vir | 223.6669961598 | 2.6951966095 | 12.9339 | 6828 | 1.5108 | 2735.24 | -0.76 |
| 493 | DF Boo | 223.7969669153 | 28.0421792144 | 14.6795 | 6514 | 1.7092 | 6243.40 | -1.01 |
| 494 | UX Lib | 223.8129270834 | -6.2924766228 | 14.4439 | 6163 | 1.7413 | 4893.07 | -0.75 |
| 495 | EU Lup | 224.3710323415 | -43.4141501228 | 14.2867 | 6473 | 1.5530 | 4636.88 | -0.60 |
| 496 | V1366 Cen | 224.4371507614 | -30.4441157134 | 12.8856 | 6179 | 1.6400 | 3005.07 | -1.14 |
| 497 | V1367 Cen | 224.5150245735 | -35.8179905683 | 15.0616 | 6386 | 1.5800 | 5614.41 | -0.26 |
| 498 | UY Aps | 224.8933781652 | -71.7983261215 | 13.4155 | 5865 | 1.7190 | 2906.63 | -0.62 |
| 499 | V0347 Aps | 224.9082168544 | -72.7650705875 | 13.7255 | 5034 | 0.3040 | 3909.51 | 0.46 |
| 500 | RZ Lib | 224.9373784473 | -15.2707339654 | 13.1560 | 5800 | 1.8497 | 2590.72 | -0.76 |
| 501 | HS Boo | 225.0261138898 | 21.7951853987 | 14.2523 | 6961 | 1.3245 | 5682.41 | -0.84 |
| 502 | GU Vir | 225.5144881045 | 7.0559790065 | 14.8750 | 6645 | 0.2860 | 6438.88 | 0.54 |
| 503 | V0339 Lup | 225.8642975021 | -47.9343783499 | 11.8931 | 5810 | 1.8760 | 1463.64 | -0.81 |
| 504 | DX Vir | 226.5458717041 | 2.9092493794 | 15.9825 | 6359 | 1.3430 | 6254.91 | 0.66 |
| 505 | V0604 Vir | 226.5530754989 | 2.9158626827 | 13.2328 | 6550 | 1.5700 | 2953.04 | -0.69 |
| 506 | HT Boo | 226.6920581224 | 21.4379371003 | 14.7209 | 7158 | 1.2917 | 4807.60 | 0.02 |
| 507 | XX Lib | 226.7917737026 | -25.9976404577 | 12.4085 | 6126 | 1.6450 | 2261.40 | -1.01 |
| 508 | OW Boo | 226.8914483402 | 10.0468476854 | 14.7463 | 6794 | 1.3905 | 6922.90 | -0.85 |
| 509 | XZ Lib | 227.0743584718 | -25.1256609695 | 14.6779 | 6301 | 1.7480 | 6263.85 | -1.05 |
| 510 | YZ Aps | 227.5032170303 | -78.3842643831 | 14.0503 | 6160 | 0.9610 | 1322.21 | 2.48 |
| 511 | V0414 Ser | 228.0915546561 | 11.9098339130 | 13.4644 | 6323 | 1.6328 | 5832.10 | -2.00 |
| 512 | BH Ser | 228.7544980342 | 19.4431581294 | 12.7761 | 6150 | 1.6450 | 2321.44 | -0.70 |
| 513 | DF Ser | 228.8294947579 | 18.6577402454 | 12.9704 | 6383 | 1.8550 | 2556.19 | -0.92 |
| 514 | GL Aps | 229.4015236481 | -79.2103141809 | 15.2360 | 6272 | 1.6385 | 5715.99 | -0.19 |
| 515 | DG Boo | 229.5084648443 | 46.7015735826 | 12.5510 | 6403 | 1.5530 | 2290.41 | -0.80 |
| 516 | AA Aps | 229.5489863008 | -77.6066397206 | 14.9162 | 6697 | 1.7405 | 5392.70 | -0.48 |
| 517 | V0422 Ser | 229.5567395478 | -1.3189145909 | 13.9472 | 5833 | 1.9327 | 3646.93 | -0.80 |
| 518 | AV Lib | 229.8053576097 | -24.7660707026 | 14.0209 | 6344 | 1.5810 | 4621.18 | -0.88 |
| 519 | EZ Ser | 229.8125586695 | -0.4969446919 | 14.1920 | 6241 | 1.8880 | 4838.76 | -1.12 |
| 520 | V0423 Ser | 229.8282793006 | 7.8849855586 | 12.8923 | 7048 | 1.5773 | 3214.28 | -1.22 |
| 521 | AB Aps | 229.9095445411 | -78.6774379057 | 14.0059 | 6668 | 1.6505 | 4740.04 | -1.02 |
| 522 | BG Lib | 231.2199169471 | -25.7248817062 | 14.3171 | 6072 | 1.5990 | 5050.45 | -0.80 |
| 523 | BH Lib | 231.2788605673 | -23.5253935250 | 15.1470 | 5827 | 1.6420 | 4900.30 | 0.05 |
| 524 | BI Lib | 231.3963934899 | -20.6687092012 | 14.5779 | 5869 | 1.7530 | 7115.15 | -1.44 |
| 525 | LQ Lib | 231.5046926777 | -15.5461286474 | 12.5134 | 6417 | 1.6170 | 2686.79 | -1.25 |
| 526 | AE Aps | 231.6325063724 | -77.0647361870 | 15.4609 | 6261 | 1.7720 | 6075.07 | -0.23 |
| 527 | AD Aps | 231.6460245981 | -78.5129975687 | 14.8091 | 6305 | 1.8237 | 6551.50 | -1.10 |
| 528 | BN Lib | 231.8354976939 | -22.2220071869 | 15.2645 | 6172 | 1.8700 | 5773.85 | -0.41 |
| 529 | V0443 Ser | 232.1634035359 | 5.0140240341 | 13.6819 | 6064 | 1.5968 | 3600.11 | -0.70 |
| 530 | BR Lib | 232.5323990188 | -20.7171607337 | 13.1801 | 6538 | 1.8160 | 2834.96 | -0.90 |
| 531 | ST Boo | 232.6634590882 | 35.7845318116 | 11.0971 | 6876 | 1.6350 | 1301.18 | -1.11 |
| 532 | V0450 Ser | 233.2777123857 | 11.4156683506 | 13.6697 | 6332 | 1.5830 | 3514.99 | -0.64 |
| 533 | QU Lib | 233.2999644642 | -17.9728737122 | 14.8026 | 6409 | 1.4990 | 6376.39 | -0.72 |
| 534 | QR Boo | 233.3348594543 | 43.7408503294 | 14.8757 | 6961 | 1.5285 | 5286.27 | -0.27 |
| 535 | AR Ser | 233.3783851134 | 2.7771909557 | 11.9012 | 6769 | 1.5640 | 1969.83 | -1.13 |
| 536 | V0451 Ser | 233.4459071397 | -1.8856768025 | 15.4286 | 5857 | 1.9590 | 5954.58 | -0.40 |

| # | RRab | R.A. (J2000) | Decl. (J2000) | $G_{mag}$ | $T_{eff}$ | $A_G$ (mag) | d (pc) | $M_G$ |
|---|---|---|---|---|---|---|---|---|
| 537 | CC Lib | 233.4635968800 | -20.3735608449 | 15.0625 | 6265 | 1.7595 | 5068.21 | -0.22 |
| 538 | QS Boo | 233.6762246281 | 48.8695437123 | 14.6806 | 7168 | 1.2458 | 6843.61 | -0.74 |
| 539 | AK Aps | 234.0119742619 | -74.3841292804 | 14.8118 | 5798 | 1.7860 | 4680.60 | -0.33 |
| 540 | AI Aps | 234.0384537387 | -75.0729991603 | 15.3222 | 5332 | 0.5170 | 1213.96 | 4.38 |
| 541 | V0455 Ser | 234.2578187303 | -1.2188811335 | 15.5199 | 6809 | 1.3840 | 6236.97 | 0.16 |
| 542 | V0458 Ser | 234.7994849625 | 3.9234875935 | 13.3596 | 6945 | 1.2620 | 4079.21 | -0.96 |
| 543 | VV Lib | 234.9453128177 | -21.0135843064 | 13.2659 | 5953 | 1.9488 | 2828.22 | -0.94 |
| 544 | AM Aps | 234.9736108991 | -74.3579975106 | 15.5151 | 6600 | 1.5243 | 7024.60 | -0.24 |
| 545 | V0460 Ser | 235.3438708761 | -0.4914550135 | 16.9614 | 5843 | 0.3007 | 3193.01 | 4.14 |
| 546 | EW Lup | 235.7135050473 | -30.1997498273 | 15.3747 | 6050 | 1.1830 | 5570.80 | 0.46 |
| 547 | BN CrB | 235.7955985197 | 36.8915802102 | 13.1579 | 6483 | 1.6645 | 2823.55 | -0.76 |
| 548 | V0464 Ser | 235.8731908786 | -0.1198423349 | 16.0092 | 6121 | 1.6130 | 5853.58 | 0.56 |
| 549 | V0384 Lup | 235.9363293069 | -30.7917264849 | 14.4277 | 6267 | 1.6660 | 4477.88 | -0.49 |
| 550 | GV Aps | 236.1535404533 | -73.9208341987 | 15.1644 | 6822 | 1.6130 | 6164.30 | -0.40 |
| 551 | V0385 Lup | 236.1990055387 | -33.9262170892 | 13.6067 | 5363 | 2.0900 | 3180.45 | -1.00 |
| 552 | V0345 Lib | 236.3512155300 | -19.7016449556 | 14.5234 | 5731 | 1.8630 | 5535.44 | -1.06 |
| 553 | SZ CrB | 236.8577180111 | 29.6617572536 | 13.8169 | 6352 | 1.6270 | 3947.54 | -0.79 |
| 554 | V0471 Ser | 236.8834811808 | -0.2768946977 | 16.0830 | 6026 | 1.8407 | 6562.14 | 0.16 |
| 555 | V0340 Boo | 237.2486554417 | 40.2356775916 | 13.5590 | 6617 | 1.5800 | 4633.31 | -1.35 |
| 556 | BP CrB | 237.2507647883 | 35.2664023618 | 13.7583 | 6898 | 1.4843 | 3616.04 | -0.52 |
| 557 | AU Aps | 237.3116523211 | -76.4570008729 | 14.3018 | 6357 | 1.6913 | 4616.77 | -0.71 |
| 558 | V0478 Ser | 237.6924822428 | -1.3263796302 | 14.9966 | 5798 | 1.9830 | 7182.63 | -1.27 |
| 559 | V0765 Sco | 237.7288272931 | -24.8639878621 | 13.1681 | 5847 | 1.8290 | 2792.83 | -0.89 |
| 560 | VY Lib | 237.8207887142 | -15.7509284302 | 11.7083 | 5646 | 2.1030 | 1229.66 | -0.84 |
| 561 | V0483 Ser | 238.1036161330 | 19.6087361527 | 14.4873 | 6500 | 1.5418 | 4781.81 | -0.45 |
| 562 | V0484 Ser | 238.1221645708 | 10.1427494246 | 12.9654 | 6344 | 1.4180 | 3505.98 | -1.18 |
| 563 | V0487 Ser | 238.3152686871 | 17.5698840389 | 13.9099 | 6935 | 1.5480 | 6070.69 | -1.55 |
| 564 | AN Ser | 238.3793892821 | 12.9611425060 | 11.0441 | 6568 | 1.7245 | 1022.21 | -0.73 |
| 565 | V0353 Lib | 238.5586734055 | -13.5277799728 | 14.1107 | 6018 | 1.8675 | 4318.85 | -0.93 |
| 566 | V0488 Ser | 238.6171806767 | 15.3560496840 | 13.5459 | 7390 | 1.3410 | 5966.48 | -1.67 |
| 567 | AX Aps | 238.7882181106 | -74.0398356639 | 15.5347 | 7055 | 1.5100 | 7039.20 | -0.21 |
| 568 | DU Lib | 238.8084913495 | -13.3184603828 | 14.8087 | 6197 | 1.6040 | 5978.53 | -0.68 |
| 569 | AT Ser | 238.9182155256 | 7.9887773905 | 11.4283 | 6828 | 1.6337 | 1657.61 | -1.30 |
| 570 | FF Lib | 239.2068399259 | -14.6227570856 | 15.7957 | 5630 | 1.9937 | 6478.72 | -0.26 |
| 571 | V0498 Ser | 239.8999233970 | 12.7731075793 | 13.0107 | 6837 | 1.5360 | 3103.05 | -0.98 |
| 572 | FP Lib | 240.0392220476 | -14.3507328938 | 15.9520 | 5588 | 1.7740 | 3627.76 | 1.38 |
| 573 | AV Ser | 240.9243078488 | 0.5991292400 | 11.5071 | 6518 | 1.8708 | 1223.02 | -0.80 |
| 574 | V0865 Sco | 241.5177054728 | -25.8748751609 | 14.5421 | 5890 | 1.8003 | 4613.45 | -0.58 |
| 575 | VY CrB | 241.5481720239 | 33.3709903876 | 14.2160 | 7253 | 1.3545 | 5282.16 | -0.75 |
| 576 | AW Ser | 241.6199675402 | 15.3682831339 | 12.8013 | 6606 | 1.3223 | 3040.77 | -0.94 |
| 577 | V0782 Sco | 241.7187514331 | -12.3820853947 | 16.1813 | 5818 | 1.9940 | 5718.70 | 0.40 |
| 578 | V0677 Her | 242.0172989709 | 24.9889441427 | 13.7224 | 6802 | 1.3300 | 3739.95 | -0.47 |
| 579 | V0513 Ser | 242.0299427883 | 11.3745140882 | 14.1593 | 6207 | 1.5280 | 4682.65 | -0.72 |
| 580 | V0682 Oph | 242.2180020427 | -7.3357227440 | 14.7297 | 5802 | 1.6030 | 4389.19 | -0.09 |
| 581 | V0586 Her | 242.2592553042 | 17.3880278656 | 15.7222 | 6815 | 1.5573 | 5672.49 | 0.40 |
| 582 | BF Aps | 242.3140306498 | -81.0291889246 | 15.1593 | 5463 | 2.2442 | 5586.16 | -0.82 |
| 583 | BI Aps | 242.3906309492 | -74.2443795514 | 15.3702 | 5801 | 1.6935 | 8510.85 | -0.97 |
| 584 | V0790 Sco | 242.4523642262 | -14.6212143652 | 14.4955 | 5689 | 1.8285 | 4337.82 | -0.52 |
| 585 | CF Ser | 242.5300464680 | -3.1306325124 | 14.1324 | 5815 | 1.7535 | 4180.75 | -0.73 |
| 586 | V0793 Sco | 242.6604181210 | -10.2700891613 | 15.3354 | 5380 | 2.0460 | 5477.07 | -0.40 |
| 587 | EE Her | 243.1775394998 | 17.9874571287 | 13.2637 | 6499 | 1.5590 | 3134.66 | -0.78 |
| 588 | V0801 Sco | 243.2234772839 | -12.1623379575 | 15.1426 | 5601 | 1.7850 | 8587.68 | -1.31 |
| 589 | BT Sco | 243.2318003957 | -8.4577502421 | 13.1762 | 5795 | 2.1935 | 2765.83 | -1.23 |
| 590 | V0537 Her | 243.4102323409 | 22.6955746018 | 15.1852 | 6529 | 1.6270 | 5556.79 | -0.17 |
| 591 | V0686 Her | 243.5968781496 | 17.9431115694 | 14.2729 | 6300 | 1.8560 | 4143.23 | -0.67 |
| 592 | V0559 Sco | 243.7667978130 | -9.7853210036 | 13.1930 | 5744 | 2.0647 | 3071.51 | -1.31 |
| 593 | V0685 Oph | 244.0415027487 | -6.7874624019 | 15.4324 | 4969 | 0.8347 | 1421.74 | 3.83 |
| 594 | VZ CrB | 244.1202907044 | 29.9384699674 | 15.5449 | 6855 | 1.5425 | 6021.59 | 0.10 |
| 595 | V0688 Oph | 244.4358953540 | -5.0468230507 | 14.3885 | 5766 | 1.0215 | 2937.28 | 1.03 |
| 596 | BN Aps | 244.5567628810 | -74.2124208127 | 15.6876 | 6336 | 1.8118 | 7744.11 | -0.57 |

| # | RRab | R.A. (J2000) | Decl. (J2000) | $G_{mag}$ | $T_{eff}$ | $A_G$ (mag) | d (pc) | $M_G$ |
|---|------|--------------|---------------|-----------|-----------|-------------|--------|-------|
| 597 | CT CrB | 244.6430746322 | 27.4703383272 | 14.6072 | 6640 | 1.6047 | 5344.66 | -0.64 |
| 598 | V1168 Her | 245.1854227722 | 9.7407269124 | 14.6388 | 6379 | 1.6913 | 4945.17 | -0.52 |
| 599 | BS Aps | 245.2145997769 | -71.6710334822 | 12.1207 | 6287 | 1.6697 | 1772.80 | -0.79 |
| 600 | V1021 Oph | 245.2525377761 | -4.2657551537 | 13.1375 | 5666 | 2.0390 | 2944.80 | -1.25 |
| 601 | BP Aps | 245.2701474813 | -77.0282019865 | 14.6973 | 6027 | 1.5400 | 5471.38 | -0.53 |
| 602 | WY CrB | 245.3716702653 | 29.3336872896 | 15.9819 | 6716 | 1.5680 | 7223.67 | 0.12 |
| 603 | V0693 Oph | 245.4291847556 | -5.3893546939 | 15.4534 | 6130 | 1.6470 | 4376.24 | 0.60 |
| 604 | V1023 Oph | 245.4468996188 | -4.0982717628 | 12.3353 | 5529 | 2.0315 | 1856.73 | -1.04 |
| 605 | EI Nor | 245.5404400709 | -59.6076558152 | 13.6794 | 5810 | 1.6363 | 2502.34 | 0.05 |
| 606 | V0634 Her | 245.7251925128 | 44.6614156733 | 14.9140 | 7554 | 1.2490 | 6535.01 | -0.41 |
| 607 | IY Nor | 245.8178846272 | -58.3726935669 | 15.7383 | 5473 | 1.8287 | 5734.17 | 0.12 |
| 608 | SU CrB | 245.9013078501 | 36.4226642471 | 14.2985 | 7061 | 1.3695 | 5416.23 | -0.74 |
| 609 | V0541 Her | 246.0213268650 | 18.8069811530 | 15.1474 | 6789 | 1.4762 | 4906.21 | 0.22 |
| 610 | V0696 Oph | 246.1269828823 | -8.5354518899 | 15.0234 | 5753 | 1.5745 | 4433.84 | 0.22 |
| 611 | CE TrA | 246.1965494033 | -64.2944679913 | 15.2908 | 5809 | 1.9150 | 5459.03 | -0.31 |
| 612 | V0697 Oph | 246.2290962918 | -1.3628905678 | 15.1556 | 6156 | 1.8898 | 5433.34 | -0.41 |
| 613 | V0413 Oph | 246.2962674631 | -10.5237497840 | 12.0331 | 6248 | 2.1045 | 1139.12 | -0.35 |
| 614 | V0695 Her | 246.4943849789 | 17.7144433941 | 14.1070 | 6508 | 1.4760 | 7053.59 | -1.61 |
| 615 | V0847 Her | 246.9389638945 | 41.6730677049 | 15.6841 | 7118 | 1.3155 | 6446.24 | 0.32 |
| 616 | V0349 Her | 247.0392194707 | 44.3618977001 | 14.9443 | 6532 | 1.6100 | 7444.48 | -1.02 |
| 617 | GR TrA | 247.0562889972 | -69.1395326339 | 13.6790 | 5870 | 0.2300 | 785.75 | 3.97 |
| 618 | V0997 Oph | 247.1023061619 | -11.6342867096 | 15.6120 | 5195 | 2.1470 | 8386.97 | -1.15 |
| 619 | GS Her | 247.1108215986 | 32.1353523150 | 14.1565 | 7205 | 1.6375 | 4565.01 | -0.78 |
| 620 | V0650 Her | 247.2823139490 | 19.1062359663 | 14.3140 | 6140 | 1.8860 | 7223.33 | -1.87 |
| 621 | V1188 Her | 247.3745440686 | 4.4880318932 | 13.4374 | 6529 | 1.3878 | 4320.04 | -1.13 |
| 622 | V0698 Her | 247.3798231577 | 18.4955303236 | 14.6153 | 6330 | 1.5307 | 6906.64 | -1.11 |
| 623 | V0714 Oph | 247.5128270462 | -0.9990190924 | 14.2208 | 6121 | 1.8870 | 5096.50 | -1.20 |
| 624 | GT Her | 247.5774622846 | 34.4596467240 | 13.8589 | 6870 | 1.4385 | 3957.92 | -0.57 |
| 625 | VX Her | 247.6700111496 | 18.3668254110 | 10.8582 | 6477 | 1.7363 | 994.37 | -0.87 |
| 626 | CU TrA | 247.7607387548 | -65.0348367452 | 14.0276 | 6443 | 1.6157 | 4548.25 | -0.88 |
| 627 | V0593 Her | 247.9269052796 | 18.2106580060 | 15.9969 | 6232 | 1.8407 | 5807.20 | 0.34 |
| 628 | V1042 Oph | 248.4545377702 | -1.5491822201 | 14.8669 | 5372 | 2.1570 | 4711.58 | -0.66 |
| 629 | IN Aps | 248.6410167469 | -76.3691329470 | 16.4634 | 6276 | 1.6717 | 5504.68 | 1.09 |
| 630 | V1048 Oph | 248.7693537016 | -3.3658057336 | 15.8324 | 5056 | 2.3657 | 7482.28 | -0.90 |
| 631 | V0850 Her | 248.7967651116 | 42.7738696859 | 15.1243 | 6249 | 1.5365 | 8396.43 | -1.03 |
| 632 | V1201 Her | 248.9219183686 | 28.4133171419 | 14.6996 | 6163 | 1.6273 | 4924.58 | -0.39 |
| 633 | V0601 Her | 249.2035836239 | 24.1674177001 | 14.9992 | 6937 | 1.1780 | 7767.78 | -0.63 |
| 634 | GX Her | 249.2806775539 | 34.9486170357 | 14.7243 | 6684 | 1.5590 | 5564.55 | -0.56 |
| 635 | V0855 Her | 249.6159465984 | 41.1949895905 | 15.7716 | 6567 | 1.6245 | 5994.95 | 0.26 |
| 636 | V0709 Her | 249.6218575081 | 19.8153989641 | 16.6076 | 6614 | 1.3140 | 4644.69 | 1.96 |
| 637 | GZ Her | 249.6306813333 | 33.0418585662 | 13.7587 | 7030 | 1.2350 | 3704.56 | -0.32 |
| 638 | V0545 Her | 249.7060828589 | 24.8630480878 | 14.3868 | 6794 | 1.6535 | 5186.18 | -0.84 |
| 639 | AF Her | 249.9090558625 | 41.1127579894 | 12.9293 | 6384 | 1.6745 | 2666.03 | -0.87 |
| 640 | V0603 Her | 249.9605374001 | 19.7779320931 | 15.1955 | 6020 | 1.6565 | 7752.60 | -0.91 |
| 641 | AG Her | 250.1368895210 | 40.6183599353 | 12.7150 | 6598 | 1.3963 | 3137.09 | -1.16 |
| 642 | CK Aps | 250.5270429091 | -74.2175291063 | 13.4341 | 6347 | 1.8200 | 3857.17 | -1.32 |
| 643 | FP TrA | 250.5364180013 | -69.4126040746 | 14.2942 | 5058 | 0.4130 | 595.18 | 5.01 |
| 644 | V0635 Her | 250.6236188118 | 49.7162635132 | 15.0196 | 6477 | 1.6123 | 5569.36 | -0.32 |
| 645 | CL Aps | 250.6328936027 | -74.5948368120 | 14.8438 | 6736 | 1.5380 | 5144.11 | -0.25 |
| 646 | V0448 Her | 250.6610676350 | 39.8792329433 | 13.4061 | 7024 | 1.6437 | 3756.65 | -1.11 |
| 647 | GT TrA | 250.7111120717 | -67.5657013397 | 14.3073 | 5048 | 0.1120 | 4461.11 | 0.95 |
| 648 | V1007 Oph | 250.9053074397 | -13.8597833792 | 15.0026 | 4813 | 0.4765 | 4897.67 | 1.08 |
| 649 | HI Her | 250.9591129885 | 37.4617806505 | 15.0199 | 6982 | 1.3205 | 5921.72 | -0.16 |
| 650 | CH Aps | 250.9847792426 | -80.9383702931 | 14.2263 | 6201 | 1.9250 | 4094.60 | -0.76 |
| 651 | V0365 Ara | 251.0356963961 | -60.4550569681 | 15.1775 | 5610 | 1.9722 | 6627.93 | -0.90 |
| 652 | V0372 Ara | 251.3687115286 | -57.7148828481 | 15.2851 | 5055 | 2.5295 | 4630.03 | -0.57 |
| 653 | LR Aps | 251.6901826417 | -72.0733916174 | 14.5191 | 6863 | 1.2760 | 6003.93 | -0.65 |
| 654 | V0711 Her | 252.1395480565 | 23.5841323889 | 14.3908 | 6304 | 1.5510 | 7647.40 | -1.58 |
| 655 | BW Ara | 252.2715773752 | -59.1970222497 | 14.9443 | 5838 | 1.8950 | 6506.49 | -1.02 |
| 656 | V0381 Ara | 252.3300121044 | -60.2648605924 | 14.8044 | 5919 | 1.5890 | 6117.10 | -0.72 |

| # | RRab | R.A. (J2000) | Decl. (J2000) | $G_{mag}$ | $T_{eff}$ | $A_G$ (mag) | d (pc) | $M_G$ |
|---|---|---|---|---|---|---|---|---|
| 657 | V0384 Ara | 252.3715732676 | -55.8641466815 | 15.0635 | 5873 | 1.7923 | 5937.56 | -0.60 |
| 658 | CO Aps | 252.4329626554 | -75.6098511952 | 15.5077 | 6600 | 1.6920 | 6891.25 | -0.38 |
| 659 | V0859 Her | 252.4344311851 | 39.6492708403 | 15.7103 | 6814 | 1.6090 | 6624.20 | 0.00 |
| 660 | V0778 Her | 252.4590924071 | 42.7149598833 | 16.3780 | 5017 | 0.1175 | 889.58 | 6.51 |
| 661 | BY Ara | 252.4846983426 | -58.8872069353 | 15.0753 | 6371 | 1.6570 | 6167.28 | -0.53 |
| 662 | V2509 Oph | 252.8743730937 | 6.3739342824 | 13.1914 | 6016 | 1.6962 | 3578.61 | -1.27 |
| 663 | V0399 Ara | 253.3274129727 | -61.6884425341 | 15.5664 | 5600 | 1.8083 | 5096.94 | 0.22 |
| 664 | V0692 Ara | 253.3916566158 | -56.9129005026 | 15.2956 | 5250 | 2.2957 | 6107.11 | -0.93 |
| 665 | V0401 Ara | 253.4320664162 | -56.0631582818 | 14.6895 | 5550 | 1.9857 | 4648.23 | -0.63 |
| 666 | V0697 Ara | 253.5296610012 | -56.1868729319 | 15.5957 | 5457 | 2.2105 | 5960.64 | -0.49 |
| 667 | CN Oph | 253.7900660223 | -29.1568557214 | 13.6757 | 5785 | 1.8635 | 4264.16 | -1.34 |
| 668 | V0406 Ara | 253.9343005717 | -56.9152871708 | 16.3876 | 5857 | 1.9107 | 6834.31 | 0.30 |
| 669 | V0703 Ara | 254.0859251997 | -56.0189276669 | 15.0879 | 6273 | 1.7500 | 5707.09 | -0.44 |
| 670 | V0408 Ara | 254.1288594055 | -57.6709680917 | 13.5562 | 6043 | 1.6803 | 3886.83 | -1.07 |
| 671 | V0407 Ara | 254.1682230807 | -59.5357311686 | 15.8140 | 6160 | 1.4130 | 8112.02 | -0.14 |
| 672 | V0409 Ara | 254.1731324405 | -56.1757837143 | 15.7281 | 5640 | 1.9283 | 6103.49 | -0.13 |
| 673 | DH Oph | 254.1778442755 | -28.4732031306 | 14.0543 | 5985 | 1.8410 | 3855.75 | -0.72 |
| 674 | V1132 Oph | 254.2527508178 | -18.6446860917 | 14.4282 | 5324 | 2.2110 | 3367.67 | -0.42 |
| 675 | V1135 Oph | 254.2652381293 | -18.4894169856 | 14.5595 | 5292 | 2.2440 | 4726.22 | -1.06 |
| 676 | HO Her | 254.3367044538 | 30.3576154432 | 13.3745 | 6846 | 1.4717 | 3267.49 | -0.67 |
| 677 | V0707 Ara | 254.4201083422 | -57.2159377857 | 14.4290 | 5633 | 1.9783 | 4235.56 | -0.68 |
| 678 | V0414 Ara | 254.4614069584 | -59.1229801340 | 13.4247 | 5954 | 1.9200 | 3037.36 | -0.91 |
| 679 | V0415 Ara | 254.4907337721 | -58.6159010836 | 15.4532 | 5785 | 1.8625 | 7237.36 | -0.71 |
| 680 | HR Her | 254.6831963023 | 33.5851950366 | 15.7498 | 6967 | 1.4343 | 7454.38 | -0.05 |
| 681 | V1164 Oph | 254.7626887443 | -20.1954816499 | 15.2301 | 5306 | 2.2640 | 5820.41 | -0.86 |
| 682 | V0418 Ara | 254.8550901131 | -63.3246709762 | 16.0983 | 6488 | 1.5550 | 6597.40 | 0.45 |
| 683 | V0424 Ara | 254.9664417986 | -62.8030534662 | 15.8569 | 6270 | 1.8820 | 5965.44 | 0.10 |
| 684 | V0434 Ara | 255.0762524180 | -59.1678582518 | 15.4311 | 5536 | 0.3655 | 1546.97 | 4.12 |
| 685 | OY Oph | 255.1491580803 | -28.7510827302 | 15.0772 | 5365 | 2.0955 | 5276.82 | -0.63 |
| 686 | V1211 Oph | 255.1973872583 | -19.7024514934 | 15.7090 | 5859 | 1.6820 | 6028.59 | 0.13 |
| 687 | V1218 Oph | 255.3054654100 | -19.1362038830 | 15.7491 | 5640 | 0.5428 | 9743.28 | 0.26 |
| 688 | FO Aps | 255.3548593390 | -71.6768203478 | 14.8766 | 6044 | 1.9035 | 6592.04 | -1.12 |
| 689 | V1228 Oph | 255.3605757144 | -18.4673322332 | 14.8135 | 4969 | 0.8530 | 905.73 | 4.18 |
| 690 | PX Oph | 255.3966958768 | -30.0187498547 | 15.5671 | 5023 | 2.6228 | 7954.18 | -1.56 |
| 691 | QQ Oph | 255.4627285788 | -29.7360565043 | 15.8886 | 4951 | 0.9345 | 1692.18 | 3.81 |
| 692 | V0443 Ara | 255.5083260299 | -59.1289251778 | 15.3045 | 5861 | 1.9133 | 5999.95 | -0.50 |
| 693 | V0786 Her | 255.5921133418 | 45.9466156734 | 15.1205 | 7008 | 1.6430 | 5341.59 | -0.16 |
| 694 | V1256 Oph | 255.6182069911 | -18.3532015329 | 16.1552 | 4820 | 0.3180 | 7393.96 | 1.49 |
| 695 | V0370 Oph | 255.7111792926 | 1.7541755485 | 14.4456 | 5714 | 1.8950 | 5588.02 | -1.19 |
| 696 | V1292 Oph | 255.9693396488 | -20.0630448659 | 16.0715 | 5335 | 2.2980 | 7745.79 | -0.67 |
| 697 | V0453 Her | 256.0432497574 | 22.8653196276 | 16.5642 | 6588 | 1.6270 | 6183.88 | 0.98 |
| 698 | V0450 Ara | 256.1230249113 | -61.6580197811 | 14.8220 | 5735 | 1.7160 | 5640.79 | -0.65 |
| 699 | V2123 Oph | 256.2149489869 | -24.7729855723 | 16.2925 | 4858 | 0.6470 | 1170.40 | 5.30 |
| 700 | V0458 Ara | 256.2939040903 | -58.7647681963 | 14.9264 | 6221 | 1.8910 | 5267.88 | -0.57 |
| 701 | V0465 Ara | 256.3961309435 | -55.3623396825 | 15.4513 | 5455 | 1.8250 | 7036.53 | -0.61 |
| 702 | V1345 Oph | 256.4060444194 | -18.1035123330 | 16.7782 | 5360 | 2.0867 | 6680.62 | 0.57 |
| 703 | V0463 Ara | 256.4108295512 | -57.8253783166 | 15.1335 | 6021 | 1.8460 | 6491.81 | -0.77 |
| 704 | V0365 Her | 256.4162417171 | 21.5162233175 | 13.1618 | 6563 | 1.6533 | 4147.66 | -1.58 |
| 705 | CX Aps | 256.4288036001 | -79.6368580399 | 16.0869 | 6359 | 0.1657 | 7872.99 | 1.44 |
| 706 | V0462 Ara | 256.4308725764 | -58.6340480383 | 15.8050 | 6465 | 1.6978 | 8333.15 | -0.50 |
| 707 | V0464 Ara | 256.5070059884 | -59.6719640526 | 15.1353 | 5975 | 1.7778 | 8416.37 | -1.27 |
| 708 | V1366 Oph | 256.5876838509 | -19.2778081715 | 15.1093 | 5804 | 1.9110 | 8393.58 | -1.42 |
| 709 | V0619 Her | 256.5881558372 | 31.8884733251 | 14.2752 | 6354 | 1.6120 | 4729.27 | -0.71 |
| 710 | V1386 Oph | 256.7421911513 | -20.0555748035 | 16.3882 | 5741 | 2.0313 | 5976.69 | 0.47 |
| 711 | V1395 Oph | 256.8073333182 | -21.6226569145 | 16.0290 | 4865 | 2.4890 | 5810.64 | -0.28 |
| 712 | V1398 Oph | 256.8206043754 | -20.1728862647 | 16.1854 | 5354 | 2.2463 | 8025.64 | -0.58 |
| 713 | V1416 Oph | 256.9402239866 | -18.6888108949 | 15.9075 | 5748 | 2.0633 | 8256.21 | -0.74 |
| 714 | V0473 Ara | 257.0135455350 | -58.6073412694 | 15.6641 | 6121 | 1.8077 | 7294.01 | -0.46 |
| 715 | V0472 Ara | 257.1099823691 | -62.5806096904 | 14.2090 | 5843 | 1.6868 | 4046.46 | -0.51 |
| 716 | V1438 Oph | 257.1178220533 | -20.5685599737 | 16.2527 | 5362 | 2.1497 | 8010.85 | -0.42 |

| #   | RRab     | R.A. (J2000)   | Decl. (J2000)   | $G_{mag}$ | $T_{eff}$ | $A_G$ (mag) | d (pc)  | $M_G$ |
|-----|----------|----------------|-----------------|-----------|-----------|-------------|---------|-------|
| 717 | V1462 Oph | 257.2645089288 | -18.8003668678  | 16.3990   | 5306      | 2.0573      | 7419.60 | -0.01 |
| 718 | V0459 Her | 257.2846032270 | 27.7419293726   | 15.4413   | 6199      | 1.6710      | 8047.14 | -0.76 |
| 719 | V1473 Oph | 257.3599956918 | -21.5719235068  | 16.8496   | 4932      | 0.2520      | 6755.45 | 2.45  |
| 720 | V0790 Her | 257.4220093895 | 46.3534432920   | 16.2809   | 6540      | 0.6443      | 6273.88 | 1.65  |
| 721 | V1485 Oph | 257.4519816603 | -21.8154077038  | 17.0295   | 4865      | 2.6980      | 6132.47 | 0.39  |
| 722 | V1494 Oph | 257.4806650941 | -18.7413726870  | 15.9214   | 5640      | 2.0710      | 6501.05 | -0.21 |
| 723 | V1504 Oph | 257.5689246258 | -19.4873404787  | 13.6730   | 4258      | 1.0302      | 2158.10 | 0.97  |
| 724 | V0791 Her | 257.5689378424 | 46.8363555733   | 16.6457   | 7135      | 0.9107      | 3837.72 | 2.81  |
| 725 | V0488 Ara | 257.6497914461 | -58.9267023543  | 15.6116   | 6496      | 1.6130      | 6504.85 | -0.07 |
| 726 | V0721 Her | 257.7026614969 | 41.6188282191   | 15.2182   | 6869      | 1.6543      | 6147.25 | -0.38 |
| 727 | V1531 Oph | 257.7537982354 | -19.8803770736  | 16.0705   | 5762      | 1.8050      | 5730.25 | 0.47  |
| 728 | V0724 Her | 257.7901791781 | 42.5665418294   | 14.4923   | 6841      | 1.4170      | 6603.86 | -1.02 |
| 729 | V0794 Her | 257.8377325500 | 44.9001476159   | 15.4828   | 6565      | 1.1682      | 7509.63 | -0.06 |
| 730 | V1550 Oph | 257.9043112935 | -18.1978305334  | 15.9251   | 5339      | 2.3580      | 5330.68 | -0.07 |
| 731 | V1570 Oph | 257.9983896666 | -18.1912622326  | 15.2357   | 5550      | 1.8058      | 4685.78 | 0.08  |
| 732 | V1574 Oph | 258.0187275781 | -18.3640846764  | 15.8876   | 5356      | 2.1457      | 8305.34 | -0.85 |
| 733 | V1576 Oph | 258.0420853594 | -18.7914872021  | 15.3577   | 5121      | 2.3810      | 7041.71 | -1.26 |
| 734 | V0370 Her | 258.1375104703 | 20.5398693083   | 15.2661   | 6396      | 1.5665      | 5969.21 | -0.18 |
| 735 | V1595 Oph | 258.1777083297 | -21.3177051357  | 16.5462   | 4956      | 0.7040      | 8581.98 | 1.17  |
| 736 | V0467 Her | 258.2116553573 | 25.0301499419   | 14.2508   | 6816      | 1.6135      | 4774.88 | -0.76 |
| 737 | V1634 Oph | 258.3994825615 | -18.7621982794  | 15.8664   | 5512      | 1.9210      | 6792.51 | -0.21 |
| 738 | V0468 Her | 258.4167448352 | 20.9801789970   | 14.8325   | 6294      | 1.6553      | 6729.13 | -0.96 |
| 739 | V1649 Oph | 258.4764433577 | -19.2198407272  | 15.5339   | 5359      | 2.2185      | 5229.68 | -0.28 |
| 740 | V0632 Ara | 258.6005690246 | -67.5581808162  | 15.5881   | 6373      | 1.0810      | 9954.03 | -0.48 |
| 741 | V0379 Her | 258.6472862938 | 22.3949168848   | 14.4635   | 6519      | 1.7060      | 6302.74 | -1.24 |
| 742 | V1674 Oph | 258.6473631866 | -18.8717871417  | 16.7648   | 5712      | 1.5895      | 6483.42 | 1.12  |
| 743 | V0503 Ara | 258.8534440209 | -59.6170042520  | 14.9174   | 5999      | 1.8550      | 5414.87 | -0.61 |
| 744 | V0506 Ara | 258.9117650434 | -56.8284060049  | 15.4480   | 5349      | 2.1033      | 5371.90 | -0.31 |
| 745 | V1722 Oph | 258.9826797469 | -16.7762166025  | 16.0464   | 5049      | 2.5450      | 5669.59 | -0.27 |
| 746 | V1719 Oph | 258.9970684657 | -18.7665935188  | 15.8607   | 5332      | 2.2020      | 5329.67 | 0.03  |
| 747 | V0736 Ara | 259.0663753339 | -59.7068348584  | 13.9928   | 5761      | 1.5952      | 3337.73 | -0.22 |
| 748 | DH Ara    | 259.1119002674 | -55.1011857817  | 14.4210   | 5459      | 1.9710      | 4287.92 | -0.71 |
| 749 | V1740 Oph | 259.1271960115 | -18.6310682266  | 16.1948   | 5306      | 2.3770      | 5890.48 | -0.03 |
| 750 | V0386 Her | 259.3516679862 | 26.8115981220   | 16.0072   | 6709      | 1.3660      | 6522.87 | 0.57  |
| 751 | V1787 Oph | 259.4084385896 | -20.3277078373  | 16.0306   | 4785      | 0.3475      | 6634.64 | 1.57  |
| 752 | DP Aps    | 259.4125691507 | -74.4520225239  | 16.0372   | 6358      | 1.1245      | 8258.76 | 0.33  |
| 753 | V0452 Oph | 259.4986550369 | 11.0744391193   | 12.2052   | 5488      | 2.1040      | 1768.45 | -1.14 |
| 754 | V1812 Oph | 259.5552790820 | -20.4444476616  | 15.8485   | 5136      | 2.2933      | 5486.67 | -0.14 |
| 755 | FR Aps    | 259.6185031397 | -68.4726772257  | 15.2107   | 6392      | 1.6400      | 7651.37 | -0.85 |
| 756 | V0800 Her | 259.7196807323 | 42.4793932856   | 15.7428   | 6533      | 1.4893      | 5695.14 | 0.48  |
| 757 | V1843 Oph | 259.7412742724 | -18.2884561422  | 16.0059   | 5370      | 2.1653      | 6264.57 | -0.14 |
| 758 | V0517 Ara | 259.7480631557 | -55.3230291618  | 14.0515   | 5842      | 1.7860      | 7525.18 | -2.12 |
| 759 | MV Ara    | 259.8303160518 | -57.6805213209  | 15.3693   | 6086      | 1.8470      | 5986.66 | -0.36 |
| 760 | V0476 Her | 260.2033019082 | 22.0977384002   | 15.6795   | 6882      | 1.1360      | 7755.48 | 0.10  |
| 761 | V0392 Her | 260.2142249278 | 26.5389881736   | 13.1012   | 6389      | 1.4913      | 2920.80 | -0.72 |
| 762 | V1921 Oph | 260.2584647643 | -18.1965172188  | 16.6332   | 5211      | 2.3485      | 6324.84 | 0.28  |
| 763 | V1923 Oph | 260.2663274459 | -18.9846199581  | 16.5840   | 4492      | 1.1610      | 1519.75 | 4.51  |
| 764 | V1946 Oph | 260.3888893053 | -18.4480312782  | 15.9867   | 5810      | 2.0325      | 5945.31 | 0.08  |
| 765 | V2639 Oph | 260.5296031207 | 3.1596958347    | 13.5748   | 5802      | 1.9605      | 3284.20 | -0.97 |
| 766 | V1971 Oph | 260.5442773111 | -17.0987015570  | 16.7766   | 5262      | 2.1637      | 6425.33 | 0.57  |
| 767 | V1968 Oph | 260.5677101353 | -20.4320824153  | 15.1302   | 4574      | 0.9295      | 5528.17 | 0.49  |
| 768 | LS Aps    | 260.5788932433 | -71.0391573244  | 15.2959   | 6017      | 1.3907      | 9195.61 | -0.91 |
| 769 | V0756 Oph | 260.6266378686 | 1.7799032267    | 13.7179   | 5600      | 1.9470      | 2953.75 | -0.58 |
| 770 | V0734 Her | 260.7283562523 | 46.2067533122   | 13.6515   | 6507      | 1.5633      | 3998.39 | -0.92 |
| 771 | V0398 Her | 260.7516175891 | 27.7337506859   | 16.0916   | 6614      | 1.6270      | 6778.50 | 0.31  |
| 772 | V0397 Her | 260.7843677975 | 22.6582124250   | 13.5643   | 6926      | 1.4360      | 4439.43 | -1.11 |
| 773 | V0399 Her | 260.8875470785 | 26.1493531030   | 15.9173   | 6772      | 0.8680      | 8363.89 | 0.44  |
| 774 | V2022 Oph | 260.9864785263 | -20.5924588849  | 16.4139   | 4861      | 0.2110      | 6534.80 | 2.13  |
| 775 | V0736 Her | 261.0935879070 | 41.3255589036   | 15.2975   | 6668      | 1.6225      | 5239.01 | 0.08  |
| 776 | DR Aps    | 261.2859876887 | -80.6128762387  | 14.7167   | 6302      | 1.8510      | 5943.09 | -1.00 |

| # | RRab | R.A. (J2000) | Decl. (J2000) | $G_{mag}$ | $T_{eff}$ | $A_G$ (mag) | d (pc) | $M_G$ |
|---|---|---|---|---|---|---|---|---|
| 777 | V0486 Her | 261.6598643773 | 26.9382572070 | 13.0825 | 6318 | 1.6805 | 3470.11 | -1.30 |
| 778 | V0402 Her | 261.8089895803 | 24.0454892785 | 15.7799 | 6299 | 1.6355 | 6651.54 | 0.03 |
| 779 | DZ Aps | 262.2384378412 | -73.1898144164 | 15.3277 | 6910 | 1.6148 | 7757.88 | -0.74 |
| 780 | EY Ara | 262.3744641002 | -56.3389156879 | 14.8346 | 6176 | 1.6710 | 8139.54 | -1.39 |
| 781 | V0735 Her | 262.5616073461 | 39.6833198802 | 16.2639 | 7325 | 1.4497 | 6798.59 | 0.65 |
| 782 | FM Ara | 262.8903428679 | -58.6226844267 | 14.9049 | 6640 | 1.5700 | 5659.88 | -0.43 |
| 783 | LT Aps | 263.0524562730 | -72.1449325412 | 14.5771 | 5391 | 0.4575 | 1037.02 | 4.04 |
| 784 | V0418 Her | 263.0541952125 | 18.2335123913 | 12.5966 | 6377 | 1.5945 | 2408.73 | -0.91 |
| 785 | V0419 Her | 263.1904856385 | 18.9412524021 | 15.6297 | 6758 | 1.6200 | 6023.94 | 0.11 |
| 786 | ST Oph | 263.4973849199 | -1.0808519210 | 12.3152 | 5279 | 2.1110 | 1270.68 | -0.32 |
| 787 | V0813 Her | 263.7243952027 | 46.6775506713 | 14.8816 | 6566 | 1.5687 | 6996.37 | -0.91 |
| 788 | V0425 Her | 263.8402260113 | 26.8724880941 | 14.5735 | 6806 | 1.3873 | 5590.29 | -0.55 |
| 789 | EM Aps | 263.8552753660 | -75.0808532541 | 15.6908 | 6341 | 1.5907 | 6732.07 | -0.04 |
| 790 | EL Aps | 263.9218226570 | -76.2210390396 | 11.8084 | 5370 | 2.0287 | 1356.83 | -0.88 |
| 791 | V0788 Oph | 264.0376416451 | 8.1649002155 | 14.3029 | 5829 | 1.8040 | 4130.82 | -0.58 |
| 792 | V0520 Sco | 264.1896988366 | -41.3034047791 | 14.8494 | 3975 | 1.2555 | 4176.45 | 0.49 |
| 793 | GL Ara | 264.3867685648 | -63.4555383688 | 14.7935 | 6579 | 1.3107 | 6060.02 | -0.43 |
| 794 | V0506 Her | 264.5471535922 | 18.8881073473 | 14.1274 | 6957 | 1.5213 | 7433.68 | -1.75 |
| 795 | V0507 Her | 264.6184247651 | 18.2567867449 | 14.2972 | 5939 | 1.5900 | 5152.43 | -0.85 |
| 796 | V0494 Sco | 265.2019897769 | -31.5421603082 | 11.3714 | 6192 | 1.8660 | 1055.41 | -0.61 |
| 797 | V2646 Oph | 265.2285357046 | 4.6911714171 | 14.5991 | 5304 | 2.2340 | 4228.45 | -0.77 |
| 798 | V0514 Her | 265.2373330748 | 24.0489220173 | 13.8469 | 6489 | 1.6090 | 4207.91 | -0.88 |
| 799 | V0866 Oph | 265.2908741084 | -0.1590985025 | 13.9157 | 5669 | 1.9128 | 4011.54 | -1.01 |
| 800 | V0743 Her | 265.2996734323 | 42.2106719354 | 14.2566 | 6576 | 1.5917 | 4457.41 | -0.58 |
| 801 | V0555 Oph | 265.5598944952 | 5.3997471049 | 14.7616 | 5332 | 2.1700 | 5220.02 | -1.00 |
| 802 | V0817 Oph | 265.6858577475 | 8.1464362893 | 15.7849 | 5781 | 1.9753 | 6069.38 | -0.11 |
| 803 | LU Aps | 265.7941381470 | -70.3960158525 | 12.7659 | 6302 | 1.5460 | 2803.87 | -1.02 |
| 804 | LX Her | 265.8280515907 | 28.2543354834 | 14.4678 | 6855 | 1.4250 | 6505.29 | -1.02 |
| 805 | V0521 Her | 265.9334996850 | 23.0035264989 | 14.1519 | 5568 | 1.8650 | 4181.92 | -0.82 |
| 806 | V0534 Sco | 265.9472736520 | -44.4336383459 | 15.6229 | 5844 | 1.8523 | 5833.38 | -0.06 |
| 807 | LZ Ara | 266.0252800907 | -46.5017177252 | 15.5488 | 5584 | 1.9630 | 6364.21 | -0.43 |
| 808 | V0536 Sco | 266.0433547830 | -43.7367373536 | 13.7815 | 5599 | 2.0753 | 2562.73 | -0.34 |
| 809 | EU Aps | 266.2312448227 | -74.8687221781 | 13.8601 | 6304 | 1.4740 | 4287.99 | -0.78 |
| 810 | HL Ara | 266.4554122315 | -56.9965593906 | 14.6403 | 6448 | 1.7393 | 6205.23 | -1.06 |
| 811 | V0829 Oph | 267.3718739311 | 12.2317705219 | 13.9735 | 6483 | 1.8240 | 6440.25 | -1.90 |
| 812 | OU Ara | 267.4056967868 | -56.6469281966 | 15.2067 | 5734 | 0.4347 | 1449.40 | 3.97 |
| 813 | MP Ara | 267.6719056495 | -46.8947613712 | 15.1086 | 6352 | 1.5530 | 6187.84 | -0.40 |
| 814 | VW Pav | 268.1057379328 | -58.8773527564 | 13.8247 | 6748 | 1.4600 | 4157.10 | -0.73 |
| 815 | MQ Ara | 268.2924878073 | -46.8009960363 | 14.9380 | 6704 | 1.6660 | 5562.15 | -0.45 |
| 816 | V0554 Sco | 268.2949441559 | -40.4988491200 | 14.5010 | 5788 | 1.1380 | 3879.19 | 0.42 |
| 817 | VX Pav | 268.3591213356 | -61.8962599457 | 13.9585 | 6263 | 1.6953 | 3850.67 | -0.66 |
| 818 | V0762 Sco | 268.4040134016 | -43.3293818520 | 13.9509 | 4037 | 0.8243 | 5398.76 | -0.53 |
| 819 | V0387 Dra | 268.5538336561 | 51.0229483656 | 12.0803 | 6896 | 1.6135 | 1985.74 | -1.02 |
| 820 | VZ Pav | 268.7633506980 | -61.6015532711 | 14.2643 | 6312 | 1.8250 | 5291.96 | -1.18 |
| 821 | V0775 Sgr | 268.7792814177 | -28.6994995662 | 14.2357 | 4339 | 1.2235 | 2707.29 | 0.85 |
| 822 | EP Her | 268.7890695818 | 26.6051894164 | 13.2245 | 6448 | 1.6595 | 3013.03 | -0.83 |
| 823 | WW Pav | 268.8925580078 | -59.5746664450 | 14.3364 | 6546 | 1.6657 | 4180.93 | -0.44 |
| 824 | WY Pav | 269.0515090685 | -57.1620624790 | 11.9771 | 6093 | 1.7767 | 1557.54 | -0.76 |
| 825 | V0682 Sco | 269.1812113676 | -39.0946777566 | 15.5242 | 3928 | 1.0673 | 6076.41 | 0.54 |
| 826 | V2033 Oph | 269.4075755239 | 4.8545875243 | 15.5393 | 5365 | 2.2850 | 4705.81 | -0.11 |
| 827 | V1013 Oph | 269.4941904231 | 5.5767749144 | 14.6871 | 6004 | 1.8383 | 5980.87 | -1.04 |
| 828 | V0694 Sco | 269.5813027434 | -38.7038853552 | 14.3996 | 4025 | 1.0763 | 3981.07 | 0.32 |
| 829 | V0452 CrA | 269.8566143454 | -44.6851984728 | 14.5335 | 5522 | 0.6540 | 1499.11 | 3.00 |
| 830 | V0459 CrA | 270.0401290288 | -40.3594825904 | 15.5845 | 5811 | 1.6620 | 3412.48 | 1.26 |
| 831 | V0946 Oph | 270.2680208397 | 1.6456625367 | 14.7254 | 5345 | 1.9380 | 4124.23 | -0.29 |
| 832 | LM CrA | 270.4104731230 | -39.8774534719 | 15.6414 | 6114 | 1.7450 | 6990.43 | -0.33 |
| 833 | V0468 CrA | 270.4453289098 | -42.5985371934 | 13.9420 | 6240 | 1.6283 | 4625.11 | -1.01 |
| 834 | V0394 Dra | 270.4684286994 | 60.1120284324 | 12.1684 | 6938 | 1.5785 | 1827.17 | -0.72 |
| 835 | V1329 Sgr | 270.5342043518 | -29.7843231594 | 14.7669 | 3635 | 1.5050 | 5145.09 | -0.30 |
| 836 | OT Her | 270.5367252952 | 40.2176753486 | 16.0881 | 6866 | 1.6520 | 7117.86 | 0.17 |

| # | RRab | R.A. (J2000) | Decl. (J2000) | $G_{mag}$ | $T_{eff}$ | $A_G$ (mag) | d (pc) | $M_G$ |
|---|---|---|---|---|---|---|---|---|
| 837 | EP CrA | 270.5598786121 | -37.3597673468 | 14.5196 | 5869 | 1.8342 | 4725.01 | -0.69 |
| 838 | LS CrA | 270.5679245450 | -38.7180475889 | 14.8479 | 6301 | 1.7270 | 8344.45 | -1.49 |
| 839 | V0667 Her | 270.6282503089 | 35.3725002909 | 15.8093 | 6600 | 1.6460 | 6277.77 | 0.17 |
| 840 | V0527 Her | 270.6440424449 | 28.5454914727 | 15.3729 | 7103 | 1.3470 | 6598.44 | -0.07 |
| 841 | V0472 CrA | 270.7432192335 | -40.7442267848 | 13.6191 | 4893 | 0.5025 | 362.28 | 5.32 |
| 842 | OW Her | 270.7535992232 | 36.4793280507 | 13.9056 | 6721 | 1.4550 | 5208.10 | -1.13 |
| 843 | IK Ara | 270.7589529901 | -54.6029333030 | 14.4489 | 6172 | 1.7590 | 4576.55 | -0.61 |
| 844 | EZ CrA | 270.8389223868 | -37.4909720899 | 15.0990 | 4476 | 0.4270 | 8021.04 | 0.15 |
| 845 | OX Her | 270.8598965891 | 38.6947526265 | 13.0114 | 6500 | 1.7633 | 3512.17 | -1.48 |
| 846 | V0629 Her | 270.9618674329 | 32.0993745035 | 16.0816 | 6400 | 0.5873 | 6482.30 | 1.44 |
| 847 | V0396 CrA | 270.9753368502 | -43.6142534944 | 13.9170 | 6424 | 1.5592 | 4498.71 | -0.91 |
| 848 | V0575 Sgr | 270.9817668317 | -36.8293846580 | 15.8564 | 5951 | 1.9630 | 7901.63 | -0.60 |
| 849 | V1007 Sgr | 270.9958840492 | -34.9363194919 | 15.3322 | 5313 | 2.2540 | 7677.06 | -1.35 |
| 850 | V1514 Sgr | 271.1151882482 | -29.8270745842 | 15.1658 | 4013 | 2.4107 | 4510.87 | -0.52 |
| 851 | V0481 CrA | 271.1259430056 | -40.7310876190 | 15.4801 | 5785 | 0.4850 | 2021.16 | 3.47 |
| 852 | V0482 CrA | 271.1331264570 | -41.2834678668 | 14.6948 | 6416 | 1.8470 | 4982.69 | -0.64 |
| 853 | V2336 Oph | 271.1432836730 | 8.3301724400 | 13.7454 | 5368 | 2.2313 | 3459.24 | -1.18 |
| 854 | WW CrA | 271.4033052226 | -43.8326255210 | 11.9858 | 6600 | 1.6603 | 1669.68 | -0.79 |
| 855 | V0600 Sgr | 271.4706971574 | -36.1221302965 | 15.6594 | 5669 | 2.0390 | 5977.83 | -0.26 |
| 856 | OV CrA | 271.5261103117 | -38.4883824502 | 13.3226 | 4933 | 0.3562 | 4350.30 | -0.23 |
| 857 | V0398 CrA | 271.6778578840 | -41.0644463308 | 15.5911 | 6518 | 1.6930 | 7711.89 | -0.54 |
| 858 | IN Ara | 271.9175211708 | -54.6079867805 | 13.2167 | 5852 | 1.8960 | 2747.46 | -0.87 |
| 859 | GK CrA | 272.0016782122 | -37.7541378026 | 15.1113 | 6041 | 1.5550 | 6854.45 | -0.62 |
| 860 | V4252 Sgr | 272.0821210215 | -30.9863142445 | 13.1821 | 3600 | 0.6400 | 7647.37 | -1.88 |
| 861 | PY CrA | 272.1779658059 | -38.9327589980 | 14.4816 | 6514 | 1.7485 | 5774.36 | -1.07 |
| 862 | V0497 CrA | 272.2459042573 | -44.5891779669 | 14.9056 | 6272 | 1.6810 | 5967.87 | -0.65 |
| 863 | QS CrA | 272.2699607358 | -38.8921181887 | 15.5036 | 4392 | 0.4520 | 6428.26 | 1.01 |
| 864 | V0499 CrA | 272.2843220858 | -40.9661038866 | 15.4061 | 6021 | 1.6550 | 6410.95 | -0.28 |
| 865 | V4342 Sgr | 272.2896439463 | -31.9223324068 | 16.0859 | 4575 | 1.2042 | 1655.02 | 3.79 |
| 866 | V0400 CrA | 272.3846885391 | -41.7445060979 | 15.1791 | 6600 | 1.3378 | 9639.13 | -1.08 |
| 867 | RV Pav | 272.3913168047 | -59.4562306847 | 13.6050 | 6060 | 1.9130 | 4425.18 | -1.54 |
| 868 | V4293 Sgr | 272.4916331057 | -31.0790304169 | 15.7005 | 5084 | 0.0170 | 917.45 | 5.87 |
| 869 | V4347 Sgr | 272.5534682259 | -32.0292101764 | 16.4921 | 4981 | 1.2540 | 3161.18 | 2.74 |
| 870 | V0642 Sgr | 272.5936280729 | -36.4937441775 | 15.4053 | 4321 | 0.6117 | 6774.30 | 0.64 |
| 871 | V0674 Her | 272.6131701414 | 39.8707746215 | 15.7283 | 6505 | 1.8530 | 5778.03 | 0.07 |
| 872 | V0571 Her | 272.6372573035 | 36.6380184651 | 15.7195 | 6392 | 1.5980 | 7817.62 | -0.34 |
| 873 | V0337 CrA | 272.6720873756 | -38.7073772368 | 15.0592 | 5450 | 1.4135 | 1840.82 | 2.32 |
| 874 | V1332 Her | 272.7032761719 | 17.2083048548 | 12.8232 | 6195 | 1.7260 | 2445.01 | -0.84 |
| 875 | V4319 Sgr | 272.7088922988 | -30.7495627510 | 16.0107 | 5067 | 0.8920 | 1848.43 | 3.78 |
| 876 | V4322 Sgr | 272.7422422665 | -31.1880093153 | 16.1579 | 4837 | 1.7190 | 2121.41 | 2.81 |
| 877 | V4611 Sgr | 272.7621730222 | -32.3708557034 | 15.9243 | 5122 | 0.0730 | 6499.46 | 1.79 |
| 878 | MQ Her | 272.8447586346 | 29.1860348205 | 15.0060 | 5683 | 0.8450 | 1603.96 | 3.14 |
| 879 | V0888 Her | 272.8780492803 | 27.9942303319 | 16.8600 | 7008 | 1.4433 | 6663.90 | 1.30 |
| 880 | V2626 Sgr | 273.0586562134 | -35.3174174899 | 14.9431 | 5013 | 0.2457 | 734.91 | 5.37 |
| 881 | V2639 Sgr | 273.1562159439 | -35.1326019445 | 15.1899 | 5394 | 2.0090 | 5708.16 | -0.60 |
| 882 | V0961 Oph | 273.1852164230 | 4.3265400532 | 14.3526 | 6301 | 1.3883 | 3607.97 | 0.18 |
| 883 | V2645 Sgr | 273.1993721070 | -35.0933674664 | 15.8279 | 5602 | 1.9660 | 5468.31 | 0.17 |
| 884 | QR Her | 273.2268263986 | 33.5927283562 | 14.8495 | 6610 | 1.6460 | 5198.61 | -0.38 |
| 885 | V0442 Her | 273.2434735567 | 42.0627638993 | 12.8724 | 7137 | 1.1320 | 2609.33 | -0.34 |
| 886 | V2663 Sgr | 273.2773846197 | -35.7578899910 | 15.6097 | 5437 | 2.2040 | 5414.28 | -0.26 |
| 887 | V2689 Sgr | 273.3578399158 | -33.5379734541 | 16.0680 | 5492 | 2.0620 | 6405.90 | -0.03 |
| 888 | V0675 Sgr | 273.3975378457 | -34.3171814217 | 10.3024 | 6064 | 1.5930 | 881.66 | -1.02 |
| 889 | V0517 CrA | 273.4242392042 | -40.9016078707 | 14.9248 | 6614 | 1.7170 | 9550.79 | -1.69 |
| 890 | V1283 Sgr | 273.4475517250 | -29.5277992733 | 13.5880 | 5542 | 1.8733 | 3193.45 | -0.81 |
| 891 | V0519 CrA | 273.4586076359 | -39.0173191236 | 16.3726 | 5343 | 1.3357 | 2923.01 | 2.71 |
| 892 | HV CrA | 273.4842523456 | -37.5368991655 | 15.2191 | 6381 | 0.5987 | 10438.93 | -0.47 |
| 893 | V2721 Sgr | 273.4899947921 | -33.2943353098 | 15.1400 | 5822 | 1.8052 | 6020.29 | -0.56 |
| 894 | HX CrA | 273.5135644874 | -38.0654961050 | 14.7350 | 6270 | 1.0470 | 10434.01 | -1.40 |
| 895 | BC Dra | 273.5794528793 | 76.6856608874 | 11.5101 | 6040 | 1.7002 | 1519.41 | -1.10 |
| 896 | V2754 Sgr | 273.6410225469 | -31.1520860126 | 15.2659 | 5403 | 2.2065 | 4693.95 | -0.30 |

| # | RRab | R.A. (J2000) | Decl. (J2000) | $G_{mag}$ | $T_{eff}$ | $A_G$ (mag) | d (pc) | $M_G$ |
|---|---|---|---|---|---|---|---|---|
| 897 | V2746 Sgr | 273.6585308031 | -35.2765948537 | 16.7632 | 6654 | 1.7930 | 7411.63 | 0.62 |
| 898 | LQ Aps | 273.6983508988 | -79.0821433124 | 15.9110 | 6121 | 1.7395 | 7695.70 | -0.26 |
| 899 | V2760 Sgr | 273.7322710969 | -33.9553418267 | 15.4569 | 5797 | 1.8365 | 6500.00 | -0.44 |
| 900 | V2770 Sgr | 273.7770654151 | -34.9031638751 | 16.7147 | 5541 | 1.7537 | 4461.53 | 1.71 |
| 901 | V2767 Sgr | 273.7831895956 | -36.7232495420 | 15.4463 | 6426 | 1.5975 | 6518.68 | -0.22 |
| 902 | V0364 CrA | 273.7936813991 | -38.3914400476 | 15.8048 | 5355 | 0.4795 | 1492.69 | 4.46 |
| 903 | V2777 Sgr | 273.8081242931 | -33.2387530085 | 15.6942 | 5214 | 2.1950 | 5692.49 | -0.28 |
| 904 | V0455 Oph | 273.8271984804 | 12.3523578688 | 12.2357 | 6567 | 1.7705 | 1822.03 | -0.84 |
| 905 | EY Aps | 273.8671678578 | -78.6713079346 | 14.7019 | 5655 | 2.0117 | 4529.33 | -0.59 |
| 906 | V2782 Sgr | 273.8856395446 | -36.8228695194 | 15.6000 | 6225 | 1.6785 | 7047.76 | -0.32 |
| 907 | V0528 CrA | 273.9053249341 | -40.9108523758 | 15.6753 | 5800 | 1.8270 | 5993.59 | -0.04 |
| 908 | XX CrA | 273.9339037617 | -37.4684978611 | 14.3412 | 5778 | 1.9800 | 4799.20 | -1.04 |
| 909 | V0527 CrA | 273.9361244028 | -42.4688433404 | 13.5537 | 6160 | 1.6178 | 3938.35 | -1.04 |
| 910 | HW Lyr | 274.0383071455 | 40.7122440816 | 15.0131 | 6143 | 1.8300 | 6329.07 | -0.82 |
| 911 | V2827 Sgr | 274.0634391109 | -34.8521094884 | 16.1894 | 6114 | 1.6667 | 7090.92 | 0.27 |
| 912 | V0700 Sgr | 274.0732776939 | -36.9406959722 | 14.9236 | 6291 | 1.7002 | 5980.65 | -0.66 |
| 913 | V0701 Sgr | 274.0797830900 | -36.4922037653 | 13.6501 | 5727 | 0.0877 | 547.53 | 4.87 |
| 914 | V0532 CrA | 274.0890587792 | -42.1598393642 | 15.3544 | 6006 | 1.6040 | 6837.40 | -0.42 |
| 915 | V0706 Sgr | 274.1210302884 | -35.2749469975 | 15.3136 | 5721 | 1.8693 | 7186.26 | -0.84 |
| 916 | V0403 CrA | 274.1650666471 | -43.8419494598 | 14.5505 | 5334 | 2.0630 | 5222.74 | -1.10 |
| 917 | V2852 Sgr | 274.1802263937 | -34.0228825405 | 15.8496 | 5431 | 1.7930 | 7358.25 | -0.28 |
| 918 | V0535 CrA | 274.1841449940 | -39.0307421201 | 15.6707 | 5972 | 1.7440 | 7322.41 | -0.40 |
| 919 | V0373 CrA | 274.2015584202 | -39.6525454795 | 14.1215 | 6608 | 1.7020 | 4692.11 | -0.94 |
| 920 | V1182 Sgr | 274.2095209266 | -27.7979880489 | 14.5680 | 3778 | 1.1590 | 4996.51 | -0.08 |
| 921 | V0534 CrA | 274.2246470442 | -42.3360761616 | 15.0971 | 5862 | 1.7670 | 5554.55 | -0.39 |
| 922 | CP Tel | 274.2390047041 | -54.9162207133 | 14.1986 | 6204 | 1.5653 | 3957.70 | -0.35 |
| 923 | V0536 CrA | 274.2398390577 | -38.3025247040 | 15.2285 | 5785 | 1.8100 | 8722.02 | -1.28 |
| 924 | V2878 Sgr | 274.2778526922 | -34.2705157467 | 15.3807 | 5852 | 1.8365 | 7938.98 | -0.95 |
| 925 | V2921 Sgr | 274.4656403223 | -34.4567035714 | 15.5225 | 7111 | 1.5100 | 7157.42 | -0.26 |
| 926 | V2931 Sgr | 274.4809764629 | -32.8935386582 | 15.2472 | 5504 | 0.5300 | 1528.59 | 3.80 |
| 927 | V2946 Sgr | 274.5317976151 | -31.8687467231 | 14.9409 | 5348 | 2.0740 | 6031.36 | -1.04 |
| 928 | V2933 Sgr | 274.5325579228 | -36.2332136940 | 15.1282 | 5926 | 1.7520 | 5286.91 | -0.24 |
| 929 | V2515 Sgr | 274.5381612757 | -32.5823358411 | 15.7439 | 5397 | 2.2685 | 8378.64 | -1.14 |
| 930 | V2948 Sgr | 274.5408923660 | -32.2896909598 | 16.0665 | 5098 | 1.2593 | 4605.08 | 1.49 |
| 931 | V2954 Sgr | 274.5778289290 | -33.6615207786 | 14.1809 | 4130 | 0.8340 | 5203.82 | -0.23 |
| 932 | V2950 Sgr | 274.6028910836 | -36.6202684404 | 15.3168 | 5876 | 1.7765 | 4958.41 | 0.06 |
| 933 | V2965 Sgr | 274.6129758769 | -32.1780037200 | 15.9595 | 5309 | 0.1657 | 7496.91 | 1.42 |
| 934 | V0545 CrA | 274.6741484675 | -38.9261954362 | 14.5154 | 6028 | 1.3610 | 1655.12 | 2.06 |
| 935 | V0378 CrA | 274.7046288562 | -39.0789945882 | 14.9852 | 6190 | 1.5890 | 6794.17 | -0.76 |
| 936 | IL Lyr | 274.7555165877 | 35.4346558703 | 16.2340 | 6069 | 1.6603 | 6469.53 | 0.52 |
| 937 | V2991 Sgr | 274.7848001691 | -35.3717939955 | 16.1779 | 6747 | 0.3823 | 3206.38 | 3.27 |
| 938 | V3013 Sgr | 274.8027512071 | -32.1285620340 | 16.2382 | 5340 | 2.1410 | 6729.98 | -0.04 |
| 939 | V3011 Sgr | 274.8061159322 | -33.3974615800 | 15.3909 | 5575 | 1.9940 | 4815.71 | -0.02 |
| 940 | V2522 Sgr | 274.8927220263 | -29.0736814767 | 15.3095 | 5504 | 2.1180 | 4701.54 | -0.17 |
| 941 | V0380 CrA | 274.9784708665 | -38.8463756753 | 14.9124 | 5871 | 1.8510 | 6500.66 | -1.00 |
| 942 | V0552 CrA | 274.9931158486 | -39.3361903653 | 15.2975 | 4690 | 0.6170 | 5384.44 | 1.02 |
| 943 | V3054 Sgr | 275.0066875025 | -31.7662861401 | 16.2257 | 5428 | 1.8765 | 8092.17 | -0.19 |
| 944 | V0381 CrA | 275.0188129679 | -38.5045332652 | 15.5533 | 5852 | 1.7830 | 5540.88 | 0.05 |
| 945 | V3053 Sgr | 275.0318789150 | -34.7662333455 | 15.9784 | 6600 | 1.5933 | 7517.84 | 0.00 |
| 946 | V3066 Sgr | 275.0414193142 | -30.8782605076 | 15.7505 | 5512 | 2.0620 | 7264.91 | -0.62 |
| 947 | V2523 Sgr | 275.0501394100 | -35.8103745721 | 14.4868 | 5876 | 1.6780 | 4550.51 | -0.48 |
| 948 | V0557 CrA | 275.0664587763 | -38.8543179935 | 15.1334 | 5624 | 1.9910 | 4453.88 | -0.10 |
| 949 | V3060 Sgr | 275.0729834657 | -35.4561106311 | 15.2531 | 5901 | 0.3460 | 1716.87 | 3.73 |
| 950 | V0554 CrA | 275.0754551943 | -43.5474485438 | 15.2820 | 6229 | 1.8987 | 7497.69 | -0.99 |
| 951 | V0382 CrA | 275.0916556846 | -39.1685189689 | 14.5424 | 7201 | 1.4340 | 7309.43 | -1.21 |
| 952 | V0418 CrA | 275.1893483920 | -44.7471959846 | 14.2865 | 6734 | 1.8000 | 4322.54 | -0.69 |
| 953 | V0559 CrA | 275.1924806832 | -40.8678628515 | 15.4990 | 6016 | 1.7570 | 6644.95 | -0.37 |
| 954 | V3090 Sgr | 275.2217709039 | -35.8599439674 | 15.5646 | 5570 | 1.9850 | 5896.69 | -0.27 |
| 955 | V2532 Sgr | 275.2797796754 | -31.1015947585 | 15.8519 | 5709 | 1.9625 | 6474.20 | -0.17 |
| 956 | V3110 Sgr | 275.2964169720 | -31.2880925236 | 15.3976 | 5455 | 2.1360 | 5772.98 | -0.55 |

| # | RRab | R.A. (J2000) | Decl. (J2000) | $G_{mag}$ | $T_{eff}$ | $A_G$ (mag) | d (pc) | $M_G$ |
|---|---|---|---|---|---|---|---|---|
| 957 | V3102 Sgr | 275.3035115156 | -34.9632338102 | 15.7965 | 5355 | 0.1657 | 9717.79 | 0.69 |
| 958 | V3112 Sgr | 275.3725755507 | -36.7071223602 | 15.7108 | 6498 | 1.5490 | 3974.29 | 1.17 |
| 959 | V0562 CrA | 275.3793294678 | -44.3453287171 | 14.7329 | 7071 | 1.4747 | 5848.60 | -0.58 |
| 960 | V3121 Sgr | 275.3816259589 | -34.8903937473 | 16.1415 | 5785 | 1.5560 | 6306.61 | 0.59 |
| 961 | V3131 Sgr | 275.4380407756 | -35.0902367401 | 16.0515 | 6234 | 1.5605 | 9063.49 | -0.30 |
| 962 | V3134 Sgr | 275.4484852238 | -34.5752083877 | 15.3609 | 5757 | 1.9370 | 8210.99 | -1.15 |
| 963 | V3146 Sgr | 275.4951772452 | -33.6004769072 | 15.8504 | 5701 | 2.0710 | 6480.29 | -0.28 |
| 964 | V0564 CrA | 275.5238060058 | -39.8845475012 | 14.1236 | 4344 | 0.4570 | 5410.19 | 0.00 |
| 965 | V3153 Sgr | 275.5407899726 | -32.2731515002 | 15.7303 | 5399 | 2.2140 | 8446.93 | -1.12 |
| 966 | V3164 Sgr | 275.5933708380 | -32.0470350736 | 16.0115 | 5354 | 1.0800 | 2266.56 | 3.15 |
| 967 | V3165 Sgr | 275.6541231914 | -36.3559697408 | 15.5519 | 6302 | 1.4900 | 4404.81 | 0.84 |
| 968 | IO Lyr | 275.6583141805 | 32.9590561228 | 11.7637 | 6482 | 1.6660 | 1476.38 | -0.75 |
| 969 | V0407 CrA | 275.6640555140 | -41.9795181064 | 15.3664 | 6464 | 1.4550 | 3653.30 | 1.10 |
| 970 | V0408 CrA | 275.6728096463 | -39.9730505261 | 15.5982 | 6276 | 1.8210 | 8293.46 | -0.82 |
| 971 | V3174 Sgr | 275.6796819156 | -31.8979877727 | 15.8936 | 5302 | 2.0627 | 7006.15 | -0.40 |
| 972 | V3168 Sgr | 275.6884927508 | -36.0588243329 | 15.4688 | 6158 | 1.7150 | 8091.56 | -0.79 |
| 973 | V3180 Sgr | 275.7591972238 | -33.9565994648 | 16.1843 | 6227 | 1.7080 | 8079.68 | -0.06 |
| 974 | V3183 Sgr | 275.7721574482 | -34.5507805229 | 15.8251 | 5915 | 1.7995 | 5796.89 | 0.21 |
| 975 | V1295 Sgr | 275.8073483793 | -34.2715238211 | 15.0054 | 6131 | 1.6617 | 5801.89 | -0.47 |
| 976 | V3191 Sgr | 275.8221649837 | -32.9077916630 | 16.0610 | 6125 | 1.8887 | 6889.65 | -0.02 |
| 977 | V3197 Sgr | 275.8247043107 | -31.2822054991 | 15.8459 | 5530 | 1.9250 | 6272.34 | -0.07 |
| 978 | V2539 Sgr | 275.9121563523 | -34.1966153529 | 14.8889 | 5860 | 1.5425 | 6239.37 | -0.63 |
| 979 | V3206 Sgr | 275.9645910348 | -36.4724413439 | 16.0132 | 6417 | 1.8560 | 6706.21 | 0.02 |
| 980 | V3209 Sgr | 275.9703861991 | -36.3125207648 | 16.0630 | 6311 | 1.4860 | 7222.92 | 0.28 |
| 981 | V0574 CrA | 275.9783047842 | -39.2025306319 | 15.1878 | 6498 | 1.5810 | 6173.15 | -0.35 |
| 982 | UX Lyr | 276.0157890249 | 39.0744698083 | 14.9809 | 6333 | 1.6260 | 5444.72 | -0.32 |
| 983 | V3248 Sgr | 276.0989229076 | -32.6533146922 | 15.9635 | 5512 | 0.3890 | 2037.47 | 4.03 |
| 984 | V0432 CrA | 276.1496764764 | -43.1578793334 | 15.1168 | 6666 | 1.6130 | 6429.67 | -0.54 |
| 985 | V3255 Sgr | 276.1872323584 | -35.7675656572 | 15.5428 | 6063 | 1.7930 | 6409.61 | -0.28 |
| 986 | V3266 Sgr | 276.1974268114 | -32.8418131342 | 15.3918 | 4332 | 0.8225 | 5558.81 | 0.84 |
| 987 | V0585 CrA | 276.2435522227 | -37.7054667748 | 14.5441 | 4962 | 0.3080 | 6342.59 | 0.22 |
| 988 | V0586 CrA | 276.3035100352 | -40.8952426244 | 13.4885 | 6206 | 1.7803 | 3941.83 | -1.27 |
| 989 | V3286 Sgr | 276.3461414345 | -36.1260463330 | 14.9694 | 6310 | 1.5440 | 5929.23 | -0.44 |
| 990 | V0435 CrA | 276.3791689856 | -37.8703864331 | 14.9538 | 5908 | 1.7497 | 5211.42 | -0.38 |
| 991 | V3306 Sgr | 276.3868328648 | -30.8490783253 | 15.9426 | 5370 | 1.2707 | 8680.63 | -0.02 |
| 992 | LZ Tel | 276.3924233890 | -45.7369977782 | 13.5495 | 5994 | 1.7555 | 4772.84 | -1.60 |
| 993 | V3310 Sgr | 276.4208112286 | -31.8107040008 | 15.0508 | 5761 | 1.6920 | 6534.01 | -0.72 |
| 994 | V1607 Sgr | 276.4513053606 | -31.6632224602 | 15.1652 | 5479 | 2.0430 | 5634.64 | -0.63 |
| 995 | V3313 Sgr | 276.4699427673 | -34.7853521729 | 15.7753 | 6126 | 1.8410 | 6686.63 | -0.19 |
| 996 | V3314 Sgr | 276.4758277619 | -34.9450467813 | 15.3186 | 6284 | 1.8710 | 5948.59 | -0.42 |
| 997 | V0371 Tel | 276.5344493290 | -56.0357303568 | 13.6621 | 6541 | 1.5820 | 3514.65 | -0.65 |
| 998 | V0596 CrA | 276.6203295613 | -37.9992412998 | 14.1870 | 5861 | 1.8550 | 4300.02 | -0.84 |
| 999 | V0598 CrA | 276.6865254207 | -40.1642989391 | 15.5084 | 7031 | 1.5577 | 6976.15 | -0.27 |
| 1000 | V0595 CrA | 276.7050609191 | -44.3953078071 | 15.4750 | 7303 | 1.2730 | 7547.19 | -0.19 |
| 1001 | V0599 CrA | 276.8043367692 | -42.7238515567 | 13.6904 | 6398 | 1.6048 | 3795.96 | -0.81 |
| 1002 | V1302 Sgr | 276.8260726335 | -28.9934431595 | 15.6526 | 4133 | 1.0590 | 5397.33 | 0.93 |
| 1003 | V3357 Sgr | 276.8495316856 | -32.2358016348 | 15.7236 | 5354 | 2.1800 | 6334.85 | -0.47 |
| 1004 | V2040 Oph | 276.8783643044 | 10.1558262437 | 15.3083 | 4908 | 0.9165 | 1145.76 | 4.10 |
| 1005 | V3363 Sgr | 276.8895728099 | -33.3986432454 | 15.0923 | 5868 | 1.7777 | 6016.07 | -0.58 |
| 1006 | V0609 CrA | 277.0451334393 | -38.8993176984 | 15.5542 | 6214 | 1.8710 | 6395.22 | -0.35 |
| 1007 | V3379 Sgr | 277.0578752282 | -35.5513653370 | 15.5119 | 6165 | 1.5400 | 6551.70 | -0.11 |
| 1008 | V2555 Sgr | 277.1023263227 | -29.2245336819 | 14.8083 | 5361 | 1.8700 | 4224.31 | -0.19 |
| 1009 | V3399 Sgr | 277.1334194832 | -31.1896759156 | 16.8228 | 4394 | 1.2990 | 1649.66 | 4.44 |
| 1010 | V0610 CrA | 277.1574495524 | -39.2033575013 | 14.8884 | 6254 | 1.6177 | 5433.32 | -0.40 |
| 1011 | EE Tel | 277.2436375616 | -56.2325045886 | 14.1094 | 6179 | 1.9083 | 3676.66 | -0.63 |
| 1012 | V2560 Sgr | 277.3221150332 | -33.6563147987 | 15.3004 | 5529 | 2.0160 | 4877.50 | -0.16 |
| 1013 | V0614 CrA | 277.3913824047 | -39.8631220278 | 13.9246 | 5735 | 1.8520 | 4109.00 | -1.00 |
| 1014 | IZ Lyr | 277.4257808100 | 39.7533347578 | 14.6399 | 6480 | 1.6553 | 5737.94 | -0.81 |
| 1015 | V3448 Sgr | 277.5264887092 | -33.7961014051 | 15.5055 | 5850 | 1.6720 | 5986.16 | -0.05 |
| 1016 | V3451 Sgr | 277.5427164554 | -33.7739935505 | 15.8949 | 6557 | 0.1657 | 10081.31 | 0.71 |

| # | RRab | R.A. (J2000) | Decl. (J2000) | $G_{mag}$ | $T_{eff}$ | $A_G$ (mag) | d (pc) | $M_G$ |
|---|---|---|---|---|---|---|---|---|
| 1017 | V1306 Sgr | 277.5462735568 | -34.2369010418 | 15.1037 | 5848 | 1.9957 | 5720.81 | -0.68 |
| 1018 | V1620 Sgr | 277.5693855133 | -28.5873356665 | 15.3276 | 5356 | 2.3003 | 7850.04 | -1.45 |
| 1019 | V3464 Sgr | 277.6425463811 | -36.3515122559 | 15.8602 | 6591 | 1.6840 | 6522.02 | 0.10 |
| 1020 | V2041 Oph | 277.6725690270 | 9.6278588539 | 14.9225 | 6029 | 1.6673 | 5201.28 | -0.33 |
| 1021 | KN Lyr | 277.6859325545 | 38.3986591549 | 13.4865 | 6565 | 1.6740 | 3977.51 | -1.19 |
| 1022 | V3477 Sgr | 277.7030172717 | -31.1651312756 | 15.5548 | 5846 | 1.9555 | 6586.37 | -0.49 |
| 1023 | KR Lyr | 277.7377911201 | 37.7415358815 | 14.2251 | 6895 | 1.5670 | 6677.67 | -1.47 |
| 1024 | V3482 Sgr | 277.7494208387 | -31.8949434963 | 15.3929 | 4890 | 0.1727 | 7764.71 | 0.77 |
| 1025 | V3481 Sgr | 277.8031010108 | -35.9185346155 | 15.4726 | 6638 | 1.5530 | 8437.28 | -0.71 |
| 1026 | V0610 Oph | 277.8490887572 | 9.1695275062 | 14.0587 | 6320 | 1.8200 | 3987.15 | -0.76 |
| 1027 | V3502 Sgr | 277.9337455351 | -32.5712864919 | 15.8126 | 6544 | 1.5720 | 8181.45 | -0.32 |
| 1028 | V2574 Sgr | 277.9577340244 | -31.2598193501 | 15.0854 | 6220 | 1.6465 | 8680.78 | -1.25 |
| 1029 | KS Lyr | 278.0488327090 | 33.2946140946 | 14.8515 | 6965 | 1.0557 | 6361.70 | -0.22 |
| 1030 | V3509 Sgr | 278.0514987422 | -35.6014619079 | 15.3577 | 5702 | 1.7410 | 9942.20 | -1.37 |
| 1031 | V2577 Sgr | 278.0800398309 | -34.2412430626 | 15.4716 | 5737 | 2.1150 | 5963.77 | -0.52 |
| 1032 | V3520 Sgr | 278.1345160671 | -35.2705229388 | 14.8666 | 6268 | 1.6123 | 5598.99 | -0.49 |
| 1033 | V1625 Sgr | 278.1348181476 | -31.1567382809 | 15.1054 | 6157 | 1.7250 | 5308.01 | -0.24 |
| 1034 | V2581 Sgr | 278.1390040210 | -32.4223342367 | 14.4532 | 6513 | 1.5930 | 5050.22 | -0.66 |
| 1035 | V2583 Sgr | 278.1671219846 | -31.5964716509 | 15.3007 | 5805 | 1.9290 | 6192.73 | -0.59 |
| 1036 | V3527 Sgr | 278.1869381029 | -35.9311847262 | 16.3957 | 6055 | 0.7890 | 3641.07 | 2.80 |
| 1037 | V3544 Sgr | 278.2819396412 | -31.7526971261 | 15.7603 | 6448 | 1.8760 | 7303.56 | -0.43 |
| 1038 | V3548 Sgr | 278.3006882208 | -31.6350026319 | 15.7313 | 5947 | 1.9392 | 6620.46 | -0.31 |
| 1039 | KX Lyr | 278.3134192565 | 40.1729961540 | 11.0372 | 6356 | 1.4080 | 1046.75 | -0.47 |
| 1040 | EU Tel | 278.3751315379 | -49.9055722176 | 14.0298 | 6748 | 1.6233 | 4235.39 | -0.73 |
| 1041 | V3554 Sgr | 278.3800012768 | -34.4759499651 | 15.1937 | 6686 | 1.7845 | 7650.95 | -1.01 |
| 1042 | V0630 CrA | 278.3927539212 | -42.6545017004 | 14.7757 | 6313 | 1.8435 | 8080.37 | -1.60 |
| 1043 | EV Tel | 278.4805936633 | -51.3531002621 | 13.2272 | 6853 | 1.6660 | 2856.38 | -0.72 |
| 1044 | V0633 CrA | 278.5006051868 | -38.0076802958 | 14.7366 | 4932 | 0.3228 | 4497.95 | 1.15 |
| 1045 | V3573 Sgr | 278.5370659112 | -31.2885617390 | 15.8770 | 5860 | 1.5745 | 5872.72 | 0.46 |
| 1046 | BH Pav | 278.6690282985 | -65.4508880569 | 12.5500 | 6472 | 1.5930 | 2285.08 | -0.84 |
| 1047 | AQ Lyr | 278.7126620802 | 26.5949295958 | 13.0470 | 6310 | 1.5915 | 2464.92 | -0.50 |
| 1048 | V1198 Sgr | 278.8249997494 | -34.8146210882 | 15.4046 | 6294 | 1.6298 | 7124.42 | -0.49 |
| 1049 | V3602 Sgr | 278.8452842025 | -36.3739220882 | 15.5120 | 6276 | 1.7722 | 8095.64 | -0.80 |
| 1050 | V3613 Sgr | 278.8576678344 | -32.1640732478 | 15.9472 | 5917 | 1.7233 | 8399.69 | -0.40 |
| 1051 | V3638 Sgr | 279.0203502865 | -31.7754618882 | 15.6405 | 5873 | 1.6797 | 6600.52 | -0.14 |
| 1052 | V3639 Sgr | 279.0718914653 | -34.7822615902 | 15.4948 | 6321 | 1.4027 | 8217.56 | -0.48 |
| 1053 | V2593 Sgr | 279.0940807335 | -32.8168175561 | 15.0440 | 6543 | 1.6747 | 6617.96 | -0.73 |
| 1054 | V3644 Sgr | 279.1035011231 | -31.6440495143 | 15.9449 | 5876 | 1.9235 | 7818.20 | -0.44 |
| 1055 | V2594 Sgr | 279.1370938732 | -30.2768698379 | 14.7535 | 5621 | 1.9210 | 4907.13 | -0.62 |
| 1056 | V3653 Sgr | 279.2415769704 | -35.2516145040 | 15.6651 | 6197 | 1.5443 | 9349.14 | -0.73 |
| 1057 | V3666 Sgr | 279.3478523667 | -34.7035368609 | 15.9434 | 6273 | 1.5890 | 6994.85 | 0.13 |
| 1058 | BR CrA | 279.3827908171 | -39.8445374468 | 14.9324 | 5937 | 1.8870 | 6559.58 | -1.04 |
| 1059 | KK Sgr | 279.3882096620 | -33.6169848093 | 15.1365 | 6597 | 1.7130 | 5157.97 | -0.14 |
| 1060 | V3676 Sgr | 279.4302276748 | -34.6896006000 | 15.1143 | 6269 | 1.8837 | 5604.85 | -0.51 |
| 1061 | V3677 Sgr | 279.4389896707 | -35.7958550074 | 15.6219 | 6598 | 1.4130 | 8753.07 | -0.50 |
| 1062 | V0408 Dra | 279.4844761828 | 56.8291641799 | 13.9414 | 6306 | 1.8160 | 3830.82 | -0.79 |
| 1063 | V3682 Sgr | 279.4923421400 | -33.4901185311 | 15.4616 | 5763 | 1.8915 | 7233.27 | -0.73 |
| 1064 | V2374 Sgr | 279.5399424078 | -27.6908062483 | 14.4030 | 5297 | 2.4277 | 3315.05 | -0.63 |
| 1065 | V1633 Sgr | 279.5611668153 | -32.5956406016 | 15.7131 | 5423 | 0.5310 | 9333.31 | 0.33 |
| 1066 | BU CrA | 279.7937548087 | -37.6439890060 | 14.9140 | 5733 | 1.7810 | 5617.76 | -0.61 |
| 1067 | CM Her | 279.7996928672 | 23.4943222414 | 14.0708 | 6036 | 1.8610 | 3589.44 | -0.57 |
| 1068 | V3704 Sgr | 279.8088542405 | -36.3863575733 | 15.5024 | 5662 | 2.0080 | 6701.27 | -0.64 |
| 1069 | V3715 Sgr | 279.8668811744 | -32.1523311410 | 16.1911 | 6820 | 1.2910 | 8086.32 | 0.36 |
| 1070 | V1204 Sgr | 279.8924859962 | -33.4900547364 | 14.3363 | 6093 | 1.6780 | 4236.04 | -0.48 |
| 1071 | V0409 Dra | 279.9084862483 | 58.0999083697 | 14.1686 | 6343 | 1.5175 | 4494.46 | -0.61 |
| 1072 | AN CrA | 279.9242992891 | -41.6358228555 | 13.6082 | 6639 | 1.6897 | 4082.14 | -1.14 |
| 1073 | CL Lyr | 279.9626179520 | 31.3922559169 | 15.0638 | 6313 | 1.8753 | 6430.50 | -0.85 |
| 1074 | V1205 Sgr | 279.9648678449 | -32.7545790962 | 15.3813 | 6020 | 1.7940 | 5694.29 | -0.19 |
| 1075 | AP CrA | 279.9818475792 | -37.5662437093 | 14.6285 | 6600 | 1.5310 | 5172.35 | -0.47 |
| 1076 | V0642 CrA | 279.9885534342 | -37.1689823972 | 15.8577 | 5572 | 2.1430 | 6279.03 | -0.27 |

| # | RRab | R.A. (J2000) | Decl. (J2000) | $G_{mag}$ | $T_{eff}$ | $A_G$ (mag) | d (pc) | $M_G$ |
|---|---|---|---|---|---|---|---|---|
| 1077 | V3732 Sgr | 280.0621778812 | -36.8260864694 | 15.5561 | 5282 | 1.9015 | 8328.72 | -0.95 |
| 1078 | LX Lyr | 280.0971410365 | 41.0396558217 | 13.2015 | 6429 | 1.5450 | 3326.69 | -0.95 |
| 1079 | V3747 Sgr | 280.1224041328 | -32.0510201175 | 15.9372 | 6357 | 1.8355 | 6438.58 | 0.06 |
| 1080 | V3750 Sgr | 280.1257129107 | -31.1388953885 | 15.4820 | 4426 | 0.8580 | 4603.56 | 1.31 |
| 1081 | V3740 Sgr | 280.1397112471 | -36.4413513248 | 15.1584 | 5941 | 1.7140 | 5706.34 | -0.34 |
| 1082 | V3751 Sgr | 280.1592324818 | -33.8083490563 | 15.6464 | 5936 | 1.6650 | 6676.86 | -0.14 |
| 1083 | V1208 Sgr | 280.2493605172 | -29.5941470853 | 14.9803 | 4308 | 0.8160 | 3776.52 | 1.28 |
| 1084 | CO Lyr | 280.4860989543 | 31.6432615769 | 13.8619 | 5945 | 1.6153 | 4072.44 | -0.80 |
| 1085 | V3779 Sgr | 280.5020601393 | -32.4354590598 | 15.7122 | 6905 | 1.6925 | 6853.15 | -0.16 |
| 1086 | AW Lyr | 280.5582074720 | 28.3789980441 | 13.4903 | 6268 | 1.7103 | 3449.19 | -0.91 |
| 1087 | MM Lyr | 280.6553125093 | 32.5974908301 | 15.9861 | 6328 | 1.5592 | 6622.23 | 0.32 |
| 1088 | FT Tel | 280.6724321402 | -52.8476773783 | 13.7634 | 7037 | 1.6810 | 4065.49 | -0.96 |
| 1089 | V0369 Sgr | 280.7690269943 | -34.9955129239 | 15.9976 | 5607 | 2.0585 | 6689.50 | -0.19 |
| 1090 | AS CrA | 280.7916561913 | -39.1034009509 | 15.3509 | 6615 | 1.6100 | 6200.80 | -0.22 |
| 1091 | CR Lyr | 280.8424862773 | 27.8456076749 | 14.2062 | 5794 | 1.9020 | 4099.73 | -0.76 |
| 1092 | V0372 Sgr | 280.9885986260 | -32.8834627883 | 14.6068 | 5964 | 1.7593 | 6518.79 | -1.22 |
| 1093 | V2613 Sgr | 281.2137924923 | -29.6585513294 | 15.0510 | 6507 | 1.6107 | 6228.28 | -0.53 |
| 1094 | AU CrA | 281.4536157166 | -40.9266683607 | 15.2976 | 6346 | 1.7028 | 6782.23 | -0.56 |
| 1095 | OV Tel | 281.4590022047 | -47.9320395592 | 13.5781 | 6740 | 1.3080 | 4617.44 | -1.05 |
| 1096 | V0373 Sgr | 281.5005137443 | -33.4353714485 | 15.5147 | 5784 | 0.9168 | 10094.68 | -0.42 |
| 1097 | V0645 CrA | 281.5705657900 | -38.0479023536 | 15.1000 | 6138 | 1.5770 | 8637.50 | -1.16 |
| 1098 | V0374 Sgr | 281.5976198259 | -36.8639433652 | 15.2176 | 6596 | 1.5880 | 6728.85 | -0.51 |
| 1099 | V0453 Sgr | 281.7545835494 | -35.8401672973 | 14.5468 | 5933 | 1.7390 | 4362.09 | -0.39 |
| 1100 | GR Sct | 281.9437852381 | -6.3414291588 | 14.1393 | 5358 | 2.2440 | 3985.12 | -1.11 |
| 1101 | V0413 CrA | 281.9900817417 | -37.7395920652 | 10.4774 | 5921 | 1.8630 | 864.08 | -1.07 |
| 1102 | CG CrA | 282.1982848255 | -39.4181452624 | 15.3178 | 6499 | 1.1705 | 4321.95 | 0.97 |
| 1103 | V0381 Sgr | 282.2641805365 | -26.2849350297 | 14.8715 | 5480 | 2.0603 | 4021.66 | -0.21 |
| 1104 | GW Sct | 282.3239722999 | -10.2087319141 | 13.7474 | 4118 | 1.3100 | 2701.07 | 0.28 |
| 1105 | V0456 Sgr | 282.3351424274 | -32.9300978766 | 14.6162 | 5995 | 1.7103 | 5348.81 | -0.74 |
| 1106 | V0808 Sgr | 282.4920657602 | -28.2154613810 | 15.2807 | 5801 | 1.5697 | 4770.82 | 0.32 |
| 1107 | V0880 Aql | 282.4980718734 | 11.4250600711 | 15.1196 | 4739 | 3.1130 | 3869.73 | -0.93 |
| 1108 | V0807 Sgr | 282.4992557017 | -31.0830306801 | 15.3161 | 4892 | 0.1290 | 6846.47 | 1.01 |
| 1109 | V0811 Sgr | 282.6505382602 | -30.2233731248 | 15.1910 | 6020 | 1.7490 | 6268.26 | -0.54 |
| 1110 | V0813 Sgr | 282.7819292328 | -27.2139309731 | 15.1189 | 5940 | 1.8200 | 5665.11 | -0.47 |
| 1111 | CL CrA | 282.8329849008 | -40.9660343424 | 14.8950 | 6314 | 1.2520 | 10553.91 | -1.47 |
| 1112 | CM CrA | 282.8436794679 | -40.5431240447 | 15.5169 | 5839 | 1.3120 | 4760.26 | 0.82 |
| 1113 | AZ CrA | 282.9362059115 | -38.9860904987 | 15.2858 | 7040 | 1.2970 | 9258.54 | -0.84 |
| 1114 | V0385 Sgr | 282.9834541089 | -23.3189655925 | 15.2352 | 5361 | 2.1270 | 5829.06 | -0.72 |
| 1115 | UW CrA | 283.0890781258 | -37.3441304168 | 14.3415 | 6860 | 1.6205 | 4250.33 | -0.42 |
| 1116 | GR Tel | 283.0927833349 | -53.1639256078 | 12.9240 | 6512 | 1.6023 | 2794.77 | -0.91 |
| 1117 | V0458 Sgr | 283.0959095663 | -35.9503162566 | 15.5302 | 5973 | 1.6825 | 7224.54 | -0.45 |
| 1118 | V0816 Sgr | 283.1459252277 | -28.3866440362 | 15.0155 | 5477 | 2.0415 | 4307.71 | -0.20 |
| 1119 | GQ Tel | 283.1644515206 | -56.3009020649 | 14.2533 | 6119 | 1.8693 | 5274.08 | -1.23 |
| 1120 | V0460 Sgr | 283.3036156347 | -34.6772175191 | 14.8120 | 5889 | 1.6730 | 5668.51 | -0.63 |
| 1121 | V0818 Sgr | 283.3126539377 | -28.7174693522 | 15.2250 | 6088 | 1.8300 | 6882.66 | -0.79 |
| 1122 | BC CrA | 283.3293549396 | -40.4049789451 | 14.8456 | 6845 | 1.3630 | 6163.70 | -0.47 |
| 1123 | V0355 Lyr | 283.3579775612 | 43.1545601217 | 14.4378 | 6139 | 1.8560 | 4405.53 | -0.64 |
| 1124 | V0386 Sgr | 283.3861159126 | -30.5293301523 | 15.1353 | 5759 | 1.9460 | 5777.79 | -0.62 |
| 1125 | V0387 Sgr | 283.5010097595 | -34.5152867347 | 13.7736 | 5675 | 2.1095 | 3916.34 | -1.30 |
| 1126 | V0388 Sgr | 283.5022883981 | -32.7442080766 | 15.3589 | 5687 | 1.9930 | 6171.69 | -0.59 |
| 1127 | V0820 Sgr | 283.5053635092 | -27.4801974877 | 14.1050 | 5276 | 2.0440 | 2931.54 | -0.27 |
| 1128 | BD CrA | 283.6244725510 | -39.5444054926 | 14.9530 | 6388 | 1.5520 | 6270.00 | -0.59 |
| 1129 | AY Sct | 283.6732856402 | -10.1760654085 | 14.0657 | 4250 | 0.8608 | 3831.22 | 0.29 |
| 1130 | V0822 Sgr | 283.6822033376 | -28.1306524372 | 15.4619 | 5722 | 1.9795 | 4676.58 | 0.13 |
| 1131 | V0464 Sgr | 283.6980992164 | -35.5741634724 | 14.9410 | 5873 | 1.8550 | 5334.77 | -0.55 |
| 1132 | V0823 Sgr | 283.7120171862 | -29.2793232388 | 15.2628 | 6065 | 1.1335 | 4297.41 | 0.96 |
| 1133 | V1031 Sgr | 283.7342329340 | -22.8725488688 | 15.5143 | 5372 | 2.0727 | 6424.74 | -0.60 |
| 1134 | V0466 Sgr | 283.8077609090 | -33.4720881127 | 15.3652 | 6670 | 0.4930 | 8730.13 | 0.17 |
| 1135 | BH Lyr | 283.9337486678 | 33.5672696910 | 14.7136 | 6334 | 1.6420 | 7881.74 | -1.41 |
| 1136 | V0470 Sgr | 284.1551194916 | -32.0373388202 | 14.6557 | 6476 | 1.6250 | 4783.89 | -0.37 |

| # | RRab | R.A. (J2000) | Decl. (J2000) | $G_{mag}$ | $T_{eff}$ | $A_G$ (mag) | d (pc) | $M_G$ |
|---|---|---|---|---|---|---|---|---|
| 1137 | BH CrA | 284.1843499936 | -39.7190299057 | 14.8306 | 6580 | 1.6148 | 7785.69 | -1.24 |
| 1138 | V0389 Sgr | 284.2185380551 | -35.1538507038 | 14.8991 | 6498 | 1.4097 | 6514.54 | -0.58 |
| 1139 | BX Vul | 284.3143248279 | 24.0422520554 | 14.4945 | 5321 | 2.3260 | 4584.55 | -1.14 |
| 1140 | V0831 Sgr | 284.4006354081 | -23.1473749366 | 15.0378 | 5602 | 1.8993 | 5569.35 | -0.59 |
| 1141 | V0830 Sgr | 284.4547516170 | -29.3617905684 | 10.9514 | 4113 | 0.7550 | 1323.60 | -0.41 |
| 1142 | V1044 Sgr | 284.5074749185 | -19.4120605421 | 15.3204 | 5512 | 2.0867 | 6426.72 | -0.81 |
| 1143 | V0833 Sgr | 284.5140715103 | -27.3098828319 | 13.6711 | 5604 | 1.7220 | 3478.28 | -0.76 |
| 1144 | EM Sct | 284.5799433883 | -7.2030340304 | 15.2890 | 5185 | 2.2040 | 4757.79 | -0.30 |
| 1145 | V0835 Sgr | 284.6244479974 | -25.2360364205 | 15.1928 | 5785 | 1.9590 | 6968.15 | -0.98 |
| 1146 | V0475 Sgr | 284.6709258775 | -32.4097601891 | 15.0504 | 5873 | 1.9045 | 5553.22 | -0.58 |
| 1147 | V0927 Aql | 284.7118161370 | -8.2107692854 | 13.8829 | 4186 | 1.0475 | 3856.93 | -0.10 |
| 1148 | V0393 Sgr | 285.0014490715 | -33.9292255002 | 14.8712 | 6515 | 1.1633 | 8708.72 | -0.99 |
| 1149 | V1055 Sgr | 285.0713417985 | -15.1019177686 | 12.1758 | 5309 | 2.2467 | 1572.98 | -1.05 |
| 1150 | V1054 Sgr | 285.1005084043 | -19.7623773383 | 15.4421 | 6344 | 1.8510 | 6746.04 | -0.55 |
| 1151 | V0396 Sgr | 285.1865591847 | -34.2561705680 | 15.1771 | 5675 | 0.8380 | 3443.86 | 1.65 |
| 1152 | DD CrA | 285.2134206165 | -40.3402796164 | 14.5193 | 6277 | 1.8010 | 6015.66 | -1.18 |
| 1153 | V0409 Lyr | 285.2329110861 | 26.3386153920 | 15.8210 | 5721 | 0.8990 | 9508.40 | 0.03 |
| 1154 | V0846 Sgr | 285.2989709796 | -26.7509127212 | 14.1385 | 4126 | 0.5590 | 5751.01 | -0.22 |
| 1155 | V0476 Sgr | 285.3482893841 | -34.4461482903 | 14.8633 | 6266 | 1.8335 | 6162.10 | -0.92 |
| 1156 | V2399 Sgr | 285.3576996502 | -34.7273711822 | 15.2732 | 6845 | 1.4780 | 6455.68 | -0.25 |
| 1157 | DQ Lyr | 285.3896783705 | 31.1996720055 | 14.7525 | 5790 | 1.6573 | 6012.63 | -0.80 |
| 1158 | V0847 Sgr | 285.4158948146 | -26.9356073755 | 14.9569 | 5770 | 1.9280 | 7270.28 | -1.28 |
| 1159 | V0479 Sgr | 285.5066067082 | -34.9204414905 | 13.9295 | 6259 | 1.8380 | 5019.66 | -1.41 |
| 1160 | V0480 Sgr | 285.5313056092 | -34.0513285021 | 14.9875 | 5973 | 1.6730 | 5749.31 | -0.48 |
| 1161 | V1065 Sgr | 285.6687506708 | -16.0760408393 | 14.6266 | 5058 | 2.3263 | 3609.13 | -0.49 |
| 1162 | V1067 Sgr | 285.8143103747 | -21.0459218008 | 14.7699 | 5611 | 1.8885 | 6171.33 | -1.07 |
| 1163 | V0405 Sgr | 285.8779378874 | -22.2680842734 | 14.9689 | 5801 | 1.8442 | 7226.94 | -1.17 |
| 1164 | DV Lyr | 286.0536647343 | 31.3300189813 | 14.0371 | 5874 | 1.8930 | 4878.74 | -1.30 |
| 1165 | V0406 Sgr | 286.0751098490 | -31.1870750784 | 15.4301 | 6341 | 1.5550 | 6233.73 | -0.10 |
| 1166 | V0481 Sgr | 286.1029504234 | -35.4870223290 | 14.4577 | 6263 | 1.7080 | 7682.10 | -1.68 |
| 1167 | V0407 Sgr | 286.2813187335 | -34.0436769984 | 15.0370 | 5897 | 1.8587 | 3987.66 | 0.17 |
| 1168 | V0362 Lyr | 286.2940961430 | 44.0949417561 | 14.9428 | 6796 | 1.5180 | 7341.02 | -0.90 |
| 1169 | V0482 Sgr | 286.2967472051 | -35.8943307759 | 14.0626 | 7042 | 1.3705 | 5542.71 | -1.03 |
| 1170 | V1244 Sgr | 286.3174223664 | -30.9090093962 | 13.0952 | 4841 | 0.2980 | 2750.79 | 0.60 |
| 1171 | V1073 Sgr | 286.3206712327 | -17.2594607574 | 15.0352 | 5336 | 0.5148 | 1072.86 | 4.37 |
| 1172 | V0485 Sgr | 286.3975835792 | -35.1532676401 | 14.5652 | 6258 | 1.8120 | 6144.74 | -1.19 |
| 1173 | ZZ Lyr | 286.4714126849 | 26.5585162959 | 13.7896 | 5670 | 1.8685 | 3305.64 | -0.68 |
| 1174 | V0486 Sgr | 286.5086111863 | -35.1984013665 | 15.4338 | 6632 | 1.5440 | 6087.89 | -0.03 |
| 1175 | V0487 Sgr | 286.5131982087 | -35.3735901328 | 15.1163 | 6878 | 1.6707 | 6583.97 | -0.65 |
| 1176 | V0853 Sgr | 286.6921662760 | -29.9413532726 | 14.2111 | 6255 | 1.6980 | 4396.26 | -0.70 |
| 1177 | V0397 Lyr | 286.7958188383 | 30.9510730334 | 16.1357 | 5891 | 1.9070 | 6726.47 | 0.09 |
| 1178 | V0419 Dra | 286.8216629447 | 55.3707596681 | 14.0511 | 6357 | 1.7200 | 5576.10 | -1.40 |
| 1179 | V0412 Sgr | 286.8510855124 | -33.3900800483 | 15.1957 | 5959 | 1.6673 | 5727.35 | -0.26 |
| 1180 | ST UMi | 286.8665619060 | 88.7704279692 | 14.7886 | 6202 | 1.9470 | 4767.16 | -0.55 |
| 1181 | NQ Lyr | 286.9515795257 | 42.2985330878 | 13.3360 | 7105 | 1.6390 | 3054.48 | -0.73 |
| 1182 | V1083 Sgr | 286.9808675476 | -19.6087509527 | 13.9393 | 4603 | 0.4440 | 3444.47 | 0.81 |
| 1183 | V1084 Sgr | 287.0567093087 | -19.8658271976 | 15.0365 | 6516 | 1.6440 | 5902.91 | -0.46 |
| 1184 | V1086 Sgr | 287.0728027163 | -16.2774424219 | 15.0085 | 5318 | 0.5440 | 1126.12 | 4.21 |
| 1185 | NR Lyr | 287.1134658593 | 38.8128131530 | 12.6505 | 5952 | 1.6000 | 2889.84 | -1.25 |
| 1186 | V0418 Lyr | 287.1394189757 | 33.3096555145 | 15.5402 | 6568 | 1.6620 | 7722.33 | -0.56 |
| 1187 | BQ Lyr | 287.1501461160 | 26.9577972830 | 13.1814 | 4482 | 0.5660 | 2756.28 | 0.41 |
| 1188 | V1088 Sgr | 287.1544597225 | -18.6640596178 | 15.1043 | 5852 | 1.9630 | 5625.29 | -0.61 |
| 1189 | V0490 Sgr | 287.2735800553 | -34.3976241157 | 15.1074 | 6600 | 1.6460 | 5706.01 | -0.32 |
| 1190 | V1092 Sgr | 287.2885644602 | -15.9692862360 | 15.0579 | 5542 | 0.4720 | 1412.17 | 3.84 |
| 1191 | V0859 Sgr | 287.3936114596 | -24.4056957921 | 15.2768 | 7008 | 1.5170 | 8018.85 | -0.76 |
| 1192 | V1094 Sgr | 287.4334837676 | -18.7713275348 | 13.8699 | 5811 | 1.8060 | 4059.91 | -0.98 |
| 1193 | V0861 Sgr | 287.7034960788 | -29.3027621644 | 14.9704 | 6565 | 1.8555 | 4726.87 | -0.26 |
| 1194 | DV Pav | 287.8414188269 | -72.4341759227 | 15.2984 | 6421 | 1.6420 | 6306.15 | -0.34 |
| 1195 | V0863 Sgr | 287.8878052995 | -25.7811829111 | 15.5799 | 5790 | 1.8867 | 7248.49 | -0.61 |
| 1196 | BK CrA | 287.8908152818 | -41.3642533603 | 13.4041 | 6560 | 1.5590 | 3377.65 | -0.80 |

| # | RRab | R.A. (J2000) | Decl. (J2000) | $G_{mag}$ | $T_{eff}$ | $A_G$ (mag) | d (pc) | $M_G$ |
|---|---|---|---|---|---|---|---|---|
| 1197 | V0864 Sgr | 287.9383150437 | -26.8477858201 | 15.6730 | 5835 | 1.8958 | 8236.16 | -0.80 |
| 1198 | V1102 Sgr | 287.9650494316 | -22.4883846984 | 16.0835 | 6550 | 0.0260 | 3329.70 | 3.45 |
| 1199 | DN CrA | 287.9790223969 | -39.1981447546 | 14.8439 | 6284 | 1.9510 | 4516.08 | -0.38 |
| 1200 | V0866 Sgr | 288.0721781934 | -26.9356005073 | 12.7634 | 5852 | 1.8885 | 3000.81 | -1.51 |
| 1201 | V1105 Sgr | 288.1306435089 | -19.2727319161 | 14.7446 | 5150 | 0.2090 | 674.71 | 5.39 |
| 1202 | V1107 Sgr | 288.1597439640 | -19.2720923547 | 15.6775 | 6415 | 0.5477 | 4633.07 | 1.80 |
| 1203 | V0424 Lyr | 288.2165460176 | 26.2651219915 | 15.2609 | 5352 | 0.0160 | 663.85 | 6.13 |
| 1204 | V0867 Sgr | 288.2404265964 | -28.5254790673 | 15.8461 | 6200 | 0.6010 | 9720.75 | 0.31 |
| 1205 | SS Pav | 288.3564553584 | -66.3582502757 | 13.1924 | 6322 | 1.6650 | 3236.27 | -1.02 |
| 1206 | V0423 Sgr | 288.5436453970 | -26.2401268049 | 14.8315 | 6612 | 1.4750 | 10104.30 | -1.67 |
| 1207 | V0874 Sgr | 288.6103405632 | -23.7000801093 | 15.2318 | 6284 | 1.7550 | 6273.51 | -0.51 |
| 1208 | WW Oct | 288.6484170361 | -74.9375769336 | 14.7510 | 6115 | 1.6430 | 5096.38 | -0.43 |
| 1209 | V1117 Sgr | 288.6987186289 | -14.3633310862 | 15.1372 | 5785 | 1.9217 | 8743.80 | -1.49 |
| 1210 | EN Lyr | 288.7262918886 | 34.7672183213 | 13.3010 | 6099 | 1.5647 | 3773.47 | -1.15 |
| 1211 | V0876 Sgr | 288.7661272866 | -28.3009851844 | 15.6481 | 6629 | 1.5327 | 8038.74 | -0.41 |
| 1212 | DY Pav | 288.8662945500 | -69.8471379977 | 15.3331 | 6049 | 1.8740 | 5664.79 | -0.31 |
| 1213 | V2410 Sgr | 289.0305763103 | -27.7680007053 | 13.5858 | 5859 | 1.8630 | 4149.33 | -1.37 |
| 1214 | EE Pav | 289.0473535685 | -70.8217475951 | 15.1172 | 5865 | 1.8930 | 6600.70 | -0.87 |
| 1215 | V0372 Lyr | 289.0517528749 | 41.9055710338 | 11.9723 | 6249 | 0.0555 | 321.78 | 4.38 |
| 1216 | V0879 Sgr | 289.1237666698 | -23.1778148337 | 14.7003 | 4586 | 0.3120 | 6977.45 | 0.17 |
| 1217 | V1123 Sgr | 289.2785059957 | -20.3445512778 | 15.4921 | 5217 | 0.3387 | 1255.86 | 4.66 |
| 1218 | V0498 Sgr | 289.3342935174 | -35.5191651744 | 14.8661 | 6214 | 1.7690 | 8097.24 | -1.44 |
| 1219 | V0430 Sgr | 289.3832938485 | -15.6389807291 | 14.8327 | 5599 | 1.8430 | 6765.40 | -1.16 |
| 1220 | V0883 Sgr | 289.4317551035 | -22.7204831617 | 14.4110 | 5815 | 1.6033 | 3382.99 | 0.16 |
| 1221 | BK Dra | 289.5861269221 | 66.4133229341 | 11.2797 | 5995 | 1.6053 | 1352.06 | -0.98 |
| 1222 | BL CrA | 289.5963495971 | -39.5759109117 | 15.0966 | 6099 | 1.9040 | 5320.66 | -0.44 |
| 1223 | PT Lyr | 289.6225328022 | 27.9281773129 | 15.9922 | 5070 | 0.2300 | 7221.73 | 1.47 |
| 1224 | V0885 Sgr | 289.6289313228 | -23.5288414386 | 14.2844 | 6111 | 1.7340 | 6955.60 | -1.66 |
| 1225 | PU Lyr | 289.6445334997 | 34.1039393738 | 15.1285 | 5612 | 2.0390 | 6049.03 | -0.82 |
| 1226 | V0887 Sgr | 289.6516709718 | -24.3760154735 | 14.3925 | 6534 | 1.7480 | 4298.84 | -0.52 |
| 1227 | V0884 Sgr | 289.6537292315 | -28.3562046267 | 14.8168 | 6333 | 1.7017 | 5015.28 | -0.39 |
| 1228 | V0500 Sgr | 289.7627415084 | -39.4017234014 | 13.7288 | 5552 | 1.7620 | 7808.66 | -2.50 |
| 1229 | V0433 Sgr | 289.8756584943 | -30.3261148736 | 14.8743 | 6268 | 1.7720 | 5324.76 | -0.53 |
| 1230 | V0890 Sgr | 289.8985644290 | -31.5480002361 | 14.4991 | 6197 | 1.6840 | 5097.48 | -0.72 |
| 1231 | V0893 Sgr | 289.9993465682 | -24.4643369975 | 14.8381 | 5778 | 1.7675 | 4997.03 | -0.42 |
| 1232 | V0891 Sgr | 290.0301555687 | -30.3799790283 | 14.9522 | 5862 | 1.8170 | 5441.05 | -0.54 |
| 1233 | V0894 Sgr | 290.0590424297 | -25.9843374266 | 15.4775 | 6330 | 1.4590 | 8873.44 | -0.72 |
| 1234 | V0895 Sgr | 290.1369548424 | -26.4106231879 | 15.2762 | 5438 | 2.0103 | 4608.90 | -0.05 |
| 1235 | V1131 Sgr | 290.1683602687 | -20.2715145604 | 15.6994 | 6158 | 1.5870 | 7415.31 | -0.24 |
| 1236 | PZ Lyr | 290.2076645649 | 32.3551945219 | 15.1202 | 5262 | 2.0838 | 6144.25 | -0.91 |
| 1237 | EH Pav | 290.2171139474 | -70.9967978276 | 14.8441 | 6602 | 1.1000 | 4287.57 | 0.58 |
| 1238 | WX Oct | 290.6407105584 | -74.8538450032 | 14.7785 | 6317 | 1.7535 | 8964.66 | -1.74 |
| 1239 | LM Pav | 290.6535325196 | -70.0457504673 | 14.9556 | 6514 | 1.3825 | 8385.03 | -1.04 |
| 1240 | V1111 Cyg | 290.7881412210 | 30.1813658869 | 15.6392 | 5528 | 2.0460 | 6767.58 | -0.56 |
| 1241 | V0900 Sgr | 290.8924049313 | -28.1746250236 | 14.8213 | 6229 | 1.6867 | 4836.92 | -0.29 |
| 1242 | QV Lyr | 290.9439366863 | 34.6319892747 | 14.5824 | 5761 | 2.0947 | 5248.05 | -1.11 |
| 1243 | V1141 Sgr | 291.1064774042 | -15.2111105713 | 14.0992 | 5584 | 2.1477 | 3364.82 | -0.68 |
| 1244 | V1142 Sgr | 291.1959077189 | -18.9715479246 | 15.1241 | 4640 | 0.2580 | 8435.88 | 0.24 |
| 1245 | QZ Lyr | 291.2731391773 | 36.4823999309 | 16.6258 | 6544 | 1.7020 | 6621.59 | 0.82 |
| 1246 | V1144 Sgr | 291.4753443125 | -20.3224779774 | 14.3009 | 6190 | 1.6385 | 4760.50 | -0.73 |
| 1247 | V0862 Cyg | 291.7771783800 | 34.2243757055 | 15.7256 | 5822 | 1.5850 | 7317.12 | -0.18 |
| 1248 | V0904 Sgr | 292.0296987878 | -28.7140724259 | 14.6456 | 6319 | 1.6350 | 9021.73 | -1.77 |
| 1249 | LQ Pav | 292.4841525566 | -71.2154141357 | 15.0685 | 6078 | 1.8453 | 7009.74 | -1.01 |
| 1250 | EP Pav | 292.5064727677 | -69.4652644154 | 15.7732 | 6302 | 1.4955 | 7963.37 | -0.23 |
| 1251 | V1949 Cyg | 292.5519526281 | 50.8058735374 | 13.1492 | 6279 | 1.5810 | 2987.55 | -0.81 |
| 1252 | ES Pav | 292.9540367346 | -57.3437248489 | 12.8311 | 6262 | 1.6170 | 3072.88 | -1.22 |
| 1253 | XZ Cyg | 293.1221061805 | 56.3881928948 | 9.9222 | 6598 | 1.7273 | 625.09 | -0.78 |
| 1254 | HH Tel | 293.3345788021 | -45.6636479144 | 12.2511 | 6287 | 1.2710 | 1680.23 | -0.15 |
| 1255 | WY Dra | 293.3376953489 | 80.9284903538 | 12.6458 | 6605 | 1.6305 | 2811.57 | -1.23 |
| 1256 | V0794 Cyg | 293.3416095932 | 32.8562410049 | 12.9615 | 6339 | 0.8043 | 800.45 | 2.64 |

| # | RRab | R.A. (J2000) | Decl. (J2000) | $G_{mag}$ | $T_{eff}$ | $A_G$ (mag) | d (pc) | $M_G$ |
|---|---|---|---|---|---|---|---|---|
| 1257 | ET Pav | 294.0291359383 | -71.0265832817 | 15.1625 | 6255 | 1.5465 | 7140.31 | -0.65 |
| 1258 | V0639 Aql | 294.4218487078 | 8.2799904496 | 16.0590 | 4189 | 1.1820 | 5617.39 | 1.13 |
| 1259 | BN Pav | 294.5140906720 | -60.6109973263 | 12.5160 | 6528 | 1.6550 | 2158.75 | -0.81 |
| 1260 | V0799 Cyg | 294.5504960327 | 39.1471488685 | 14.6612 | 5601 | 1.9477 | 5386.44 | -0.94 |
| 1261 | EV Pav | 294.7169270417 | -71.7509493505 | 14.3355 | 6575 | 1.6545 | 6856.91 | -1.50 |
| 1262 | EW Pav | 295.2805558011 | -69.9904336801 | 15.1494 | 5999 | 1.6400 | 6355.86 | -0.51 |
| 1263 | BI Oct | 295.2923948681 | -75.7168829810 | 15.3930 | 6346 | 1.4875 | 6314.50 | -0.10 |
| 1264 | V0715 Cyg | 295.5333416241 | 38.9117575088 | 16.5720 | 5671 | 1.9663 | 7818.39 | 0.14 |
| 1265 | V1152 Aql | 295.8402343172 | 0.0833075545 | 15.4848 | 5357 | 1.9940 | 5674.14 | -0.28 |
| 1266 | V1025 Aql | 296.0699255641 | 11.3589180276 | 15.7802 | 5490 | 2.2947 | 6083.33 | -0.44 |
| 1267 | FF Pav | 296.2796804220 | -69.7412516139 | 15.3727 | 5461 | 1.9648 | 5617.44 | -0.34 |
| 1268 | V0672 Aql | 296.4070762220 | 8.0214044785 | 11.9906 | 5328 | 2.1860 | 1443.64 | -0.99 |
| 1269 | V0808 Cyg | 296.4126064273 | 39.5148562303 | 15.3520 | 6122 | 1.7988 | 8357.12 | -1.06 |
| 1270 | FG Pav | 296.9444046839 | -73.7678368749 | 15.6480 | 6536 | 1.3015 | 3942.97 | 1.37 |
| 1271 | V1306 Aql | 297.3228327072 | 4.2778021052 | 15.5283 | 5212 | 1.1800 | 1888.50 | 2.97 |
| 1272 | FM Pav | 297.5237949005 | -68.3749204512 | 13.7830 | 6591 | 1.7247 | 4294.34 | -1.11 |
| 1273 | V0691 Aql | 297.6088242611 | -1.7340351734 | 14.7468 | 5292 | 2.2930 | 4834.54 | -0.97 |
| 1274 | V0695 Aql | 297.6312047531 | 6.0450551524 | 13.8998 | 5757 | 2.0630 | 4044.78 | -1.20 |
| 1275 | V1324 Aql | 297.9087517525 | 5.1343159319 | 16.0451 | 6642 | 1.6040 | 6369.27 | 0.42 |
| 1276 | DN Pav | 298.0455984376 | -63.6734717203 | 12.6805 | 6253 | 1.6495 | 2112.91 | -0.59 |
| 1277 | V0894 Aql | 298.1488923044 | 4.3176718687 | 15.3524 | 5550 | 2.0840 | 8286.12 | -1.32 |
| 1278 | FN Pav | 298.1605825043 | -69.9300423721 | 14.7790 | 6887 | 1.6550 | 6389.41 | -0.90 |
| 1279 | V0783 Cyg | 298.2197393273 | 40.7931888524 | 14.7197 | 5262 | 2.1950 | 4616.63 | -0.80 |
| 1280 | V0381 Cyg | 298.6241754608 | 46.0642593939 | 14.1538 | 5517 | 2.1895 | 4144.92 | -1.12 |
| 1281 | LX Pav | 298.7900816250 | -57.0509988555 | 14.2650 | 6541 | 1.5520 | 5009.11 | -0.79 |
| 1282 | V0717 Aql | 298.7989842316 | 0.2219731319 | 14.5273 | 5745 | 1.6743 | 5471.69 | -0.84 |
| 1283 | FY Sge | 299.0458175779 | 16.4548390457 | 14.8979 | 5349 | 2.3965 | 7770.10 | -1.95 |
| 1284 | V0784 Cyg | 299.0954377631 | 41.3398408819 | 15.6585 | 4970 | 2.3120 | 5226.36 | -0.24 |
| 1285 | V0896 Aql | 299.2405462393 | 1.5299065974 | 15.4162 | 5766 | 1.7190 | 5337.58 | 0.06 |
| 1286 | V0785 Cyg | 299.5304008891 | 35.8678168662 | 16.3573 | 4279 | 0.8717 | 6710.17 | 1.35 |
| 1287 | V0740 Aql | 299.6310934745 | 5.7346772423 | 14.1777 | 6255 | 1.5550 | 5192.30 | -0.95 |
| 1288 | AC Oct | 299.7055747358 | -75.4843845293 | 14.0780 | 6418 | 1.8695 | 3838.98 | -0.71 |
| 1289 | V1026 Cyg | 299.8679714045 | 57.4569001011 | 15.9672 | 6192 | 1.8885 | 6782.56 | -0.08 |
| 1290 | V0429 Dra | 299.8839661358 | 61.5225144745 | 14.8991 | 6254 | 1.7645 | 5138.96 | -0.42 |
| 1291 | V0753 Aql | 300.0339589432 | 10.2559988665 | 15.3210 | 5536 | 2.0235 | 4761.14 | -0.09 |
| 1292 | HT Tel | 300.0717558594 | -45.3148360035 | 13.5637 | 6543 | 1.4768 | 3834.69 | -0.83 |
| 1293 | V5661 Sgr | 300.6161434634 | -37.0046994346 | 13.7534 | 6526 | 1.5590 | 3893.38 | -0.76 |
| 1294 | V1031 Cyg | 300.6173893530 | 56.8862049972 | 16.9589 | 5716 | 0.1657 | 6575.35 | 2.70 |
| 1295 | MX Tel | 300.6739293131 | -53.9659590720 | 15.9416 | 5774 | 0.3380 | 2001.75 | 4.10 |
| 1296 | V1085 Aql | 300.9409783626 | 3.0554060656 | 15.5192 | 5839 | 1.5520 | 6067.64 | 0.05 |
| 1297 | HV Vul | 301.3607231811 | 22.2225817830 | 15.0750 | 4883 | 0.4820 | 5914.49 | 0.73 |
| 1298 | HU Vul | 301.3631939379 | 22.4172670066 | 15.2799 | 4934 | 2.5525 | 4437.30 | -0.51 |
| 1299 | V1035 Cyg | 301.4226286776 | 58.0469138292 | 15.7620 | 5446 | 1.3203 | 9009.53 | -0.33 |
| 1300 | KM Aql | 301.4844954388 | -8.5145130063 | 13.2642 | 6016 | 1.6340 | 2432.71 | -0.30 |
| 1301 | GQ Pav | 301.5110041350 | -71.7735861511 | 13.9308 | 6721 | 1.5650 | 4738.63 | -1.01 |
| 1302 | V1094 Aql | 301.6761159804 | 14.9946774808 | 13.9668 | 5370 | 2.1715 | 2706.03 | -0.37 |
| 1303 | V1307 Cyg | 302.0070310058 | 41.5465175928 | 15.0391 | 4976 | 1.9500 | 1586.05 | 2.09 |
| 1304 | HW Tel | 302.0186930626 | -56.3772865315 | 15.4363 | 6305 | 1.5800 | 5908.49 | 0.00 |
| 1305 | GV Pav | 302.1347052065 | -60.4880738198 | 13.2649 | 6661 | 1.4475 | 6193.97 | -2.14 |
| 1306 | GT Pav | 302.2926338713 | -72.8762724775 | 15.1194 | 6560 | 1.5070 | 7847.47 | -0.86 |
| 1307 | HZ Tel | 302.3900437612 | -55.4833230576 | 14.4002 | 6335 | 1.7027 | 4717.41 | -0.67 |
| 1308 | V0904 Aql | 302.5403293310 | -0.0526652732 | 15.6536 | 6213 | 1.7155 | 6076.74 | 0.02 |
| 1309 | V1177 Aql | 302.9432342524 | -3.3222380588 | 15.3413 | 5752 | 1.8890 | 5534.48 | -0.26 |
| 1310 | GX Pav | 303.1107134483 | -73.4892865683 | 15.1378 | 6963 | 1.5290 | 5752.97 | -0.19 |
| 1311 | GY Pav | 303.1595303042 | -72.9202201402 | 15.1744 | 6011 | 1.8068 | 6313.96 | -0.63 |
| 1312 | FI Sge | 303.3175486395 | 17.5102695916 | 13.8279 | 5592 | 1.9197 | 3793.75 | -0.99 |
| 1313 | HI Pav | 303.3455314231 | -59.0982821277 | 14.8441 | 6550 | 1.6733 | 5429.35 | -0.50 |
| 1314 | CM Sge | 303.3800186769 | 17.6995022524 | 14.5925 | 5869 | 1.5750 | 4939.34 | -0.45 |
| 1315 | HH Pav | 303.4949062372 | -72.4490258055 | 15.9658 | 6391 | 1.2460 | 7506.47 | 0.34 |
| 1316 | V0792 Aql | 303.7600140718 | -0.4830831033 | 14.6855 | 6669 | 1.2552 | 8612.59 | -1.25 |

| # | RRab | R.A. (J2000) | Decl. (J2000) | $G_{mag}$ | $T_{eff}$ | $A_G$ (mag) | d (pc) | $M_G$ |
|---|---|---|---|---|---|---|---|---|
| 1317 | V1102 Aql | 303.7667357943 | -1.9413070902 | 14.8666 | 5908 | 1.9675 | 4935.35 | -0.57 |
| 1318 | HM Pav | 303.9154933238 | -68.9698585492 | 15.7308 | 6777 | 0.6033 | 7366.48 | 0.79 |
| 1319 | IN Tel | 304.5919985319 | -53.1779637616 | 14.7432 | 6650 | 1.6170 | 5719.52 | -0.66 |
| 1320 | V0525 Aql | 304.9593801586 | -4.2950500680 | 12.9570 | 5900 | 1.6690 | 2494.07 | -0.70 |
| 1321 | NO Tel | 305.1547890887 | -53.7283978444 | 14.1848 | 6373 | 1.5763 | 4878.54 | -0.83 |
| 1322 | II Pav | 305.4165813591 | -56.6839984900 | 15.7130 | 6903 | 1.6405 | 6393.06 | 0.04 |
| 1323 | CN Cap | 305.4748936160 | -16.4510447791 | 14.7617 | 6224 | 1.5960 | 4961.71 | -0.31 |
| 1324 | V0909 Aql | 305.4974773001 | -4.6968328710 | 14.5378 | 6794 | 1.6040 | 5352.82 | -0.71 |
| 1325 | HW Pav | 305.5007281523 | -68.2876762387 | 15.3473 | 6815 | 1.5305 | 6463.17 | -0.24 |
| 1326 | IT Tel | 305.6682522150 | -55.4625122718 | 14.1575 | 6095 | 1.8930 | 6183.91 | -1.69 |
| 1327 | AE Oct | 305.7301569709 | -76.0680021083 | 13.9928 | 5954 | 1.7783 | 4106.03 | -0.85 |
| 1328 | V1646 Sgr | 305.7864187442 | -31.2950267633 | 12.0501 | 6512 | 1.5257 | 1824.19 | -0.78 |
| 1329 | V0910 Aql | 305.7987123632 | -1.5659571912 | 14.4477 | 6264 | 1.6233 | 4871.79 | -0.61 |
| 1330 | IM Pav | 305.9247147817 | -57.5597213412 | 15.3759 | 6669 | 1.6655 | 6649.24 | -0.40 |
| 1331 | V2273 Sgr | 306.0611676017 | -40.8697486041 | 14.1740 | 6030 | 1.6483 | 4913.73 | -0.93 |
| 1332 | IK Pav | 306.1125409522 | -71.2264996739 | 15.6606 | 6520 | 1.7060 | 7456.07 | -0.41 |
| 1333 | IV Tel | 306.1403656824 | -55.6340115996 | 15.3960 | 6394 | 1.6170 | 5199.26 | 0.20 |
| 1334 | IW Tel | 306.3089345000 | -56.2766366785 | 15.2992 | 6040 | 1.5655 | 7588.36 | -0.67 |
| 1335 | IX Tel | 306.4607126447 | -52.6974517826 | 13.7152 | 6752 | 1.5980 | 5117.17 | -1.43 |
| 1336 | V2275 Sgr | 306.5728256117 | -43.4534977010 | 13.9720 | 6459 | 1.5417 | 4361.99 | -0.77 |
| 1337 | IN Pav | 306.6560548571 | -71.1510342460 | 15.9761 | 7130 | 1.2900 | 6344.31 | 0.67 |
| 1338 | V2281 Sgr | 307.0511987661 | -42.6021049919 | 13.8059 | 6910 | 1.2710 | 4894.73 | -0.91 |
| 1339 | FG Del | 307.1059200637 | 12.3387552236 | 14.5081 | 6310 | 1.6010 | 8336.99 | -1.70 |
| 1340 | SX Cap | 307.1680191553 | -12.4754880362 | 12.9253 | 6816 | 1.2813 | 3442.30 | -1.04 |
| 1341 | V0417 Pav | 307.1840670854 | -64.7183968021 | 14.3620 | 6580 | 1.6355 | 5216.22 | -0.86 |
| 1342 | AG Oct | 307.4277577033 | -75.7018285028 | 15.8235 | 6314 | 1.6940 | 6521.97 | 0.06 |
| 1343 | V1391 Cyg | 307.5703920548 | 40.2820613405 | 15.4247 | 3640 | 2.5965 | 813.66 | 3.28 |
| 1344 | IT Pav | 307.6972625286 | -60.0047591480 | 13.7780 | 6453 | 1.3247 | 4687.42 | -0.90 |
| 1345 | CV Del | 307.7259672553 | 16.5429501675 | 14.0298 | 6217 | 1.6095 | 5433.41 | -1.26 |
| 1346 | AH Oct | 307.8453732589 | -76.3170162658 | 15.6996 | 6512 | 1.6163 | 6469.25 | 0.03 |
| 1347 | HM Del | 307.9468861252 | 18.8044873776 | 15.7249 | 6276 | 1.5007 | 7534.75 | -0.16 |
| 1348 | SZ Ind | 308.1236149622 | -55.4269940620 | 14.6366 | 6306 | 1.7440 | 6487.47 | -1.17 |
| 1349 | V0341 Aql | 308.1314932502 | 0.5852912673 | 10.9478 | 6207 | 1.6153 | 1160.02 | -0.99 |
| 1350 | IX Pav | 308.2741631762 | -61.2222368689 | 14.0611 | 6313 | 1.6370 | 5712.68 | -1.36 |
| 1351 | PU Del | 308.3653970447 | 4.6523582280 | 14.2297 | 6419 | 1.7202 | 5759.18 | -1.29 |
| 1352 | IU Pav | 308.4028332405 | -71.7796091201 | 14.4609 | 6631 | 1.5415 | 5016.54 | -0.58 |
| 1353 | IZ Pav | 308.8549176348 | -69.2242060776 | 14.5988 | 6607 | 1.4545 | 5747.69 | -0.65 |
| 1354 | DE Del | 308.8989078273 | 15.5657303492 | 15.2633 | 6279 | 1.6815 | 5958.88 | -0.29 |
| 1355 | TY Ind | 309.0756674366 | -59.0290358885 | 15.3203 | 5997 | 1.7732 | 5366.93 | -0.10 |
| 1356 | DH Del | 309.2436002973 | 11.5866444217 | 14.4312 | 6017 | 1.6955 | 4475.69 | -0.52 |
| 1357 | TT Mic | 309.3168766419 | -44.7696714954 | 14.3692 | 6139 | 1.7250 | 6156.62 | -1.30 |
| 1358 | UU Ind | 309.7248220551 | -58.0661802122 | 14.3776 | 6146 | 1.6260 | 5250.75 | -0.85 |
| 1359 | UV Ind | 309.7865434530 | -54.4216576431 | 13.9546 | 6752 | 1.6337 | 4643.72 | -1.01 |
| 1360 | BC Vul | 309.7989122361 | 25.6592140804 | 13.6247 | 6178 | 1.5400 | 3667.87 | -0.74 |
| 1361 | UX Ind | 309.9205593764 | -54.4590974649 | 14.8418 | 6453 | 1.6335 | 6038.05 | -0.70 |
| 1362 | GN Del | 310.0600434086 | 15.8565954834 | 15.6868 | 6600 | 1.6280 | 6902.21 | -0.14 |
| 1363 | VV Ind | 310.1242459269 | -53.2817434104 | 13.5787 | 6604 | 1.4550 | 4484.98 | -1.14 |
| 1364 | UV Mic | 310.4264828785 | -44.3055142342 | 14.1950 | 7114 | 1.2620 | 5474.86 | -0.76 |
| 1365 | KM Pav | 311.0694480146 | -71.1574101432 | 15.7203 | 6133 | 1.5930 | 5834.52 | 0.30 |
| 1366 | FQ Del | 311.2117384724 | 18.8934804576 | 15.2681 | 6491 | 1.5150 | 5583.13 | 0.02 |
| 1367 | DU Del | 311.4082432312 | 11.6123675165 | 15.3560 | 6295 | 1.5580 | 7064.33 | -0.45 |
| 1368 | VV Mic | 311.4328617764 | -42.2932588046 | 15.1025 | 6797 | 1.5130 | 6972.62 | -0.63 |
| 1369 | WZ Ind | 311.5805603450 | -53.8027265489 | 15.0118 | 6264 | 1.5200 | 6065.54 | -0.42 |
| 1370 | KP Pav | 311.6261457188 | -73.1372212160 | 14.1018 | 6287 | 1.7110 | 4273.54 | -0.76 |
| 1371 | VX Mic | 311.7124618466 | -44.9682997032 | 14.4566 | 7037 | 1.5730 | 6113.70 | -1.05 |
| 1372 | KQ Pav | 311.8641254915 | -73.0911748423 | 15.4958 | 6603 | 1.6210 | 7731.45 | -0.57 |
| 1373 | KT Pav | 312.0698773968 | -70.9108809242 | 16.4723 | 5751 | 1.8565 | 5424.04 | 0.94 |
| 1374 | DO Aqr | 312.2394052871 | 1.2760593886 | 14.3487 | 6517 | 1.6150 | 4390.84 | -0.48 |
| 1375 | FM Cep | 312.8010193509 | 69.2765246779 | 13.9055 | 4382 | 0.7280 | 3893.80 | 0.23 |
| 1376 | AB Ind | 312.8222140995 | -46.1294302577 | 16.0962 | 4872 | 0.1320 | 957.90 | 6.06 |

| # | RRab | R.A. (J2000) | Decl. (J2000) | $G_{mag}$ | $T_{eff}$ | $A_G$ (mag) | d (pc) | $M_G$ |
|---|---|---|---|---|---|---|---|---|
| 1377 | DP Aqr | 312.9555355382 | 2.3906678978 | 14.4875 | 6465 | 1.7530 | 5147.46 | -0.82 |
| 1378 | V0509 Vul | 313.0246592679 | 21.8227955945 | 13.4353 | 6106 | 1.7533 | 3010.84 | -0.71 |
| 1379 | LX Del | 313.0790948857 | 7.1463412064 | 13.5688 | 6379 | 1.6297 | 3396.42 | -0.72 |
| 1380 | FK Vul | 313.1291850063 | 22.4365909886 | 12.5283 | 6227 | 1.8770 | 2026.13 | -0.88 |
| 1381 | BV Del | 313.2914861930 | 16.1470157978 | 13.1727 | 5942 | 1.6657 | 2616.82 | -0.58 |
| 1382 | CG Aqr | 313.7877950000 | -12.3456641114 | 14.2800 | 6347 | 1.6670 | 4825.53 | -0.80 |
| 1383 | FS Del | 314.0612919720 | 16.6719427758 | 15.0301 | 6672 | 1.5357 | 5200.28 | -0.09 |
| 1384 | XX Mic | 314.0640407066 | -38.9745595679 | 15.8864 | 6537 | 1.5800 | 5470.84 | 0.62 |
| 1385 | FT Del | 314.1236847607 | 16.3964380022 | 14.8893 | 6427 | 1.7068 | 5509.67 | -0.52 |
| 1386 | UZ Cap | 314.1541152747 | -15.6723295028 | 13.3024 | 5451 | 0.0848 | 438.14 | 5.01 |
| 1387 | EM Del | 314.3694908903 | 10.0426185756 | 14.3001 | 5942 | 1.5120 | 5151.12 | -0.77 |
| 1388 | BY Aqr | 314.5237329019 | -11.0426392902 | 13.4134 | 6174 | 1.8590 | 3981.82 | -1.45 |
| 1389 | XZ Mic | 314.7283076696 | -38.9414647958 | 13.0498 | 6809 | 1.6275 | 3262.02 | -1.15 |
| 1390 | CH Aqr | 314.8522833204 | -14.0315066443 | 15.3136 | 6564 | 1.2630 | 6064.80 | 0.14 |
| 1391 | AG Ind | 314.9095169021 | -59.3765601784 | 13.5559 | 6449 | 1.3070 | 3639.44 | -0.56 |
| 1392 | YY Mic | 315.0026262909 | -41.8453003783 | 14.5854 | 6956 | 1.2480 | 8205.72 | -1.23 |
| 1393 | YZ Mic | 315.0277806678 | -40.5696327725 | 13.5952 | 6662 | 1.5200 | 3328.37 | -0.54 |
| 1394 | KV Pav | 315.1793796833 | -74.2678954864 | 13.7226 | 6584 | 1.6482 | 3919.75 | -0.89 |
| 1395 | KW Pav | 315.3007617354 | -72.9050080706 | 15.0331 | 6292 | 1.7993 | 7519.17 | -1.15 |
| 1396 | AD Mic | 315.4187366692 | -38.9821445488 | 14.6684 | 6688 | 1.3310 | 4764.57 | -0.05 |
| 1397 | NS Pav | 315.7176528995 | -74.3286807124 | 13.4876 | 6263 | 1.6385 | 3385.35 | -0.80 |
| 1398 | CL Aqr | 316.5189282053 | -14.4004972554 | 15.0070 | 6623 | 1.5692 | 5558.47 | -0.29 |
| 1399 | V0398 Peg | 317.2414493514 | 15.9487024719 | 13.8627 | 6389 | 1.7227 | 3786.75 | -0.75 |
| 1400 | NY Aqr | 317.2539581326 | -14.4564626027 | 13.5057 | 6508 | 1.6398 | 3382.97 | -0.78 |
| 1401 | V0420 Pav | 317.3226082203 | -74.1488429621 | 14.0314 | 6638 | 1.6050 | 4950.54 | -1.05 |
| 1402 | WW Cap | 317.7396981946 | -15.6583194626 | 15.5451 | 6209 | 1.8133 | 5915.57 | -0.13 |
| 1403 | V Ind | 317.8745917158 | -45.0745530576 | 10.0494 | 6550 | 1.5570 | 655.87 | -0.59 |
| 1404 | DH Cap | 317.9811282668 | -21.1615162682 | 14.0866 | 6647 | 1.4815 | 3236.61 | 0.05 |
| 1405 | CC Aqr | 318.3836597844 | -8.6264792116 | 14.5978 | 6749 | 1.6270 | 5011.68 | -0.53 |
| 1406 | CT Aqr | 319.0815155888 | -12.5934374577 | 15.3605 | 6956 | 1.4618 | 6222.77 | -0.07 |
| 1407 | Z Mic | 319.0946027981 | -30.2841854629 | 11.4970 | 5946 | 1.7117 | 1190.26 | -0.59 |
| 1408 | CU Aqr | 319.5977133762 | -11.2461910855 | 15.7016 | 6175 | 0.8515 | 6065.14 | 0.94 |
| 1409 | ST Equ | 320.1043870335 | 10.5668701263 | 15.7624 | 7129 | 1.2550 | 4774.63 | 1.11 |
| 1410 | V0429 Peg | 320.1425363919 | 18.6214663410 | 11.8667 | 6653 | 1.5865 | 1674.89 | -0.84 |
| 1411 | AN Ind | 320.3318433152 | -46.7350583699 | 14.6273 | 6669 | 1.5850 | 5237.01 | -0.55 |
| 1412 | BL Peg | 320.7479681669 | 23.8922474992 | 14.2540 | 6340 | 1.6667 | 5445.60 | -1.09 |
| 1413 | DK Mic | 320.9021326122 | -39.8260434510 | 14.6702 | 6600 | 1.4760 | 5758.06 | -0.61 |
| 1414 | FN Peg | 321.9581240380 | 13.5110002933 | 14.5640 | 7106 | 1.3032 | 4373.94 | 0.06 |
| 1415 | FP Peg | 322.2926770068 | 10.9870731318 | 15.2273 | 6780 | 0.4707 | 2125.88 | 3.12 |
| 1416 | V1725 Cyg | 322.4826898229 | 41.7622549167 | 15.0655 | 5341 | 2.0600 | 5437.32 | -0.67 |
| 1417 | FR Peg | 322.5060807396 | 13.2815181085 | 15.9588 | 5976 | 1.7630 | 4218.74 | 1.07 |
| 1418 | V0438 Peg | 322.5730760270 | 18.7326387419 | 13.1774 | 6058 | 1.5980 | 5007.41 | -1.92 |
| 1419 | FS Peg | 322.5862924623 | 9.5856000361 | 16.0636 | 6974 | 1.3637 | 5436.06 | 1.02 |
| 1420 | FW Peg | 323.2142269256 | 12.8918720714 | 13.9866 | 5764 | 1.8498 | 3982.28 | -0.86 |
| 1421 | DZ Gru | 323.3852264742 | -49.3105474880 | 14.0914 | 6626 | 1.5163 | 5209.17 | -1.01 |
| 1422 | RY Oct | 324.0389608982 | -77.3037602938 | 12.0773 | 6085 | 1.8067 | 1712.01 | -0.90 |
| 1423 | BT Peg | 324.0996608908 | 26.3019708488 | 12.1088 | 6182 | 1.8410 | 1676.28 | -0.85 |
| 1424 | EE Gru | 324.6235992525 | -49.0147937381 | 13.6891 | 7065 | 1.3267 | 4658.04 | -0.98 |
| 1425 | CI Peg | 325.5940589397 | 22.1971762852 | 13.9972 | 5881 | 1.6452 | 5165.73 | -1.21 |
| 1426 | CS Peg | 326.7192877557 | 24.7207929805 | 13.1486 | 6580 | 1.5467 | 3263.38 | -0.97 |
| 1427 | RS Oct | 326.8204231152 | -87.6517836218 | 12.8575 | 6053 | 1.7733 | 2475.54 | -0.88 |
| 1428 | CV Peg | 327.0534922117 | 22.2433093766 | 13.5204 | 6217 | 1.5577 | 3471.15 | -0.74 |
| 1429 | AP Oct | 327.4407818899 | -81.9730220170 | 14.1148 | 5725 | 1.8760 | 3723.59 | -0.62 |
| 1430 | V1434 Cyg | 329.5483446748 | 55.1703389478 | 14.6656 | 4228 | 0.9922 | 4770.85 | 0.28 |
| 1431 | V0684 Cyg | 329.9415044394 | 47.2677068346 | 15.0158 | 5450 | 2.2390 | 5420.51 | -0.89 |
| 1432 | AD PsA | 330.8482853075 | -29.4811660420 | 12.8440 | 6585 | 1.5307 | 2853.65 | -0.96 |
| 1433 | AI Aqr | 331.7580617474 | -20.7087781886 | 13.9632 | 6345 | 1.7878 | 3500.14 | -0.55 |
| 1434 | TT Gru | 332.6067445198 | -43.6935262106 | 13.4026 | 6979 | 1.2797 | 2988.56 | -0.25 |
| 1435 | FX Aqr | 333.2552568530 | -1.7279921396 | 12.7568 | 5986 | 1.7940 | 1976.13 | -0.52 |
| 1436 | AR Aqr | 333.3349679899 | -24.7174568649 | 13.2422 | 6134 | 0.1710 | 789.36 | 3.58 |

| # | RRab | R.A. (J2000) | Decl. (J2000) | $G_{mag}$ | $T_{eff}$ | $A_G$ (mag) | d (pc) | $M_G$ |
|---|---|---|---|---|---|---|---|---|
| 1437 | TZ Gru | 334.3357293404 | -50.5763953774 | 13.4989 | 6991 | 1.6285 | 4155.40 | -1.22 |
| 1438 | GG Aqr | 334.4708200390 | -0.0931232985 | 14.4045 | 6685 | 1.8337 | 4301.69 | -0.60 |
| 1439 | XZ Lac | 334.7427418601 | 49.7396409138 | 13.3329 | 5790 | 2.0577 | 3763.06 | -1.60 |
| 1440 | UW Gru | 335.0547116856 | -54.5580148388 | 13.3240 | 6617 | 1.5700 | 3119.73 | -0.72 |
| 1441 | CQ Lac | 335.2933456955 | 39.7201457145 | 12.4265 | 6268 | 1.6717 | 2014.34 | -0.77 |
| 1442 | GR Aqr | 336.6486660960 | -8.7606813930 | 15.7904 | 6104 | 1.8830 | 5314.96 | 0.28 |
| 1443 | AE Peg | 336.8397331863 | 16.8046398754 | 12.6502 | 6835 | 1.6810 | 2648.29 | -1.15 |
| 1444 | BN Aqr | 336.9531963960 | -7.4838142359 | 12.4511 | 7006 | 1.3260 | 2353.73 | -0.73 |
| 1445 | AA PsA | 337.5313381157 | -28.1699914437 | 13.0545 | 6510 | 1.5513 | 2825.95 | -0.75 |
| 1446 | DH Ind | 338.0483568164 | -68.3155469163 | 14.0347 | 6700 | 1.7140 | 3960.92 | -0.67 |
| 1447 | EP Tuc | 338.6114153484 | -56.5902672347 | 13.1889 | 6385 | 1.6663 | 3176.63 | -0.99 |
| 1448 | AA Aqr | 339.0160491967 | -10.0152576850 | 12.9274 | 6734 | 1.6285 | 2437.43 | -0.64 |
| 1449 | WW Gru | 339.2676781055 | -47.1852950613 | 15.6506 | 6931 | 0.1657 | 7473.04 | 1.12 |
| 1450 | BG Aqr | 339.3586362985 | -18.3716593594 | 14.6885 | 7567 | 1.2540 | 5027.83 | -0.07 |
| 1451 | BH Aqr | 339.7452815098 | -19.8074286702 | 13.6058 | 6498 | 1.5865 | 3863.40 | -0.92 |
| 1452 | HH Aqr | 340.3812406340 | -6.4774675972 | 12.2751 | 6707 | 1.5562 | 1925.46 | -0.70 |
| 1453 | AB PsA | 341.1731200105 | -31.9687583030 | 13.2620 | 7030 | 1.3560 | 2775.90 | -0.31 |
| 1454 | XX Gru | 341.2473345479 | -50.0956250641 | 16.3559 | 6673 | 0.8930 | 2783.71 | 3.24 |
| 1455 | AB Gru | 342.7740504381 | -47.3112258006 | 16.9141 | 6808 | 1.3010 | 3980.40 | 2.61 |
| 1456 | V0354 Aqr | 343.1979723043 | -24.7037743595 | 13.2990 | 6951 | 1.2760 | 3825.17 | -0.89 |
| 1457 | AD Gru | 343.2571338897 | -44.5442155008 | 16.0250 | 6581 | 1.5070 | 5315.34 | 0.89 |
| 1458 | V0550 Peg | 343.3461300742 | 8.7685287602 | 13.4180 | 6419 | 1.6827 | 3350.90 | -0.89 |
| 1459 | BO Aqr | 343.5344610709 | -12.3606062606 | 12.1691 | 6498 | 1.6177 | 2268.05 | -1.23 |
| 1460 | AO Gru | 347.0450770166 | -48.3691858775 | 13.7011 | 7222 | 1.0780 | 4303.16 | -0.55 |
| 1461 | CV Scl | 347.3771570159 | -35.7880310805 | 12.3727 | 6835 | 1.6387 | 2250.84 | -1.03 |
| 1462 | YY Tuc | 347.7523741456 | -58.3353471279 | 12.0723 | 6498 | 1.7747 | 2170.96 | -1.39 |
| 1463 | AP Gru | 348.5519644297 | -50.6533412544 | 13.3216 | 6391 | 1.6710 | 4083.16 | -1.40 |
| 1464 | V0356 Aqr | 348.9579777199 | -23.0035794510 | 12.6119 | 7062 | 1.0893 | 2014.79 | 0.00 |
| 1465 | AV Oct | 348.9635145820 | -81.4915404682 | 15.6067 | 6046 | 1.7393 | 6433.23 | -0.17 |
| 1466 | DE And | 349.3589274426 | 48.5517100997 | 13.6881 | 5770 | 1.6178 | 2927.02 | -0.26 |
| 1467 | DN Aqr | 349.8216727273 | -24.2163395015 | 11.1389 | 6668 | 1.4280 | 1482.76 | -1.14 |
| 1468 | V0408 And | 350.4809676634 | 48.9628056530 | 15.8358 | 5958 | 1.7860 | 5674.12 | 0.28 |
| 1469 | UZ Scl | 350.6959998381 | -30.1193413676 | 12.3686 | 6084 | 1.6780 | 2816.88 | -1.56 |
| 1470 | V0361 Aqr | 350.8917822570 | -8.0124929693 | 13.9579 | 6967 | 1.6248 | 3976.64 | -0.66 |
| 1471 | V0424 And | 351.1652863010 | 49.6001901961 | 15.6004 | 6053 | 1.1350 | 7663.05 | 0.04 |
| 1472 | V0410 And | 352.9883283560 | 48.9882904753 | 15.6369 | 6130 | 1.8490 | 6455.83 | -0.26 |
| 1473 | DM And | 353.0029844527 | 35.1969168476 | 11.8645 | 6516 | 1.7475 | 1572.52 | -0.87 |
| 1474 | HR Aqr | 353.9483027618 | -10.9903853836 | 13.9200 | 6562 | 1.3030 | 4176.39 | -0.49 |
| 1475 | IN Psc | 354.1559230715 | -2.2124159779 | 13.0634 | 6980 | 1.4557 | 3739.13 | -1.26 |
| 1476 | CN Scl | 355.0955002733 | -38.3161355075 | 13.3103 | 6459 | 1.6083 | 2993.91 | -0.68 |
| 1477 | BO Tuc | 355.7487070233 | -74.1347415241 | 15.8751 | 6855 | 1.6217 | 7233.00 | -0.04 |
| 1478 | V0610 Peg | 356.4616195461 | 15.9884366740 | 14.2883 | 6346 | 1.6182 | 4245.36 | -0.47 |
| 1479 | KL Peg | 356.7378584906 | 29.8503691383 | 16.9141 | 6358 | 0.2095 | 2848.55 | 4.43 |
| 1480 | XY Tuc | 357.4481291893 | -74.1927660782 | 14.2139 | 6943 | 1.5217 | 4773.33 | -0.70 |
| 1481 | V0419 Peg | 357.5209390498 | 17.8955509828 | 14.3191 | 6127 | 1.8063 | 3906.18 | -0.45 |
| 1482 | V0708 And | 357.6925815154 | 33.3512303481 | 13.1391 | 6483 | 1.7987 | 2590.86 | -0.73 |
| 1483 | QR Cas | 357.7678908993 | 55.6967364636 | 15.2275 | 5314 | 2.2393 | 4184.33 | -0.12 |
| 1484 | V0618 Peg | 359.0031558743 | 10.8884597109 | 12.0453 | 6260 | 1.7785 | 2130.96 | -1.38 |